\documentstyle[12pt]{article}

\date{}

\begin{document}
\newcommand{\ve }[1]{{\mbox{\boldmath $#1$}}}
\newcommand{\vsm }[1]{{\mbox{\footnotesize{\boldmath $#1$}}}}
\newcommand{\cross }[1]{#1 \hspace{-0.4em} /}
\newcommand{\crossm }[1]{#1 \hspace{-0.5em} /}

\def\bx{{\mbox{\boldmath$x$}}}
\def\bk{{\mbox{\boldmath$k$}}}

\def\bp{{\mbox{\boldmath$p$}}}
\def\pb{\mbox{${\bar \psi}$}}
\def\dg{\mbox{$^{\dagger}$}}
\def\bN{{\mbox{\boldmath$N$}}}
\def\bB{{\mbox{\boldmath$B$}}}
\def\bn{{\mbox{\boldmath$n$}}}
\def\bP{{\mbox{\boldmath$P$}}}
\def\bM{{\mbox{\boldmath$M$}}}
\def\bsk{{\mbox{\footnotesize{\boldmath$k$}}}}
\def\br{{\mbox{\boldmath$x$}}}
\def\f {\mbox{${\hat f}$}}
\def\bq{{\mbox{\boldmath$q$}}}
\def\m {\mbox{${\hat m}$}}

{\Large\bf Unitary Transformations in Quantum Field Theory and Bound
States}\\[2mm]
{\large\em A.V.\ Shebeko,$^{a,b}$ M.I.\ Shirokov\phantom{,}$^a$}\\[2mm]
{\small $^a$BLTP, JINR, 141980 Dubna, Russia\\
$^b$NSC Kharkov Institute of Physics \& Technology, 310108 Kharkov, Ukraine}\\[2mm]

{\bf Abstract}\\

Finding the eigenstates of the total Hamiltonian $H$ or its
diagonalization
is the important problem of quantum physics. However, in relativistic quantum
field theory (RQFT) its complete and exact solution is possible for a few
simple models only. Unitary transformations (UT's) considered in this survey
do not diagonalize $H$, but convert $H$ into a form which enables us to find
approximately some $H$ eigenstates.
During the last years there have appeared many papers
devoted to physical applications of such UT's.
Our aim is to present a systematic and self-sufficient exposition of the
UT method. The two general
kinds of UT's are pointed out, distinct variations of each kind being
possible. We consider in detail the problem of finding the simplest $H$
eigenstates for interacting mesons and nucleons using the so--called
``clothing'' UT and Okubo's UT. These UT's allow us to suggest definite
approaches
to the problem of two--particle (deuteron--like) bound states in RQFT.
The approaches are shown to yield the same two--nucleon quasipotentials in the
first nonvanishing approximation.
We demonstrate how the particle mass renormalization can be fulfilled
in the framework of the ``clothing'' procedure. Besides the UT of the
Hamiltonian we discuss the
accompanying UT of the Lorentz boost generators.

\bigskip
\centerline{\bf 1. Introduction}
\bigskip

The so-called unitary transformation (UT) method has the same age as
the quantum theory itself. Its first applications for
constructing Hermitian effective interactions (HEI) can be found in
\cite{VLECK29} and \cite{KEMBLE37} (in this connection, see a review article
\cite{KLEIN74} where the different perturbation expansions of HEI were discussed).
In quantum field theory the first considerations using the UT method were
given by Wentzel\cite{WENT46} and Heitler\cite{HEIT54}.

A number of the schemes for reduction of the exact eigenvalue problem to the
model--space problem via the various UT's were put forward
in 50's (\cite{FST54},\cite{OKU54},\cite{TANI54}) during the extensive
development of the meson theory of nuclear forces (see,e.g.,\cite{NISH56}).
Owing to the work \cite{GAHY76} this approach has proved to be very useful in
studies of electromagnetic (e.m.) interactions with nuclei for nonmesonic
channels (in particular,when constructing the effective operators of meson
exchange currents /see, e.g., \cite{MESONS79}, surveys \cite{SHE87},
\cite{KOMESHE95} and refs. therein/).

Along with this guideline the UT method was used for the formulations of
RQFT in terms of physical or "clothed" particles (\cite{GRESCH58},
\cite{SCH61}, \cite{FAD63}, \cite{FIV70}, \cite{SHIVI74}, see also our
talks at the recent conferences \cite{THES97} and \cite{GRON97}).

By using the Okubo UT the authors of \cite{GLOMU81} constructed effective
generators for the Poincar\'{e} algebra, acting on nucleonic degrees of freedom
only. It was made in the framework of perturbation theory for a simple model
of "spinless" nucleons exchanging scalar mesons. Therefore, we have an
instructive example of how some noncommuting Hermitian operators can be reduced
by one and the same UT to Okubo's block form.

Late 80's and 90's have brought a renewed interest to this area. First of
all, we mean applications (\cite{SATO91},\cite{HAGLO92}) of the UT method to
a covariant treatment of the two--body bound--state problem (cf.\cite{KO92}).
The corresponding transformed Hamiltonian and boost operators do not couple
(in the second
order in meson--nucleon coupling constants) the nucleon (no--meson) subspace
with its complement in the full Fock space  of hadron states. In the
work \cite{KOSHE93} the same method has been employed to derive the effective
nucleon-nucleon and nucleon--antinucleon interactions starting from a field
Hamiltonian with the exchange of $\pi $, $\rho $, $\omega $ and $\sigma $
mesons. Then, within the Hartree approximation,these effective interactions
have been introduced for describing the saturation properties of nuclear
matter. Note also recent explorations \cite{PLGARI92,EDGARI96} of
the hard--core problem in the theory of nuclear forces.

In papers (\cite{LESHE89},\cite{LESHE95}) some extensions of the UT method
have been suggested for constructing effective current operators in the
theory of photomeson processes on nuclei (see also ref.\cite{LESHE96} where
one can find the calculations of the four structure functions for pion
electroproduction on the deuteron near threshold). Recently,the UT method has
been used \cite{FUZH96} for deriving effective two--particle one--meson exchange
potentials in the instant and front forms of relativistic quantum mechanics.
At last, with the aid of a modification of the method the authors of \cite{SALE96}
have proposed a meson--exchange model for $\pi $N scattering and
$\gamma $N--$\pi $N reaction. Certainly, one may say that nowadays
the UT method has survived its second birth.

This review is focused upon an application of UT's in
RQFT and aimed at an approximate treatment of the physical vacuum, the observable
one--particle and two--particle bound and scattering states. These states have
a common feature,
viz., they do not change in time. The simplest example is the state without
observable particles, i.e., the physical vacuum. Other examples give free
particles with a definite four--momentum (elementary particles,
atoms and nuclei in their ground states).

Our point of departure is that such states should be eigenvectors of a
Hamiltonian $H$, which are stationary. In the context, the "bare" states, i.e.,
eigenvectors of a free part $H_0$ of the total Hamiltonian $H$, are
inappropriate for theoretical description of the physical objects. Firstly,
they change in time although the possible transitions to other states are
virtual and can be considered undetectable because of their small
probabilities. Secondly, $H_0$ has no eigenvectors which would
correspond to bound states (e.g., the hydrogen atom or the deuteron).

From this postulate it follows such a definition for a bound
state of the deuteron--type in RQFT : it should be described by
a proper eigenvector of the total Hamiltonian of the theory. In other words,
the deuteron problem should be reduced to an exact or approximate solution
of the $H$ eigenvalue problem. There are various approaches to this
problem, viz.,  within the Bethe--Salpeter formalism, via its
three--dimensional versions, etc. (see, e.g., \cite{KADMIR72}, \cite{NAKAN88},
and refs. therein). However, the links between the respective "wavefunctions"
(for instance, the Bethe--Salpeter amplitudes for bound states) and
the exact or approximate $H$ eigenvectors are very sophisticated. According
to our postulate similar links in the framework of the approach developed
below are much more direct and transparent (see Sect.4). In addition, the
description of particle scattering can be reduced to calculation of
the relevant scattering wavefunctions.

The reasons for employing just RQFT when describing bound and scattering
states are well known.
Being not satisfied with the multitude of disconnected phenomenological
explanations we strive for a unified description of Nature.
RQFT's are the best known candidates for unified theories. Firstly, they
give a qualitative and natural consideration of particle creation and
destruction. Secondly, local RQFT's ensure in a sense the relativistic
causality unlike phenomenological approaches.

General idea of constructing the UT's in question can be formulated in
the following way.

Any UT of $H$ can be considered as a transformation expressing the $H$--matrix
determined
with respect to a new basis in the Hilbert (Fock) space of a given physical
system
through the $H$--matrix with respect to the old basis in the same space
(details see in Sect. 6). Normally, the latter is composed of eigenvectors
(the "bare" states) of
the free part $H_0$ in the partition $H = H_0 + V$, where $V$ represents the
interaction between the fields involved. In general, the new $H$--matrix turns
out to be more complicated. For example, in the case with Yukawa
coupling $V \sim \int \pb(\bx) \gamma_5 \psi(\bx) \phi(\bx) d\bx $,
that conserves the baryon number and may change the meson number merely by
unit, nonzero elements of the initial $H$--matrix are either diagonal, or
near diagonal (e.g., they can be of the $ \langle N^{\prime} \mid H
\mid \pi N \rangle $--kind). In a new representation for this matrix all
its elements
can be nonzero (e.g., along with the aforementioned elements one can meet
the elements $ \left ( N^{\prime} N^{\prime} \mid H \mid NN ) \right )$,
$ \left ( \pi^{\prime} N^{\prime} \mid H \mid \pi N \right )$,
$ \left ( N^{\prime} \mid H \mid \pi \pi N \right )$, and the others in
which the meson number can be arbitrarily altered\footnote{In order to
distinguish the matrix elements with respect to the new basis we employ
round brackets.}). The elements
$ \left ( N^{\prime} N^{\prime} \mid H \mid NN \right )$ contribute to
the $NN$ scattering even in the first order of perturbation theory.
Of course, there are other elements of the transformed $H$--matrix, which
contribute to the $NN$ scattering as well, but in higher orders.

Let us impose the following constraint upon the transformation associated
with the basis change: in the new $H$--matrix the $NN$ interaction should
be described only through the elements
$ \left ( N^{\prime} N^{\prime} \mid H \mid N N \right )$,
i.e., all the other elements which could contribute to the $NN$ scattering
should be zero. This means that the $H$--matrix must be reduced to
a block diagonal form where  specific off--diagonal blocks consist of
zero elements (other clarifications can be found in Sect.6).
This requirement (the Okubo condition\cite{OKU54}) can be replaced by
the other constraints, more flexible and easier realizable.

Our aim is to present a systematic exposition of the UT method. A main attention
within the method is paid to the two approaches, viz., the clothing procedure
and the blockdiagonalization after Okubo.
We strive for a self--sufficient presentation which may be understood without
referring to original papers. As a rule, we avoid to point out mistakes and
obscurities in the latter (but sometimes allow ourselves to note their tacit
assumptions).

This review is organized as follows.

In Sect.2 we consider the problem of
obtaining the simplest $H$ eigenstates. The lowest $H$ eigenstate $\Omega$
can be juxtaposed to the state without observable particles
(the physical vacuum).
Further, we seek one--particle--like $H$ eigenstates which have "bare"
partners (e.g., in the case of the interacting pion and nucleon fields
the "bare" one--meson $a\dg (\bk) \Omega_0$ and one--nucleon
$b\dg (\bp) \Omega_0$ states, $\Omega_0$ being the "bare" vacuum).
These $H$ eigenstates may be called "clothed" \cite{GRESCH58} because
meson--nucleon interaction is taken into account when constructing such
states. At this point, instead of the usual "bare" creation--destruction
operators $a\dg(\bk)$, $a(\bk)$, ... we introduce the "clothed" operators
$a_c\dg(\bk)$, $a_c(\bk)$, ..., so that the physical one--meson state is
described by the vector $a_c\dg(\bk) \Omega$.

While finding the above $H$ eigenstates one has to determine a "clothing" UT
such that the transformed Hamiltonian does not contain the interaction terms
which correspond to some virtual energy--nonconserving processes
(e.g., $N \rightarrow \pi N$, $\pi \rightarrow N \bar N$ ). These terms are
called "bad".

As one might expect the mass of the clothed particle turns out to be inequal
to the respective bare mass. The related problem of mass renormalization is
explored in Subsect. 2.5.

The clothing transformation of the Hamiltonian results in its representation
through the clothed creation--destruction operators. A basic task of Sect.3
is to investigate how transformed generators of the Lorentz boosts depend on
the clothed operators. We show that the bad terms can be removed via
the same clothing transformation simultaneously from the total Hamiltonian and
these generators. A consequence of such reduction is that the "clothed"
vacuum $\Omega $ remains invariant under Lorentz transformations
(unlike the bare vacuum $\Omega_0$) while the clothed one--particle states
have the proper transformation properties (in particular, their momenta being
suitably changed).

Along with the clothed one--particle states the total Hamiltonian can have
one--particle--like states (e.g., bound states of the deuteron--type) without
any bare partners. In Sect.4 within our clothing procedure we suggest an
approximate way of finding such states. The resulting bound--state equations
resemble the Schroedinger equation  for stationary states in
nonrelativistic quantum mechanics, where the usual potentials are
replaced by
the interactions between clothed particles (the quasipotentials). Explicit
analytical expressions for the model nucleon--nucleon and meson--nucleon
quasipotentials are given in Subsect.4.3.

In Sect. 5 we discuss some modifications and extensions of the clothing
approach.

Sect.6 starts with a definition of the similarity transformation
$H \rightarrow U^{-1} H U$ ($U^{-1} = U\dg$) as a transformation of the
matrix $ \langle n^{\prime} \mid H \mid n \rangle$ (determined with respect
to a set of orthonormal vectors $\mid n \rangle$) into the matrix
$\left ( \nu^{\prime} \mid H \mid \nu \right )$ with respect to another
basis $\mid \nu )$.

The Okubo condition on $U$ and the corresponding decoupling equation from [7]
are presented in Subsect. 6.2.

Two operator kinds of the Okubo UT are considered in the framework of the
Okubo approach using the new particle creation--destruction
operators which correspond to the clothed operators determined in Sect. 2.
This allows us to establish the relation to the clothing procedure and its
modifications (cf. Sects. 2 and 5).

Some original results of this work are summarized in Sect. 7.

More technical details and some auxiliary calculations are deferred to
Appendix A. Appendix B is devoted to some mathematical aspects related to
the UT method. We suggest an algebraic approach which enables to treat the
clothing UT as an element of an algebra lacking any operator representation.
At last, Appendix C exemplifies an explicit solution of Okubo's decoupling
equation for a simplified field model.

\bigskip
\centerline{\bf 2. Clothed Particles in Quantum Field Theory}
\bigskip

The notion of clothed particles will be considered using the following
model: a spinor (fermion) field $\psi$ interacts with a neutral
pseudoscalar meson field $\phi$ by means of the Yukawa coupling. The
model Hamiltonian is $H = H_0 + V $ where
$$
H_0 = \int \pb(\bx)[-i\ve{\gamma} \ve{\nabla} + m_0 ] \psi(\bx) d\bx +
{1\over 2} \int\ \left[{\pi}^2(\bx) + {\ve{\nabla}\phi(\bx)}^2 +
{\mu_0}^2{\phi}^2(\bx) \right] d\bx
\eqno(2.1)
$$
$$
V = ig \int\pb(\bx)\gamma _5 \psi(\bx)\phi(\bx) d\bx. \eqno(2.2)
$$
For simplicity, we do not employ a more refined form of H properly
symmetrized in the fields involved (see, e.g., \cite{SCH61,ZUB80}).
This model has much in common with more realistic models for
the interacting fields (e.g., the nucleon isodoublet (p,n)
interacting with the meson isotriplet ($\pi ^+$, $\pi ^0$, $\pi ^-$).

The Hamiltonian can be expressed in terms of bare destruction (creation)
operators $a(\ve{k})$ $(a\dg (\ve{k}))$, $b(\ve{p},r)$ $(b\dg (\ve{p},r))$
and $d(\ve{p},r)$ $(d\dg (\ve{p},r))$ of the meson, the fermion and the
antifermion (see Eqs.(2.8) and (2.16)).
Here $\ve{k}$ and $\ve{p}$ denote the momenta; $r$ is the spin index.
An exact definition of the bare operators which we use will be given in
Subsect. 2.2. In what follows, the set of all these operators is denoted
by a symbol $a$, while $a_p$ is used for one of them. The state without
bare particles $\Omega _0$ and the bare one-particle states $a\dg (\ve{k})\
\Omega _0$, $b\dg (\ve{p},r)\ \Omega _0$ and $d\dg (\ve{p},r)\ \Omega _0$ are
not $H$ eigenvectors.

\bigskip
\centerline{\em 2.1 Clothed Particle Operators and States}
\bigskip

Now, we introduce new destruction(creation) operators
$$
a_c(\bk) (a_c\dg(\bk)),\ b_c(\bp,r) (b_c\dg(\bp,r)) \mbox{ and }\
d_c(\bp,r) (d_c\dg(\bp,r)) \quad            { \forall \bk, \bp, r}
\eqno(2.3)
$$
with the following properties:

i) The physical vacuum (the $H$ lowest eigenstate) must coincide
with a new no--particle state $\Omega$, i.e., the state that obeys the equations
$$
a_{\rm c} (\ve{k}  ) \mid \Omega \rangle =
b_{\rm c} (\ve{p},r) \mid \Omega \rangle =
d_{\rm c} (\ve{p},r) \mid \Omega \rangle = 0 \quad  { \forall \bk, \bp, r}
\eqno(2.4)
$$
$$
\langle \Omega \mid \Omega \rangle = 1
$$

ii) New one--particle states $a_c\dg(\bk) \Omega $
etc. are $H$ eigenstates as well

iii) The spectrum of indices that enumerate the new operators
must be the same as that for the bare ones (this requirement has been used
when writing Eq.(2.3))

iv) The new operators satisfy the same commutation rules as do their
bare partners. For instance,
$$
[ a_{\rm c} (\ve{k}), a_{\rm c}\dg (\ve{k'}) ] =
\delta(\ve{k}-\ve{k'})\ ,\
\left\{ b_{\rm c} (\ve{p},r), b_{\rm c}\dg (\ve{p'},r') \right\} =
\left\{ d_{\rm c} (\ve{p},r), d_{\rm c}\dg (\ve{p'},r') \right\} =
\delta _{rr'} \delta(\ve{k}-\ve{k'})\ . \eqno(2.5)
$$

Following \cite{GRESCH58}, \cite{SCH61} (Ch.XII) we shall call clothed the new
operators and states. Note that the name is sometimes used
in a sense which differs from that defined by the points i) - iv).

As one can see, the problem of clothing is equivalent to determination
of some $H$ eigenvectors. In fact, the property iii) means that we do
not pretend to find all $H$ eigenstates which are one--particle--like.
For example, $H$ may have a deuteron--like eigenstate with a mass
$<\ 2m$, where $m$ is the nucleon mass. No bare one--particle state
coresponds to such a state. Now we intend to find only those
one--particle--like eigenstates of $H$ which have bare partners.

One should stress that the clothing problem may turn out to be
unsolvable. A solvability condition will be pointed out in Subsect 2.4. Note
also that the properties i) - iv) can be supplemented by some physical
constraints which will be discussed in Sect. 3, but are not needed
here.

Some clothing procedures  have been realized within simple
field models (see, e.g., \cite{GRESCH58,SCH61,FIV70}).
In the paper, we use a kind of perturbation theory developed in
\cite{HEIT54,FAD63} and \cite{SHIVI74}. It can be
applied to any field theory to yield an approximate solution of the
problem.

\bigskip
\centerline{\em 2.2 Bare Particles with Physical Masses }
\bigskip

By definition, the bare one--fermion eigenstate $\mid \bp,r \rangle _0$
of the operator $H_0$, being simultaneously the eigenstate of total
momentum $\ve{P}$, belongs to the $H_0$ eigenvalue
$E_{\bp}^0 = \sqrt{\bp^2 + {m _0}^2}$. Let us consider an H eigenstate
$\mid \bp,r \rangle$ for which $\mid \bp,r \rangle _0$ is a zeroth
approximation (ZA). Perturbation theory shows that the corresponding
$H$ eigenvalue $E_{\bp}$ differs from $E_{\bp}^0$. In the relativistic
case the function $E_{\bp}$ must be of the form  $\sqrt{\bp^2 + m^2}$
where $m$ is the mass of an observed free fermion. We shall call it physical
mass. Analogously, one can argue appearance of the meson physical mass
$\mu$ which differs from the the trial mass $\mu _0$.

So, we expect that the physical fermion and meson masses $m$ and $\mu$
arise in a natural manner when finding $H$ eigenvalues which correspond
to the clothed one--particle states.

Such an introduction of the masses $m$ and $\mu$ can be used to divide the
total Hamiltonian into the new free part $H_F$ and the new interaction $H_I$.
Namely, let us rewrite $H = H_0 + V$ as $H = H_F + H_I$ where
$$
H_F = \int \pb(\bx)[-i\ve{\gamma} \ve{\nabla} + m ] \psi(\bx) d\bx +
{1\over 2} \int\ \left[{\pi}^2(\bx) + {\ve{\nabla}\phi(\bx)}^2 +
{\mu}^2{\phi}^2(\bx) \right] d\bx
\eqno(2.6)
$$
$$
H_I = V + (m_0 - m)\int\pb(\bx) \psi(\bx) d\bx +
{1\over 2} ({\mu _0}^2 - {\mu}^2)
\int\ {\phi}^2(\bx) d\bx \equiv V + M_{ren}\ , \eqno(2.7)
$$

The decomposition $H = H_F + H_I$ is the well-known trick(see, e.g.,
\cite {SCH61}), but this is not necessary for our clothing program:
all the following results can be obtained without
the introduction of the mass counterterms $M_{ren}$\footnote{A simple example
of the calculation of radiative correction to particle "bare" mass can be
found in App. {\rm C} }.  However, it simplifies the program
realization. In other words, the separation $H = H_F + H_I$ may be
justified not {\em ab initio} but {\em post factum}.

The operator $H_F$ can be brought to the "diagonal" form\footnote
{Nonessential c-number terms are henceforth omitted}
$$
H_F = \int \omega_{\bk} a\dg(\bk)a(\bk) d\bk + \int E_{\bp} \sum_{r} [ b\dg(\bp,r)b(\bp,r) +
d\dg(\bp,r)d(\bp,r)] d\bp\  \eqno(2.8)
$$
by means of the standard expansions
$$
\phi(\bx) = (2\pi )^{- 3/2}  \int (2\omega_{\bk} )^{- 1/2}
[ a(\bk) + a\dg(-\bk)] {\rm exp}{(i\bk\bx)} d\bk\, \eqno(2.9)
$$
$$
\pi(\bx) = -i(2\pi )^{- 3/2} \int (\omega_{\bk}/2)^{1/2}
[ a(\bk) - a\dg(-\bk)] {\rm exp}{(i\bk\bx)} d\bk\, \eqno(2.10)
$$
$$
\psi(\bx) = (2\pi )^{- 3/2}\int (m/E_{\bp})^{1/2} \sum_{r}
[ u(\bp,r)b(\bp,r) + v(-\bp,r)d\dg(-\bk,r)] {\rm exp}{(i\bp\bx)} d\bp\, \eqno(2.11)
$$
where $u(\bp,r)$ and $v(\bp,r)$ are the Dirac spinors, which satisfy the
conventional equations $(\cross{p} - m)u(\bp,r) = 0$ and
$(\cross{p} + m)v(\bp,r) = 0$  with $\cross{p}= E_{\bp} \gamma^0 - \bp \ve {\gamma } $.
In these formulae $E_{\bp} = \sqrt{\bp^2 + m^2} $ and $\omega_{\bk} =
\sqrt{\bk^2 + \mu^2} $.

\bigskip
\centerline{\em 2.3 The Unitary Transformation}
\bigskip
The operators (2.3) are the corner-stone of the clothing procedure. Our aim
is to find clothed operators which should satisfy the requirements i)--iv).
Now, the symbol $\alpha$ will be used for set (2.3) with $\alpha _p$ being
one operator of the set (cf. $a$ and $a_p$). In order to implement
the properties iii) and iv), we suppose that the clothed operators
$\alpha$ are related to bare ones $a$ via a unitary transformation
$$
\alpha _p = W\dg a_p W,\ \ W\dg W = WW\dg = 1\ , \eqno(2.12)
$$
where $W$ is a function of all the bare operators $a$. Therefore,
Eq. (2.12) represents $\alpha _p$ as a function(functional) of $a$.

Note that  $W$ is the same function of either
clothed or bare operators (see \cite{GRESCH58}).
Indeed, if $f(x)$ is a polynomial or a series of x, the relation
$f(\alpha ) = W\dg(a) f(a) W(a) $ follows from Eq.(2.12). Replacing
$f(\alpha)$ by $W$ leads to
$$
W(\alpha ) = W\dg(a)W(a)W(a) = W(a)\ , \eqno(2.13)
$$
i.e., to the above statement.
Hence, the operator $a_p$, when expressed in terms of $\alpha$, is
given by
$$
a_p = W(\alpha )\ \alpha _p \ W\dg (\alpha )\ . \eqno(2.14)
$$

Unitarity of $W$ is automatically ensured if $W$ is represented as
the exponential of an antihermitian operator $R$: $W={\rm exp}R$.
For a given $R$, the r.h.s. of Eq.(2.14) can be evaluated with the help of
$$
{\rm e}^A B {\rm e}^{-A} = B + [A,B] + {1\over 2}[A,[A,B]] +
{1\over {3!}}[A,[A,[A,B]]] + ... \eqno(2.15)
$$
and the commutation rules (2.5).

In the context, the total Hamiltonian can be written as $H = H(a) = H_F + H_I$
where $H_F(a)$ is determined by Eq.(2.8) and $H_I = V(a) + M_{ren}(a)$ with
\footnote{In cumbersome formulae summations over the dummy spin indices are
sometimes omitted}
$$
V(a) = { {ig} \over {(2\pi )^{3/2} } } \int  d\bp'\ d\bp\ d\bk\
{ {m} \over {(2\omega_{\bk} E_{\bp'} E_{\bp} )^{1/2} } }
\delta (\ve{p}+\ve{k}-\ve{p'})\
\{  \bar{u}(\ve{p'}r') \gamma _5 u(\ve{p}r)\ b\dg (\ve{p'}r')
b(\ve{p}r)\ +
$$
$$
+\ \bar{u}(\ve{p'}r') \gamma _5 v(\ve{-p}r)\ b\dg (\ve{p'}r')
d\dg (\ve{-p}r)\
+\ \bar{v}(\ve{-p'}r') \gamma _5 u(\ve{p}r)\ d (\ve{-p'}r')
b(\ve{p}r)\
+
$$
$$
+\ \bar{v}(\ve{-p'}r') \gamma _5 v(\ve{-p}r)\ d (\ve{-p'}r')
d\dg (\ve{-p}r)\}\
[ a(\ve{k}) + a\dg (-\ve{k}) ]\ .
\eqno(2.16)
$$

By using Eq.(2.14), one can replace the bare operators by the clothed ones
$$
H(a) = H(W(\alpha ) \alpha W\dg (\alpha ) ) \equiv K(\alpha)\ .
\eqno(2.17)
$$
The operator $K(\alpha )$ represents the same Hamiltonian,
but it has another dependence on its argument $\alpha$ compared to $H(a)$.
$K(\alpha )$ can be found as follows. First, Eq.(2.17) can be
written as
$$
K (\alpha ) = W (\alpha )H(\alpha )W\dg (\alpha )\ .
\eqno(2.18)
$$
Second, putting $W(\alpha ) = {\rm exp} R(\alpha ) $ and using Eq. (2.15)
we have
$$
H = K(\alpha ) = {\rm e}^R\ [ H_F+H_I ]\ {\rm e}^{-R} =
$$
$$
= H_F(\alpha ) + H_I(\alpha ) + [R, H_F ] + [ R, H_I ] + {1\over 2}
[ R, [ R, H_F ] ] + {1\over 2} [ R, [ R, H_I ] ] + ...
\eqno(2.19)
$$
Eq. (2.19) gives a practical recipe for the $K(\alpha )$ calculation:
at the beginning one replaces $a$ by $\alpha$ in the initial expression
$H(a)$ and then calculates $W (\alpha )H (\alpha )W\dg (\alpha )$
using Eqs. (2.15) and (2.5). The above transition $H(a) \rightarrow
H(\alpha )$ generates a new operator $H(\alpha )$ as compared to
$H(a)$, but Eqs. (2.17) and (2.18) show that $W (\alpha )H (\alpha )W\dg (\alpha )$
turns out to be equal to the original total Hamiltonian (cf. \cite {SATO92}).

We would like to stress that the transformation $WHW\dg$ under
consideration should not be understood here as $W(a)H(a)W\dg(a)$. The latter
would be a new operator $H'(a)$, which, in general, does not coincide
with $H$. For a detailed discussion of different unitary transformations,
see Sect. 6.

\bigskip
\centerline{\em 2.4 Elimination of ``Bad'' Terms }
\bigskip

The next step is to fulfil the requirements i) - ii). If we want
the no--clothed--particle state $\Omega$ and clothed one--particle states
to be $H$ eigenvectors, the r.h.s. of Eq. (2.19) must not contain
some undesirable terms. Particularly, $K(\alpha )$ must not contain the
$b_{\rm c}\dg d_{\rm c}\dg a_{\rm c}\dg $--type terms because they would
give rise to the f$\bar{\rm f}$m states\footnote{with a transparent
abbreviation f$\bar{\rm f}$m for a ``fermion--antifermion--meson''.}, when
acting on $\Omega$, and $K(\alpha) \Omega$ could not be proportional to
$\Omega$.  Similarly, $b_{\rm c}\dg b_{\rm c} a_{\rm c}\dg$ converts a
one-fermion state $b_{\rm c}\dg \Omega$ into a fm state. But just
terms of this kind enter into the operator $V(\alpha )$ which occurs in the
r.h.s. of Eq. (2.19). In this connection, recall that $H_I (\alpha ) = V(\alpha)
 + M_{ren}(\alpha )$ where $V(\alpha )$ is derived from $V(a)$, see Eq. (2.16),
by means of the replacement $a \rightarrow \alpha $.

As we have argued above, the terms
$$
b_{\rm c}\dg b_{\rm c} a_{\rm c}\dg ,\
b_{\rm c}\dg d_{\rm c}\dg a_{\rm c} ,\
b_{\rm c}\dg d_{\rm c}\dg a_{\rm c}\dg ,\
d_{\rm c} d_{\rm c}\dg a_{\rm c}\dg \eqno(2.20)
$$
in $V(\alpha )$ do not allow the clothed no--particle and
one--particle states to be $H$ eigenvectors. The remaining terms in
$V(\alpha )$ are Hermitian conjugate of (2.20). We shall call ``bad''
all these terms. The contribution $b_{\rm c}\dg d_{\rm c}\dg
a_{\rm c}\dg $ will be called the ``bad'' term of the class [3.0]: it
is a product of three creation operators with destruction operators
not included. The other three terms in (2.20) belong to the class [2.1]:
two creation operators and one destruction operator.

The interaction $H_I$ includes also the mass counterterm $M_{ren}$ (see Eq.(2.7)).
The latter contains bad terms of the class [2.0] (see,e.g., Eq.(2.27)).
The self-energy correction to the particle mass can be represented by a
series which starts with the terms of the $g^2$ -order(see Eq. (2.31)). So,
$M_{ren}(\alpha) \sim g^2 $ while $V(\alpha) \sim g^1$.

Let us eliminate from $K(\alpha)$ the bad terms of the $g^1$ -order. For
this purpose we choose such R that
$$
 V + [ R, H_F ] = 0\ .\eqno(2.21)
$$
One readily verifies that Eq. (2.21) cannot be satisfied until
$R(\alpha )$ is linear or bilinear in $\alpha$. To meet the equation,
$R(\alpha )$ must be a three--operator, e.g., have the structure of
$V(\alpha )$. Then, the commutator $[R,H_F]$ will also be
three-operator expression since $H_F (\alpha )$ is the two-operator
\footnote{Note that the commutator of the $m$--operator term and $n$--operator one
yields a $(m+n-2)$--operator contribution}.

Let us assume that the antihermitian $R(\alpha)$ contains the bad terms of
the same kind as $V(\alpha)$ is. Namely, we put $R (\alpha ) =
{\cal R} - {\cal R}\dg $ where (cf. Eq. (2.16))
$$
{\cal R} = \int  d\bp'\  d\bp\ d\bk\ \sum_{r'r} \{
R_{11}^{\ve{k}} (\ve{p'}r';\ve{p}r)
b_{\rm c}\dg (\ve{p'},r') b_{\rm c} (\ve{p},r)\ +\
R_{12}^{\ve{k}} (\ve{p'}r';\ve{p}r)
b_{\rm c}\dg (\ve{p'},r') d_{\rm c}\dg (-\ve{p},r)\ +\
$$
$$+\
R_{21}^{\ve{k}} (\ve{p'}r';\ve{p}r)
d_{\rm c} (-\ve{p'},r') b_{\rm c} (\ve{p},r)\ +\
R_{22}^{\ve{k}} (\ve{p'}r';\ve{p}r)
d_{\rm c} (-\ve{p'},r') d_{\rm c}\dg (-\ve{p},r)
\}\ a_{\rm c} (\ve{k})\ .\eqno(2.22)
$$
The c-number coefficients $R_{ij}^{\bk}(i,j = 1,2)$ are to be derived from
Eq. (2.21), see Appendix A. We find that the solution exists if $\mu < 2m $.
This condition has a clear physical meaning, viz., the meson can decay into
the f$\bar{\rm f}$--pair if $\mu > 2m $, and, therefore, one--meson state cannot
be stable,i.e., it cannot be an $H$ eigenvector.
Once $[R,H_F ] = -V$, Eq. (2.19) can be rewritten as
$$
K(\alpha ) = H_F (\alpha ) + M_{ren}(\alpha) + {1 \over 2} [R,V] + [R,M_{ren}] +
{1 \over 3} [ R, [R,V] ] + \ ...\ .\eqno(2.23)
$$
Thus we have removed from $K(\alpha)$ all the bad terms of the $g^1$--order.

However, the r.h.s. of Eq. (2.23) embodies  other bad terms of the $g^2$- and
higher orders. For example, $[R,V]$ contains the terms
$g^2 b_{\rm c}\dg d_{\rm c}\dg a_{\rm c}\dg a_{\rm c}\dg $ of the class [4.0],
which do not destroy the physical vacuum $\Omega$ (our evaluation of $[R, V]$
is given in Appendix A). In addition, we find in $[R,V]$ the terms
$g^2 b_{\rm c}\dg d_{\rm c}\dg a_{\rm c}\dg a_{\rm c} $ of the class [3.1],
which neither destroy $a_{\rm c}\dg \Omega $, nor retain it with a multiplicative
factor. These and similar bad terms can be eliminated in a way analogous to
the described above via one more transformation
$$
\alpha _p = W_4 (\alpha ')\alpha '_p W_4\dg (\alpha ')\ ,
\eqno(2.24)
$$
where $W_4 ={\rm exp}R_4$ and $R_4$ is an expression of the $g^2$-order, which
consists of the above bad four-operator terms. This $R_4$ should be such
that $[R_4, H_F]$ would cancel the latter terms.

Note also bad terms of the classes [2.0] and  [1.1], which are present in $M$
(see Eq. (2.27)), and similar terms, which appear after normal ordering of
$[R,V]$ (see Appendix A). The bad terms of the class [2.0 ] must be removed
from $K(\alpha)$ as well. We shall show in Subsect. 2.5 how they may be
cancelled under a condition that relates the physical masses with the input
parameters $m_0$, $\mu_0$ and $g$.

Further, the double commutator $[R, [R ,V] ]$ in Eq. (2.23) is composed
of five-operator terms (cf. the footnote 3), and there are
bad terms among them. In particular, after reshuffling the
operators into normal order, new three--operator bad terms occur. However, they
are of the $g^3$--order. This type of bad terms can also be found in $[R,M]$.
The subsequent unitary transformation makes it possible to remove all bad
terms of the $g^3$--order.

Along the guideline, one may eliminate from the Hamiltonian the bad terms of
increasing orders in the coupling constant $g$. It is assumed that in the
limit the requirements i) and ii ), which are equivalent to the absence of
bad terms in $K$, will be fulfilled.

Finally, if our clothing procedure were perfect, the resulting
representation $K$ of the total Hamiltonian would possess the property
$$
K(\alpha ) {\mid \bk \rangle}_c = H_F(\alpha ) {\mid \bk \rangle}_c =
\omega_k {\mid \bk \rangle}_c\   \eqno(2.25)
$$
with ${\mid \bk \rangle }_c = a_c\dg(\bk) \Omega $ . In other words, the
new interaction term $K_I(\alpha) = K(\alpha) - H_F(\alpha) $ would satisfy
the equation $K_I(\alpha) {\mid \bk \rangle }_c = 0 $. Analogous equations
will hold for the physical vacuum $\Omega $ and the clothed one--fermion and
one--antifermion states.

\bigskip
\centerline{\em 2.5 Particle Mass Renormalization}
\bigskip
The cancellation of the bad two--operator terms in $K(\alpha )$ will be
demonstrated for those of them which are bilinear in meson operators $a_c$
and $a_c\dg$. The latter originate , first of all, from the meson mass
counterterm
$$
M_{mes} = {1\over 2} ({\mu _0}^2 - {\mu}^2) \int\ {\phi}^2(\bx) d\bx \
\eqno(2.26)
$$
Indeed, substituting the expansion (2.9) for $\phi(\bx )$ into Eq.(2.26),
we obtain
$$
M_{mes} = \int\ {{{\mu _0}^2 - {\mu}^2}\over {4\omega_{\bk}}}
[2 {a_c}\dg(\bk)a_c(\bk) + a_c(\bk)a_c(-\bk) +
{a_c}\dg(\bk){a_c}\dg(-\bk)] d\bk \
\eqno(2.27)
$$
As shown in Appendix A, terms of the same operator structure occur in $K(\alpha)$
after normal ordering of the commutator $[R,V]$ from Eq.(2.23):
$$
\int\ {{t_{\bk}} \over { \omega_{\bk}}}
[2 {a_c}\dg(\bk)a_c(\bk) + a_c(\bk)a_c(-\bk) +
{a_c}\dg(\bk){a_c}\dg(-\bk)] d\bk \ ,
\eqno(2.28)
$$
where $t_{\bk}$ is determined by Eq.(A.20).

Now, we see that the sum of (2.27) and (2.28) gets equal to zero if
$$
{\mu _0}^2 - {\mu}^2 = - 4t_{\bk} \quad or \quad {\mu}^2 = {\mu_0}^2 + 4t_{\bk} \
\eqno(2.29)
$$
Here the quantity $t_{\bk}$ must not depend on $\bk$ along with ${\mu}^2$
(cf. our argumentation in Subsect. 2.2). Unfortunately, this independence is
not automatically provided in the Schroedinger picture that we use
throughout. The integral that determines $t_{\bk}$ (see Eq.(A.20)) is
quadratically divergent and one needs to overcome the trouble
(e.g., by introducing a cutoff factor). So, special efforts are required to
yield the proof of independence (see, e.g., \cite{HEIT54}, Ch.6).

In the same way, the fermion mass countrerterm
$$
M_{ferm} = (m_0 - m)\int\pb(\bx) \psi(\bx) d\bx \  \eqno(2.30)
$$
cancels, under a proper condition, all the terms bilinear in the fermion operators,
which arise from ${1\over 2}[R,V] $ as a result of normal ordering.

Up to now we have considered the bilinear terms of the $g^2$--order. Normal
ordering of the six--operator and other terms of $K(\alpha )$ gives bad two--operator
terms of the $g^4$-- and higher orders. To eliminate them we suppose that
${\delta {\mu}^2} \equiv {{\mu_0}^2 - {\mu}^2} $ and $\delta m \equiv
{m_0 - m} $ may be expanded in the series
$$
\delta {\mu}^2 = \sum_{n=1}^{\infty} g^{2n}({\delta {\mu}^2})_{2n}, \quad
\delta m = \sum_{n=1}^{\infty} g^{2n}({\delta m})_{2n}\ \eqno(2.31)
$$
Let us assume that $(\delta {{\mu}^2})_2 $ and $({\delta m})_2 $ are used to
remove the two--operator terms of the $g^2$--order as described above. Then,
the terms $(\delta {{\mu}^2})_4 $ and $({\delta m})_4 $ are destined to cancel
two-operator terms of the $g^4$--order, and so on.

\bigskip
\centerline{\em 2.6 Some Remarks}
\bigskip

So, the transformation realized by $W = {\rm exp}R$ fulfills
an incomplete clothing, {\it viz.}, it removes bad terms of
the least order in $g$. The no-particle $\Omega$ and one-particle
states $b_{\rm c}\dg\Omega$, $d_{\rm c}\dg\Omega$ and
$a_{\rm c}\dg\Omega$ constructed at this stage are merely approximate
$H$ eigenvectors.

We shall confine ourselves to the consideration of this
transformation only while discussing the bound state problem like the
deuteron (Sect. 4). Even  this simplest application of our approach
turns out to be rather cumbersome.

This section has been aimed to show how the unitary transformations
of the original Hamiltonian for a system of interacting fields can be
regarded as the introduction of new creation(destruction) operators instead of the
initial "bare" ones. These new operators and corresponding one--particle states
occurring at the first stage of the clothing procedure may be called
"partially clothed". We juxtapose the one--particle states to the observable
particles (say, pion and nucleon).

Let us note that the clothed states (operators) are not the in(out)--states
(operators) of
RQFT (see, e.g., \cite{SCH61}, Ch. 17). Indeed, the two--particle in--states
$ a_{in}\dg(\bk_1) a_{in}\dg(\bk_2) \Omega $ are $H$ eigenstates while
the two--particle  clothed states
$a_c\dg(\bk_1) a_c\dg(\bk_2) \Omega$ do not (simple models with noninteracting
particles are the evident exception). Within the in(out)--formalism it is
{\it supposed} that $H$ has the following eigenstates, viz., no--particle
state (physical vacuum), states with one in--particle, two--, three--, etc.,
these states being analogous to the corresponding $H_0$ eigenstates.
Meanwhile, the clothing formalism needs not such a supposition: one can find
explicit expression for no-- and one--particle clothed states in terms of
bare states (using formulae of this section under the condition $\mu < 2 m$ ).
In addition, the in(out)--formalism does not consider two--particle states
which have no $H_0$ bare partners, e.g., an in--state describing the deuteron.
In the clothing formalism the problem of deuteron--like states is subject to
further investigation (see Sect. 4).

\bigskip
\centerline{\bf 3. Generators for Space Translations and Space-Time Rotations}
\centerline{\bf within the Clothing Procedure}
\bigskip

In the previous section we have expressed the total Hamiltonian $H$ in terms of
the clothed operators. $H$ is the time translation generator of the Poincar$\acute e$
group. Here we shall discuss how the rest of the group generators (total linear
and angular momenta, and generators of Lorentz boosts) depend upon the clothed
operators. This will allow us to formulate the transformation properties of
the clothed no-particle and one-particle states under the Lorentz boosts.
First of all, let us consider various constraints imposed on clothing
transformations by the general symmetries in RQFT and those which are specific
for a given field model.

\bigskip
\centerline{\em 3.1 Total Momenta and Other Motion Integrals in Terms of
Clothed Operators}
\bigskip
Similar to the determination of the total Hamiltonian $H$ as a function
$K(\alpha )$ of the clothed operators $\alpha $ we can obtain the expression
$\ve {P_{\rm c}}(\alpha )$ for the total linear momentum $\ve {P}$ (cf. Eq.(2.17)):
$$
\ve {P}(a) = \ve {P}( W(\alpha)  \alpha W\dg (\alpha ) ) =
W(\alpha ) \ve {P}(\alpha ) W\dg (\alpha ) \equiv \ve {P_{\rm c}}(\alpha)\
\eqno(3.1)
$$
One can show that $W \ve {P} W\dg $ is simply equal to $\ve {P} $. Indeed,
this is equivalent to $W \ve {P} = \ve {P} W$ that follows , in its turn,
from $[R, \ve {P}] = 0$ (remind that $W = {\rm exp}R $). The validity
of the latter can be verified using Eq.(2.22), ({\rm A.1}), ({\rm A.2})
and the well-known
expression of the total linear momentum in terms of the creation(destruction)
operators
$$
\ve {P} = \int \bk a\dg(\bk)a(\bk) d\bk + \int \bp \sum_{r} [ b\dg(\bp,r)b(\bp,r) +
d\dg(\bp,r)d(\bp,r)] d\bp\  \eqno(3.2)
$$
(see, e.g., Eq.(7.33) in \cite{SCH61}).
Alternatively, one can take the representation (A.4) for $R$ and use the equation
$[\ve {P},V]$ = 0 that holds because of the invariance of $V$ with respect to
space translations.

So, we have found that $\ve {P_{\rm c}}(\alpha ) = \ve {P}(\alpha )$. This
means that $\ve {P}$ is the same function of the clothed operators as of
the bare ones.

Analogous statements are valid for the total angular momentum $\ve {M}$
\footnote{The field-theoretical formula for the generator can be found in
\cite{BD64}(Ch.11), see also Subsect.3.2.} and the baryon(fermion) number
operator $B = \int \psi\dg(\bx) \psi(\bx) d\bx $. This means that the clothed
states $\Omega $, $a_{\rm c}\dg\Omega$, $b_{\rm c}\dg\Omega$,
and $d_{\rm c}\dg\Omega$ have the following properties: a) they are eigenvectors
of $\ve {P}$ ; b) they are transformed under space rotations in the same
manner as the relevant bare states do ; c) they possess definite $B$ values.

Instead of verifying the properties $[R, \ve {P}] = [R, \ve {M}] = [R,B] = 0$
with
the solution $R$ of the equation $[R,H_F] + V = 0$ one may consider them as
some new requirements supplementary to those listed in Subsect. 2.1 as i) - iv).
These requirements would result in definite restrictions on the coefficients
involved in the expansion (2.22) for $R$. For example, it follows from
$[R, \ve {P}] = 0$  that $R_{ij}^{\ve{k}} (\ve{p'}r';\ve{p}r)$ must have the form
$\delta(\bp + \bk - \bp')r_{ij}^{\ve{k}} (\ve{p'}r';\ve{p}r)$. The equation
$[R, \ve {M}] = 0$ means that $R_{ij}^{\ve{k}} (\ve{p'}r';\ve{p}r)$ must depend
on rotationally invariant combinations of its arguments. Besides the condition
$[R,B] = 0$ prevents $R$ to be dependent on terms of the $b_{\rm c}\dg d_{\rm c}$-
kind (we eliminate such terms from the beginning assuming the form (2.22) for
$R$).

One may add to these restrictions those which are consequences of the following
requirements: clothed operators and clothed one-particle states must have the same
transformation properties with respect to space inversion, time reversal and
charge conjugation as their bare partners.

Let us stress that all the constraints are exact whereas the equation
$[R,H_F] + V = 0$ considered in Subsect. 2.4 is merely approximate one. However,
its solution has all the properties in question since the interaction $V$
commutes with $\ve {P}$, $\ve {M}$, $B$, etc.(see Eq.(A.4)).

\bigskip
\centerline{\em 3.2 Transformations of Bare Operators and States under
Lorentz Boosts}
\bigskip

A distinctive feature of the conventional relativistic dynamics ("instant"
form after Dirac \cite{DIR49} ) is that the generators $\ve {N} = (N^1,N^2,N^3)$
of the Lorentz boost $\Lambda =  {\rm exp}(i \ve {\beta} \ve {N})$\footnote
{Here $\mbox{\boldmath$\beta$}=\beta\bn$, $\bn = {\ve {v} \over v}$ and
$th\beta = v$, where $v$ is the velocity of a reference frame moving along
the $\bn$ direction.In this paper we use the system of units in which the light
velocity $c$ is equal to unity.} contain interaction terms while the linear
$\bP = (P^1,P^2,P^3)$ and angular $\bM = (M^1,M^2,M^3) $ momenta are determined
by the same expressions as for free fields.

In order to see this explicitly let us resort to the Lagrangian formalism
where the quantities of fundamental importance are the energy-momentum density
tensor ${\cal T}^{\mu \nu} (x) $ and the angular momentum density tensor
${\cal M}^{\lambda \mu \nu} (x)$\footnote{Greek labels run the values 0,1,2,3.}
(see, e.g., Ch.11 in \cite{BD64}). By definition, $P^\mu = (H,\bP) =
\int {\cal T}^{0 \mu}(x) d\bx $, $M^j = \epsilon_{jkl}M^{kl}$, and $N^j = M^{0j}$
where $M^{\mu \nu} = \int {\cal M}^{0\mu \nu}(x) d\bx $. According to the
Noether theorem all the ten operators are time independent, i.e., they are
the motion integrals. In other words, they can be evaluated at $t = 0$, i.e.,
they can be expressed through field operators in the Schr\"odinger
picture.

The corresponding representation of $\bN$ depends on the form of
${\cal T}^{\mu \nu} (x) $ (nonsymmetrized or
symmetrized\footnote{The symmetrized form with the Belinfante ansatz \cite{BEL40}
for Lorentz boosts has been employed in our talks \cite{THES97,GRON97}.
Another application of the form can be found in a covariant description of
electromagnetic
interactions with nuclei \cite{SHE90}.}), that is utilized. Here we shall use the
nonsymmetrized form (see Eqs.(13.45) and (13.47) in \cite{BD64}), which leads
to
$$
\bN = \bN_F - \int \bx V(\bx) d\bx + \bN_{ren}\ , \eqno(3.3)
$$
where $\bN_F$ is the free part of $\bN $:
$$
\bN_F = \bN_{ferm} + \bN_{mes}\ , \eqno(3.4)
$$
$$
\bN_{ferm} = - \int \bx \pb(\bx)[-i\ve{\gamma} \ve{\nabla} + m ] \psi(\bx) d\bx +
\frac{i}{2} \int \pb(\bx) \ve{\gamma} \psi(\bx) d\bx\, \eqno(3.5)
$$
$$
\bN_{mes} = - {1\over 2} \int \bx \left[{\pi}^2(\bx) + {\ve{\nabla}\phi(\bx)}^2 +
{\mu}^2{\phi}^2(\bx) \right] d\bx\                  \eqno(3.6)
$$
In accordance with the particle mass renormalization described in Subsect.2.5
we have separated the contribution to $\bN$ from meson and fermion mass
counterterms (cf. Eqs. (2.26) and (2.30)):
$$
\bN_{ren} = \bN^{mes}_{ren} + \bN^{ferm}_{ren}\, \eqno(3.7)
$$
$$
\bN^{mes}_{ren} = - {1\over 2} ({\mu _0}^2 - {\mu}^2) \int \bx {\phi}^2(\bx) d\bx \,
\eqno(3.8)
$$
$$
\bN^{ferm}_{ren} = - (m_0 - m)\int \bx \pb(\bx) \psi(\bx) d\bx \  \eqno(3.9)
$$

Now, to express the generators through the creation(destruction) operators
let us take the expansions (2.9) - (2.11) and employ the relation
$$
\int x^j {\rm exp}{(i \ve {q} \bx)} d\bx = - i {(2 \pi)}^3 \frac{\partial}
{\partial q^j}\delta(\ve {q})
$$
Then, e.g., we find
$$
N^j_{mes} = \frac{i}{2} \int a\dg(\bk')a(\bk) \frac{\omega_{\bk'} \omega_{\bk}
+ \bk' \bk + {\mu}^2 }{\sqrt{\omega_{\bk'} \omega_{\bk}}} \frac{\partial }
{\partial k^j} \delta(\bk' - \bk) d\bk' d\bk \   \eqno(3.10)
$$
or
$$
N^j_{mes} = \frac{i}{2} \int \omega_{\bk} \bigl[ \frac{\partial a\dg(\bk)}
{\partial k^j}a(\bk) - a\dg(\bk) \frac{\partial a(\bk)}{\partial k^j} \bigr]
d\bk \  \eqno(3.10')
$$

Simultaneously, we get (cf. Eq. (2.16))
$$
N^j_I \equiv - \int x^j V(\bx) d\bx = -
$$
$$
{ {g} \over {(2\pi )^{3/2} } } \int d\bp'\ d\bp\ d\bk\
{ {m} \over {(2\omega_{\bk} E_{\bp'} E_{\bp})^{1/2} } }\
\frac{\partial}{\partial k^j} \delta(\ve{p}+\ve{k}-\ve{p'})\
\{  \bar{u}(\ve{p'}r') \gamma _5 u(\ve{p}r)\ b\dg (\ve{p'}r') b(\ve{p}r)\ +
$$
$$
+\ \bar{u}(\ve{p'}r') \gamma _5 v(\ve{-p}r)\ b\dg (\ve{p'}r')
d\dg (\ve{-p}r)\
+\ \bar{v}(\ve{-p'}r') \gamma _5 u(\ve{p}r)\ d (\ve{-p'}r')
b(\ve{p}r)\
+
$$
$$
-\ \bar{v}(\ve{-p'}r') \gamma _5 v(\ve{-p}r)\ d\dg (\ve{-p}r) d (\ve{-p'}r')\}\
[ a(\ve{k}) + a\dg (-\ve{k}) ]\ \eqno(3.11)
$$
with the Yukawa interaction density $V(\bx)$ (see Eq. (2.2)).

These formulae enable to perform directly transformations of the bare operators
and states under the Lorentz boosts. In particular, in the infinitisemal case
with $\mid \beta^j \mid \ll 1 $ (j = 1,2,3) one has
$$
\Lambda \Omega_0 = {\rm exp}{[i \ve {\beta} (\bN_F + \bN_I + \bN_{ren})]}
\Omega_0
$$
$$
\simeq [ 1 + i \ve {\beta }(\bN_I + \bN_{ren})] \Omega_0 \simeq ( 1 + i \ve
{\beta} \bN_I ) \Omega_0 \ , \eqno(3.12)
$$
where we took advantage of the relation
$$
\bN_F \Omega_0 = 0\ , \eqno(3.13)
$$
and omit terms of the order higher than $g^1$.

Eq. (3.12) means that the bare vacuum is not invariant
with respect to $\Lambda$, viz., $\Lambda \Omega_0 \neq \Omega_0$.
A moving observer "sees"
$\Lambda \Omega_0 $ as the superposition of no - particle state $\Omega_0 $,
$N \bar {N} \pi $ states, etc.

Similarly, bare one-particle states (e.g.,$a\dg (k) \Omega_0$) are not
transformed with respect to $\Lambda$
as in the free case where along with Eq. (3.13) one has
the property \footnote{Under the discussion it is convenient to proceed with the
operators $a(k)$ which obey the covariant commutation relation
$[a(k), a\dg (k')] = \omega_{\bk} \delta (\bk - \bk')$ (cf., e.g., Eq.
(7.23) in \cite{SCH61}). In the framework of our consideration it is equivalent
to replacement of $a(\bk)$ by $a(k)/ \sqrt{\omega_{\bk}}$. It is
true for the respective clothed operators, so that, for instance, the first of
the relations (2.5) should be replaced by
$[a_{\rm c}(k), a_{\rm c}\dg (k')] = \omega_{\bk} \delta (\bk - \bk')$.}
$$
e^{i{\mbox{\boldmath$\beta$}}\bN_F} a\dg (k) e^{-i{\mbox{\boldmath$\beta$}}\bN_F}
=a\dg (Lk) \  \eqno(3.14)
$$
In Eq. (3.14) $L$ denotes a pure Lorentz transformation ("boost") with the
matrix:
$$L = \bigl[ L^{\mu }_{\nu } \bigr] = \left[\matrix {
      u^0    &\vdots & u_j\cr
      \ldots &\ldots & \ldots\cr
     - u^j   &\vdots & {\delta}^i_j - \frac{u^i u_j}{1 + u_0}
\cr} \right] \  , \eqno(3.15)
$$
where $u^{\mu } = (u^0, \ve {u}) = (ch\beta , \bn sh\beta )$ is the four
- velocity vector. The boost converts the four - momentum $k = (\omega_{\bk},\bk)$
into $k' = Lk = (\omega_{\bk'}, \bk')$.

We have $\Lambda a\dg (k) {\Omega}_0 = a_{\Lambda}\dg(k) \Lambda {\Omega}_0 $
with the transformed meson operator $a_{\Lambda}\dg (k) =
\Lambda a\dg(k) \Lambda \dg$.
For an infinitisemal boost the operator
$$
a_{\Lambda} \dg(k) =
e^{i{\mbox{\boldmath$\beta$}}\bN} a\dg (k) e^{-i{\mbox{\boldmath$\beta$}}\bN} \simeq
a\dg(k) + i \ve {\beta} [{\bN}_{mes}, a\dg (k)] +
i \ve {\beta} [{\bN}_I, a\dg (k)] \    \eqno(3.16)
$$
contains bare fermion operators due to the last ("interaction") term in the
r.h.s. of the equation. By using Eq. (3.10') one can see that the two first
terms yield
$a\dg(k) - {\beta}^j \frac{\partial a\dg (k)}{\partial k^j} {\omega}_{\bk}
\simeq a\dg ({\omega}_{\bk} - \ve {\beta} \bk, \bk - {\omega}_{\bk} \ve {\beta})$
that coincides  with the operator $a\dg (Lk)$ in Eq.(3.14) for
the infinitisemal $L$.

Now, taking into account Eqs. (3.12) and (3.16), one
can ascertain that the transformed state $\Lambda a\dg (k) {\Omega}_0 $
is composed of the one - meson state $ a\dg (k') {\Omega }_0 $ with the properly
changed momentum $\bk' = \bk - {\omega}_{\bk} \ve {\beta} $ and the states
$ \mid f \bar {f} \rangle $ and $ \mid f \bar {f} \pi \pi \rangle$
containing fermions.

\bigskip
\centerline {\em 3.3 Boost Generators for Clothed Particles. Elimination of
Bad Terms}
\bigskip

It is reasonable to anticipate that the physical vacuum
$\Omega$ and clothed one-particle states (e.g., $a_c\dg (k) \Omega$)
should be, respectively, the no-clothed-particle state and  clothed
one-particle states from the point of view of a moving observer. More exactly,
they should meet the relations
$$
\Lambda\Omega = \Omega \  \eqno(3.17)
$$
and
$$
\Lambda {a_c}\dg (k) \Omega = {a_c}\dg (Lk) \Omega  \  \eqno(3.18)
$$
The previous experience of handling with Eqs. (3.12) and (3.16) prompts that
these conditions could be provided (at least, approximately) if we shall
manage to remove bad terms from $\bN$ (in practice, some of them) while
expressing it through the clothed operators. In this connection, let us write
down
$$
\bN \equiv \bN(a) = W\bN(\alpha )W\dg \equiv \bB(\alpha)
= e^{R(\alpha)}[\bN_F (\alpha) + \bN_I (\alpha) + \bN_{ren}(\alpha)]e^{-R(\alpha)}
$$
$$
= \bN_F+\bN_I+[R,\bN_F]+[R,\bN_I]+...   \   \eqno(3.19)
$$
(cf. Eq. (2.19)) and then remove from the r.h.s. of (3.19) all the bad terms
of the $g^1$-order by requiring that
$$
[N_F^j,R]=N_I^j=-\int x^j V(\bx)d\bx \quad (j=1,2,3) \  \eqno(3.20)
$$

Now, we want to show that Eqs. (3.20) will automatically hold if $R$ satisfies
the condition (2.21). For this purpose, one can use the representation (A.4)
for such $R$:
$$
R=-i\lim_{\epsilon\to 0^+} \int_0^{\infty}dt e^{-\epsilon t}\int V(x) d\bx
\ , \eqno(A.4)
$$
where $V(x) = V(\bx,t)= e^{i H_F t}V(\bx)e^{-i H_F t}$ is the interaction
operator in the Dirac picture. Being a scalar it is transformed as
$$
e^{i{\ve \beta}{\bN}_F} V(x) e^{-i{\ve \beta}{\bN}_F} = V(Lx)\ \eqno(3.21)
$$
under the Lorentz boost $L$ determined by the matrix (3.15).  Note that one
can directly verify the validity of Eq. (3.21) in case when $N_F$ is the
Schroedinger operator (as it does in Eq. (3.19)) while $V(x)$ being any
Lorentz invariant operator in the Dirac picture.

In order to exploit this property  note that
$$
[N_F^1,R]=-i\frac{\partial}{\partial\beta^1} \bigl[ e^{i\beta^1 N_F^1} R
e^{-i\beta^1 N_F^1} \bigr] \bigr|_{\beta^1 = 0} \  \eqno(3.22)
$$
or taking into account firstly Eq.(A.4) and then Eq.(3.21),
$$
[N_F^1,R]=-\lim_{\epsilon\to 0^+}\lim_{\beta^1\to 0}
\frac{\partial}{\partial\beta^1}
\int_0^{\infty}dt e^{-\epsilon t}\int V(Lx) d\bx \  \eqno(3.23)
$$

For the infinitesimal boost we have
$$
V(Lx) = V(\bx - {\ve \beta}t,t - {\ve \beta}\bx ) = V(x^1-\beta^1 t,x^2,x^3,t-\beta^1 x^1)
\ \eqno(3.24)
$$
and
$$
[N_F^1,R] = -\lim_{\epsilon\to 0^+}\int_0^{\infty} dt e^{-\epsilon t}
\int\lim_{\beta^1\to 0}\frac{\partial}{\partial\beta^1}
V(x^1-\beta^1 t,x^2,x^3,t-\beta^1 x^1) d\bx
$$
$$
= -\lim_{\epsilon\to 0^+}\int_0^{\infty} dt e^{-\epsilon t}
\int (-t\frac{\partial}{\partial x^1}V(\bx,t)
-x^1\frac{\partial}{\partial t}V(\bx,t)) d\bx
$$
$$
= \lim_{\epsilon\to 0^+}\bigl[
\int x^1 d\bx \int_0^{\infty}e^{-\epsilon t}\frac{\partial}{\partial t}
V(\bx,t)dt
+\int_0^{\infty} t dt e^{-\epsilon t}\int \frac{\partial}{\partial x_1}
V(\bx,t) d\bx\bigr] \  \eqno(3.25)
$$

One can show (cf., the proof of the relation (A.4)) that the first term in
the square brackets yields $-\int x^1 V(\bx) d\bx$.
At the same time the second term is equal to zero since the operator
$V(\bx,t)$ (more exactly, its matrix elements) vanishes at $x^1 = \pm\infty$.
So, we get the desirable relation (3.20) with $j = 1$ (the cases $j = 2,3$
are analogous).

So, the transformation $W = {\rm exp}R$ eliminates simultaneously the
three-operator terms $ \sim g^1$ both from the total Hamiltonian $K(\alpha)$
and the boost generators $\bB(\alpha)$. One should emphasize that the proof
is valid for any Lorentz scalar function $V(\bx,t)$. Specific expressions for
$\bN_F$ have not been required as well since all we have needed is
Eq. (3.21).

After this elimination of bad terms we get by analogy to Eq. (2.23),
$$
\bB(\alpha) = \bN_F(\alpha) + {\bN}_{ren}(\alpha) + \frac{1}{2} [R,\bN_I] +
[R,{\bN}_{ren}] + \frac{1}{3} [R,[R,\bN_I]]  + ... \        \eqno(3.26)
$$
We shall not exemplify separate interaction terms in the r.h.s. of this
equation since their structure repeats that for the corresponding contributions
to $K(\alpha)$ (cf., e.g., Eqs.(3.11) and (2.16)).

\bigskip
\centerline{\bf 4. Equations for Bound and Scattering States in RQFT}
\bigskip

The clothed one--particle states are eigenstates of the the total Hamiltonian
$H$ according to their definitions (see Subsect. 2.1). There may be other
$H$ eigenstates which describe physical systems resembling one--particle states,
viz., the states with discrete values of the system mass that may be defined as
the system energy in the rest frame of reference (scattering states
belong to continuous values of the mass).

First of all, we keep in mind the simplest bound states similar to the hydrogen atom
or the deuteron. In the non--relativistic approach the wavefunction of such
two--body state is the product of a function that describes the system as a
whole and other function being dependent on the internal variables (for
instance, the relative momentun $\bp = \frac{1}{2} (\bp_1 - \bp_2)$ for two
identical particles). Therefore, the centre--of--mass motion is separated from
the internal motion. It is not the case for any relativistic model which
satisfies the Poincar\'e algebra . In fact, the coupling
between the internal and centre--of--mass motions is inherent to relativistic
theories of interacting particles (see, e.g., the papers \cite{SATO91}, \cite
{HAGLO92} where within simple field models the two--body bound states are
studied in a moving reference frame).

For the Yukawa model the corresponding states may be fermion--fermion states
(deuteron--like), meson--fermion ones, etc. Of course, we are not able to find
the exact Hamiltonian eigenstates, except some exactly solvable
models (see \cite{GRESCH58}). However, within the clothing procedure in
question one can suggest reasonable approximations to this problem.

\bigskip
\centerline{\em 4.1 New Zeroth Approximation for the Total Hamiltonian $K$}
\bigskip

Our approach is based on the choice of an appropriate zeroth approximation (ZA)
to the total Hamiltonian expressed through the clothed operators, i.e., the
operator $K$ determined by Eq. (2.17). Since its free part $K_2 = H_F(\alpha)$
has no deuteron--like eigenstates, we shall try to take
$$
K_{ZA}=K_2 + g^2K_4^{(2)}\  \eqno(4.1)
$$
by adding to the two--operator(one--body) contribution
$$
K_2 = \int \omega_{\bk} a_c\dg(\bk)a_c(\bk) d\bk + \int E_{\bp} \sum_{r} [ b_c\dg(\bp,r)b_c(\bp,r) +
d_c\dg(\bp,r)d_c(\bp,r)] d\bp \equiv K_{\pi} + K_N \  \eqno(4.2)
$$
the four--operator(two--body) contributions of the $g^2$ - order which arise from the
commutator $\frac{1}{2}[R,V] \equiv \frac{1}{2}[R_3,V]$ in the r.h.s. of Eq. (2.23).
This commutator is evaluated in Appendix A. Doing so, we obtain the decomposition
$$
K_4 \equiv g^2 K_4^{(2)} = K(NN \rightarrow NN) + K(\bar {N} \bar {N}
\rightarrow \bar{N} \bar {N}) + K(N \bar {N} \rightarrow N \bar {N})
$$
$$
+ K(\pi N \rightarrow \pi N) + K(\pi \bar{N} \rightarrow \pi \bar {N})
+ K(\pi \pi \rightarrow N \bar {N}) + K(N \bar {N} \rightarrow \pi \pi )
\  \eqno(4.3)
$$
with the separate interactions between the different clothed particles.
They are displayed (very schematically) in Fig.1
where the graph (a) represents the nucleon--nucleon interaction
$$
K(NN \rightarrow NN) = \sum_{r,r^{\prime}}
\int d\bp_{1}^{\prime} d\bp_{2}^{\prime} d\bp_{1} d\bp_{2}
$$
$$
V_{NN}(\bp_1^{\prime},r_1^{\prime},\bp_2^{\prime},r_2^{\prime};\bp_1,r_1,
\bp_2,r_2)
b_c\dg(\bp_{1}^{\prime},r_1^{\prime})
b_c\dg(\bp_{2}^{\prime},r_2^{\prime})
b_c(\bp_{1},r_1)
b_c(\bp_{2},r_2)
\  \eqno(4.4)
$$
while the pion--nucleon interaction
$$
K(\pi N \rightarrow \pi N) = \sum_{r,r^{\prime}}
\int d\bk^{\prime} d\bp^{\prime} d\bk d\bp
$$
$$
V_{\pi N}(\bk^{\prime}, \bp^{\prime},r^{\prime};\bk,\bp,r)
a_c\dg(\bk^{\prime})
b_c\dg(\bp^{\prime},r^{\prime})
a_c(\bk)
b_c(\bp,r) \  \eqno(4.5)
$$
is displayed by the graph (c).

Explicit expressions for the coefficients $V_{\pi N}$ and $V_{NN}$ will be
given in the next Subsects. Here, however, one should note that all these
terms of $K_4^{(2)}$ describe only real processes such as $N + N \rightarrow
N + N$, $\pi + N \rightarrow \pi + N$, etc. The bad terms of $[R_3,V]$ are
not included in $K_{ZA}$ for the reasons discussed in Subsects. 2.4 and 2.5.
For example,
the terms of the kind [4.0] (e.g.,$b\dg d\dg b\dg d\dg$ and $b\dg d\dg a\dg a\dg$)
and [3.1] (e.g., $b\dg d\dg b\dg b$ and $b\dg d\dg a\dg a$) must be removed during
our clothing procedure via the transformation $W_4 = {\rm exp}{R_4}$ (see Eq.
(2.24)). Recall that $[R_4,K_2]$ must cancel these bad terms.

In its turn, the transformation $W_4$ (see Eq. (2.23) for $K$)
$$
W_4 K({\alpha}^{\prime}) {W_4}\dg = W_4( K_2({\alpha}^{\prime}) +
\frac{1}{2}[R_3({\alpha}^{\prime}),V({\alpha}^{\prime})] + ... ){W_4}\dg =
$$
$$
K_2({\alpha}^{\prime}) + \frac{1}{2}{[R_3({\alpha}^{\prime}),V({\alpha}^{\prime})]}_4 +
[R_4,K_2] + \frac{1}{2}[R_4,{[R_3,V]}_4] + ... \ , \eqno(4.6)
$$
brings in the Hamiltonian new four--operator terms in addition to those mentioned
above, see Eq. (4.3). Here, ${[R_3,V]}_4$ denotes the four-operator part of
$[R_3,V]$\footnote{The two--operator terms of $[R_3,V]$ are supposed to be
cancelled with the respective mass counter terms.}. After the normal ordering,
the double commutator
$[R_4,{[R_3,V]}_4]$ yields new four--operator interactions in the total
Hamiltonian. However, they are of the $g^4$ - order whereas the interaction
terms which we have included in $K_{ZA}$ are of the $g^2$ - order. So, the latter
are not altered by $W_4$. They exhaust all the interaction terms of the $g^2$ - order, which remain
in $K_{ZA}$.

We hope that $K_{ZA}$ eigenstates are good approximations to exact $K$
eigenstates. The former should be found nonperturbatively (e.g., by means of
numerical methods). Five--operator and more complicated interaction terms can be
taken into account via perturbation theory recipes.

\bigskip
\centerline{\em 4.2 Meson--Nucleon Eigenstates of $K_{ZA}$ and Pion--Nucleon Quasipotential}
\bigskip
The operator $K_{ZA}$ has an important property: it conserves the total
number of clothed particles. In particular, $K_{ZA}$ transforms clothed
two--particle states (e.g., of the $NN$ or $\pi N$ types) to two--particle
ones. Moreover, the Fock subspace of all the clothed states can be divided into
several sectors (the $NN$ sector, the $\pi N$ sector, etc.) such that $K_{ZA}$
leaves each of them to be invariant, i.e., for any state vector $\Phi$ of
such sector $K_{ZA} \Phi$ belongs to the same sector.

Let us show in the simplest case of the $\pi N$ sector that the property
just mentioned allows us to reduce the eigenstate equation
$K_{ZA} \Phi^E = E \Phi^E $ to the related Schroedinger equation of
the particle--number--conserving quantum mechanics. For this
purpose we seek $\Phi^E $ as the following superposition of the $\pi N$
sector states $a_c\dg b_c\dg \Omega$,
$$
\Phi_{\pi N}^E = \sum_{r} \int d\bk d\bp \Phi_{\pi N}^E(\bk; \bp,r) a_c\dg (\bk) b_c\dg (\bp,r) \Omega \
\eqno(4.7)
$$
In the $\pi N$ sector $K_{ZA}$ is equal to $K_{\pi} + K_N + K(\pi N \rightarrow \pi N)$
because the rest terms of $K_{ZA}$ give zero when acting on $\Phi_{\pi N}^E$.
Thus, the equation $K_{ZA} \Phi_{\pi N}^E = E \Phi_{\pi N}^E $ reduces to
$$
[K_{\pi} + K_N + K(\pi N \rightarrow \pi N)] \Phi_{\pi N}^E = E \Phi_{\pi N}^E \
\eqno(4.8)
$$

Taking the scalar products of the both parts of (4.8) with
$\langle a_c\dg(\bk) b_c\dg(\bp, r) \Omega \mid $, we get the relevant equation for
$\Phi_{\pi N}^E(\bk; \bp,r) $,
$$
(E - \omega_{\bk} - E_{\bp}) \Phi_{\pi N}^E(\bk; \bp,r) = \\
\sum_{r^{\prime}} \int d\bk^{\prime} d\bp^{\prime} V_{\pi N}(\bk,\bp, r; \bk^{\prime}, \bp^{\prime},r^{\prime})
\Phi_{\pi N}^E(\bk^{\prime}; \bp^{\prime}, r^{\prime} ) \  \eqno(4.9)
$$

The kernel $V_{\pi N}$ of this integral equation is determined by Eq. (4.5)
(an explicit expression for it is given below).
This kernel can be called a quasipotential: in the coordinate representation
it may depend not only on the particle coordinates but on their derivatives
as well. In our opinion, a popular name  "effective Hamiltonian" is
inappropriate for $K_{ZA}$. When one deals with an effective
Hamiltonian one has to argue that its eigenvalues and eigenvectors coincide
(at least, approximately) with those of an original Hamiltonian. In the
framework of our approach we do not need such a proof since we believe that
$K_{ZA}$ is in a sense a major part of the total Hamiltonian $H$ and,
therefore, the above approximate coincidence is provided.

The solutions of Eq. (4.9), which belong to a discrete spectrum (if it exists),
describe $\pi N$ bound states. As emphasized above, the solutions should be
found nonperturbatively. Continuous $\pi N$--mass values correspond to
$\pi N$--scattering states. The respective $S$--matrix elements may be
evaluated either exactly by using numerical methods of solving
the Lippmann--Schwinger equation for the $T$--matrix
with the interaction $K(\pi N \rightarrow \pi N)$ (see,e.g., \cite{KOMESHE90}),
or approximately in the framework of old--fashioned noncovariant perturbation
theory (see, e.g., Ch.1 in \cite{MATT57} or Ch.11 in \cite{SCH61}).

To obtain the explicit expression for $K(\pi N \rightarrow \pi N)$ , one needs
to seperate out the $a_c\dg b_c\dg a_c b_c$--kind terms of the commutators
$[{\cal R},{\cal V}^{\dagger}]$ and $[{\cal R},{\cal V}^{\dagger}]^{\dagger}$
(see Eq. (A.14))
$$
K(\pi N \rightarrow \pi N) = \frac{1}{2} : \biggl\{ [{\cal R},{\cal V}^{\dagger}]_{\pi N} +
 [{\cal R},{\cal V}^{\dagger}]_{\pi N}^{\dagger} \biggr\} :
$$
$$
= -\frac{1}{2} \int d\bk_2 d\bp_2 d\bk_1 d\bp_1
$$
$$
\biggl\{ [V^{-\bk_2}, R^{\bk_1}]_{11}(\bp_2 r_2;\bp_1 r_1) +
      [V^{-\bk_1}, R^{\bk_2}]_{11}^{\dagger}(\bp_2 r_2;\bp_1 r_1) \biggr\}
a_c\dg(\bk_2)
b_c\dg(\bp_2,r_2)
a_c(\bk_1)
b_c(\bp_1,r_1) \ , \eqno(4.10)
$$
where the symbol : : denotes the normal ordering\footnote{Henceforth
summation over the dummy spin indices is implied.}.

This result has been obtained in \cite{SHIVI74}(cf. Eq.(A.7) therein). The
corresponding coefficients
$V_{\pi N}(\bk^{\prime}, \bp^{\prime},r^{\prime};\bk,\bp,r)$ that determine
the pion--nucleon quasipotential (see Eq. (4.5)) are equal to
$$
\langle a_c\dg(\bk^{\prime}) b_c\dg(\bp^{\prime},r^{\prime}) \Omega\mid
K(\pi N \rightarrow \pi N) \mid  a_c(\bk) b_c(\bp,r) \Omega \rangle \  \eqno(4.11)
$$
They can be represented in the following covariant (Feynman--like) form
$$
V_{\pi N}(\bk_2, \bp_2,r_2;\bk_1,\bp_1,r_1) =
\frac{g^2}{(2\pi)^3} \delta(\bk_2 + \bp_2 - \bk_1 - \bp_1)
{ {1}\over {2 \sqrt{ \omega_{\bk_2} \omega_{\bk_1} } } }
{ {m}\over { \sqrt{ E_{\bp_2} E_{\bp_1} } } }
$$
$$
\bar{u}(\bp_2,r_2) \Biggl\{ \frac{1}{2} \biggl
[{ {1}\over { \cross{k}_2 + \cross{p}_2 + m} } + { {1}\over { \cross{k}_1 + \cross{p}_1 + m} } \biggr]
+ \frac{1}{2} \biggl
[{ {1}\over { \cross{p}_2 - \cross{k}_1 + m} } + { {1}\over { \cross{p}_1 - \cross{k}_2 + m} } \biggr]
\Biggr\} u(\bp_1,r_1) \
\eqno(4.12)
$$
(see also Appendix in \cite{SALE96}).

In order to comment this expression, let us consider the Feynman graphs in Fig. 2
for the $S$--matrix elements of $\pi N$ scattering. According to Feynman rules
the
four--momentum of the internal nucleon line in the graph 2a equals either
the sum $k_1 + p_1$ of the incoming four--momenta  or the sum
$k_2 + p_2$ of the outgoing four--momenta. These sums are equal due to the
energy and momentum conservation, viz., the $S$--matrix contains the
multiplier
$$
\delta(k_2 + p_2 - k_1 - p_1) =
\delta(\omega_{\bk_2} + E_{\bp_2} - \omega_{\bk_1} - E_{\bp_1})
\delta(\bk_2 + \bp_2 - \bk_1 - \bp_1) \  \eqno(4.13)
$$
Therefore, the Feynman propagator corresponding  to the internal line can be
written either as
$(\cross{k}_1 + \cross{p}_1 + m)^{-1}$  or as
$(\cross{k}_2 + \cross{p}_2 + m)^{-1}$.
In the case of quasipotential the energy conservation is not assumed (only
the total three--momentun is conserved) and hence $k_1 + p_1$ is not
necessarily  equal to $k_2 + p_2$. The representation (4.12) shows that
$V_{\pi N}$ includes the contribution associated with graph 2a and it can be
obtained if we juxtapose to the internal nucleon line the half--sum
$\frac{1}{2} \biggl
[{ {1}\over { \cross{k}_2 + \cross{p}_2 + m} } + { {1}\over { \cross{k}_1 +
\cross{p}_1 + m} } \biggr]$.

In the case of $S$--matrix we juxtapose to the internal line in graph 2b
either the propagator $(\cross{p}_2 - \cross{k}_1 + m)^{-1}$  or
$(\cross{p}_1 - \cross{k}_2 + m)^{-1}$. In the case of quasipotential
$p_1 - k_2 \neq  p_2 - k_1$, in general, and  the half--sum of these propagators
does correspond to the internal line (it is the second half--sum in the
curly brackets in the r.h.s. of Eq. (4.12)).

It follows from these observations that multiplying $V_{\pi N}$ by the factor
$ - 2\pi \imath \delta(\omega_{\bk_2} + E_{\bp_2} - \omega_{\bk_1} - E_{\bp_1})$
we shall obtain the $S$--matrix elements for $\pi N$ scattering
in the $g^2$--order.

So, we have seen that the r.h.s. of Eq. (4.12) resembles the Feynman
amplitudes being different from them in the two respects: i) the multiplier
$\delta(\bk_2 + \bp_2 - \bk_1 - \bp_1)$ is substituted instead of (4.13);
ii) the above Feynman propagators are replaced by the corresponding
half--sums.

\bigskip
\centerline{\em 4.3 Clothed Nucleon--Nucleon Eigenstates and Nucleon--Nucleon Quasipotential}
\bigskip

Now, we consider the $K_{ZA}$ eigenstates which belong to the $NN$ sector,
being superpositions of the kind,
$$
\Phi_{NN} = \sum_{r} \int d\bp_{1} d\bp_{2}
\Phi_{NN}(\bp_1,r_1;\bp_2,r_2)
b_c\dg(\bp_{1},r_1) b_c\dg(\bp_{2},r_2) \Omega \  \eqno(4.14)
$$
A subset of such states with a definite momentum $\ve P$ is determined by
Eq. (4.14) with the coefficients $\Phi_{NN}(1;2) \sim \delta(\bp_1 + \bp_2 - \ve P)$
\footnote{Here and sometimes
below we use the evident abbreviations, viz., $1 = (\bp_1, r_1)$, etc.}.
Obviously, $K_{ZA}\Phi_{NN}$ is the the  state vector of the same sector.
In fact, the operators $K_{\pi}$, $K(\pi N \rightarrow \pi N)$,
$K(\pi \bar{N} \rightarrow \pi \bar {N})$,
$K(N \bar {N} \rightarrow N \bar {N})$, $K(\pi \pi \rightarrow N \bar {N})$
and $K( N \bar {N} \rightarrow \pi \pi )$
involved in $K_{ZA}$ do not contribute to $K_{ZA}\Phi_{NN}$ (see Eqs. (4.1)
and (4.3)), and we find that $K_{ZA}$ is reduced to the operator
$K_N + K(NN \rightarrow NN)$. So, the
eigenvalue equation $K_{ZA}\Phi^E_{NN} = E\Phi^E_{NN}$ yields the equation
$$
\biggl[K_N + K( NN \rightarrow NN) \biggr]\Phi^{E}_{NN}=E\Phi^{E}_{NN}\  \eqno(4.15)
$$
in the sector.

The corresponding equation for $\Phi^E_{NN}(1;2)$ (see Eq. (4.14)) can be
derived from Eq.(4.15) by taking  scalar products of the both parts of the
latter with $ \langle b_c\dg(\bp_{1},r_1) b_c\dg(\bp_{2},r_2) \Omega \mid $.
Doing so, we get
$$
(E - E_{\bp_1} - E_{\bp_2}) \Phi^{E}_{NN}(\bp_1, r_1; \bp_2, r_2)
\hskip 60mm
$$
$$
= \int d\bp_1^{\prime} d\bp_2^{\prime}
\tilde V_{NN}(\bp_1, r_1, \bp_2, r_2;
\bp_1^{\prime}, r_1^{\prime} , \bp_2^{\prime}, r_2^{\prime} )
\Phi^{E}_{NN}(\bp_1^{\prime}, r_1^{\prime}; \bp_2^{\prime}, r_2^{\prime} )\
\eqno(4.16)
$$
with the properly symmetrized interaction (the quasipotential)
$$
\tilde V_{NN}(1, 2; 1^{\prime}, 2^{\prime} ) =
- \frac{1}{2} \biggl[ V_{NN}(1, 2; 1^{\prime}, 2^{\prime}) - V_{NN}(1, 2; 2^{\prime}, 1^{\prime})
- V_{NN}(2, 1; 1^{\prime}, 2^{\prime}) + V_{NN}(2, 1; 2^{\prime}, 1^{\prime})  \biggr] \
\eqno(4.17)
$$
for the two clothed nucleons.

The two-body operator $K(NN \rightarrow NN)$ (see Eq. (4.4)) is generated by
the second term in the curly brackets of Eq. (A.14) and its H.c. :
$$
K(NN \rightarrow NN)
= \frac{1}{2} : \biggl\{ [{\cal R},{\cal V}^{\dagger}]_{NN} +
 [{\cal R},{\cal V}^{\dagger}]_{NN}^{\dagger} \biggr\} : \   \eqno(4.18)
$$

One should note that the coefficients $V_{NN}(1, 2; 1^{\prime}, 2^{\prime})$
in Eq. (4.4) are not in the one--to--one correspondence with
$K(NN \rightarrow NN)$, viz., they can be changed without altering the latter.
For instance, the property
$$
b(1) b(2) = - b(2) b(1)\   \eqno(4.19)
$$
enables one to replace  $V_{NN}(1, 2; 1^{\prime}, 2^{\prime})$ in Eq. (4.4)
by $ - V_{NN}(1, 2; 2^{\prime}, 1^{\prime})$, and so on.

Moreover, the operator $K(NN \rightarrow NN)$  remains unaltered when
adding to $V_{NN}$ arbitrary  functions $S_L(1, 2 ; 1^{\prime}, 2^{\prime})$
or $S_R(1, 2 ; 1^{\prime}, 2^{\prime})$ which are symmetrical under
the transpositions $1 \leftrightarrow 2$ or $1^{\prime} \leftrightarrow 2^{\prime}$,
respectively. The above mentioned replacement
$V_{NN}(1, 2; 1^{\prime}, 2^{\prime}) \rightarrow - V_{NN}(1, 2; 2^{\prime}, 1^{\prime})$
is equivalent to the replacement
$$
V_{NN}(1, 2; 1^{\prime}, 2^{\prime}) \rightarrow V_{NN}(1, 2; 1^{\prime}, 2^{\prime})
+ S_R(1, 2 ; 1^{\prime}, 2^{\prime}) \  \eqno(4.20)
$$
with $S_R(1, 2 ; 1^{\prime}, 2^{\prime}) = - V_{NN}(1, 2; 1^{\prime}, 2^{\prime})
- V_{NN}(1, 2; 2^{\prime}, 1^{\prime})$.

A distinctive feature of the  coefficient (4.17) is its invariance with
respect to the transformation (4.20) with arbitrary $S_R$.

After these notations, we write down one of the possible expressions for
$V_{NN}$, that can be obtained using Eq. (4.18),
$$
V_{NN}(1, 2 ; 1^{\prime}, 2^{\prime} ) =
\hskip 60mm
$$
$$
- \frac{1}{2} \int d\bk
\Biggl\{ \frac{1}{ E_{\bp_1} - E_{\bp_1^{\prime}} -\omega_{\bk}} +
         \frac{1}{ E_{\bp_2^{\prime}} - E_{\bp_2} -\omega_{\bk}} \Biggr\}
V_{11}^{ - \bk}(\bp_1 r_1 ; \bp_1^{\prime} r_1^{\prime} )
V_{11}^{\bk}( \bp_2 r_2 ; \bp_2^{\prime} r_2^{\prime} ) \  \eqno(4.21)
$$

The respective quasipotential is
$$
\tilde V_{NN}(1, 2 ; 1^{\prime}, 2^{\prime}) =
- \frac{g^2}{(2\pi)^3} \delta( \bp_1 + \bp_2 - \bp_1^{\prime} - \bp_2^{\prime})
{ {m^2}\over {2 \sqrt{ E_{\bp_1} E_{\bp_2} E_{\bp_1{\prime}} E_{\bp_2{\prime}} } } }
$$
$$
\bar{u}(1)\gamma_5 u(1^{\prime})
\frac{1}{2} \Biggl\{ { {1}\over {(p_1 - p_1^{\prime})^2 - {\mu}^2} } +
         { {1}\over {(p_2 - p_2^{\prime})^2 - {\mu}^2} } \Biggr\}
\bar{u}(2)\gamma_5 u(2^{\prime})
$$
$$
- (1 \leftrightarrow 2 ) \  \eqno(4.22)
$$
Expression (4.22) is the $NN$ part of an one--boson--exchange interaction
derived via the Okubo transformation method in \cite{KOSHE93} (cf. \cite{FUZH95}).
The potential $\tilde V_{NN}$ consists of the direct term written explicitly
and the exchange term  $(1 \leftrightarrow 2 )$. In order to obtain the latter
one needs to replace $\bp_1, r_1 $ by $\bp_2, r_2 $ and $\bp_2, r_2 $ by
$\bp_1, r_1 $ in the former.

As has been pointed out in \cite{KOSHE93}\footnote{There one can find another
representation of the nucleon--nucleon quasipotential, which resembles the
expressions of old--fashioned perturbation theory (see, e.g., Ch. 13 in \cite
{SCH61}).}, a distinctive feature of the potential is appearance of a
covariant (Feynman--like) "propagator"
$$
\frac{1}{2} \Biggl\{ { {1}\over {(p_1^{\prime} - p_1)^2 - {\mu}^2} } +
 { {1}\over {(p_2^{\prime} - p_2)^2 - {\mu}^2} } \Biggr\} \ , \eqno(4.23)
$$
where $p=(E_{\bp},\bp)$ is the nucleon four-momentum. On the energy shell
that is when
$$
E_i \equiv E_{\bp_1}+E_{\bp_2} = E_{\bp_1^{\prime}}+E_{\bp_2^{\prime}} \equiv E_f
\  \eqno(4.24)
$$
the r.h.s. of Eq. (4.23) becomes the genuine Feynman propagator which appears when
evaluating the S--matrix for $NN$ scattering in the $g^2$ - order. The respective
graphs are displayed in Fig. 3. Like $V_{\pi N}$ the quasipotential
$\tilde V_{NN}$ can be associated with  these Feynman graphs being different
from the corresponding Feynman amplitude in the two respects, viz., $\tilde V_{NN}$
does not contain
$\delta (E_{\bp_1}+E_{\bp_2} - E_{\bp_1^{\prime}} - E_{\bp_2^{\prime}})$,
and "propagator" (4.23) now corresponds to the internal meson line in graph
3a. A more extended analysis of this observation  has been been given in
\cite{KOSHE93}.

In conclusion, one should note that all the quasipotentials are nonlocal since
the vertices and propagators in
Eqs. (4.12) and (4.22) are dependent not only on the relative three--momenta
involved  but also on their total three--momentum. They include the nonstatic
(recoil) effects in all orders of the socalled $\frac{1}{c^2}$ expansion
\cite{FOLKRA75}.

\bigskip
\centerline{\em 4.4 Other Clothed Eigenstates of Meson--Fermion System }
\bigskip

Up to now we have focused upon the clothed $\pi N$ and $NN$ states. Let us
discuss other clothed states.

If we start with the same "zeroth" approximation to the total Hamiltonian,
our description of clothed $\pi \bar N$ and $\bar N \bar N$ states will be
very similar to that given for $\pi N$ and $NN$ states. Actually, it is the case
where one has to deal with the charge--conjugated states. Here, we mean the
nucleon--antinucleon conjugation.

A different situation holds in the case of clothed fermion--antifermion

and two--meson states. In fact, the operator $K_4^{(2)}$ contains
the interactions $K(\pi \pi \leftrightarrow N \bar N) $. Therefore,
superpositions of
the $\pi \pi$ states $a_c\dg a_c\dg \Omega $ and the $N \bar N$ states
$b_c\dg d_c\dg \Omega $ taken separately cannot be $K_{ZA}$ eigenvectors.
So, one has to consider the eigenstates of a mixed kind,
$$
\Phi = \int d\bk_1 d\bk_2 \Phi_{\pi \pi}(\bk_1; \bk_2) a_c\dg (\bk_1) a_c\dg (\bk_2) \Omega
$$
$$
+ \int d\bp_{1} d\bp_{2} \Phi_{N \bar  N}(\bp_1,r_1;\bp_2,r_2)
b_c\dg(\bp_{1},r_1) d_c\dg(\bp_{2},r_2) \Omega \  \eqno(4.25)
$$

Calculation of the scalar products of  $K_{ZA} \Phi = E \Phi$ with
$\langle a_c\dg a_c\dg \Omega \mid $ and $\langle b_c\dg d_c\dg \Omega \mid $
leads to a set of coupled equations for the coefficients $ \Phi_{\pi \pi} $
and $\Phi_{N \bar N} $. Of course, one may obtain separate equations for
each of them. Thereat, the eigenvalue equation for $\Phi_{N \bar N} $  will
involve some terms of the $g^4$--order. Obviously, to be consistent they
should be disregarded within the ZA considered.

In the analogous manner one can study the eigenvalue problem for
clothed three--nucleon and more complicated states. However, handling with
$K_{ZA}$, we enter into the $3N$--problem only with the two--body interaction
$K(NN \rightarrow NN) $. It would be interesting to take into account the
three--body (six--operator) interactions (irreducible to two--body ones) that
are present in the total Hamiltonian $K$ starting from the term
$[R,[R,[R,V]]] \sim g^4$ not explicitly written in Eq. (2.23).

\bigskip
\centerline{\bf 5. Possible Modifications of the Clothing Approach}
\bigskip

\bigskip
\centerline{\em 5.1 Heitler's Unitary Transformation}
\bigskip

The "clothing" in Sect.2 has been realized in the framework of the
Schr\"odinger
picture, the "bare" and "clothed" operators being the Schr\"odinger ones.
Heitler in
his book \cite{HEIT54} discussed the corresponding UT's
in the
framework of the interaction picture. To establish the relation with
Heitler's
approach we shall derive Heitler's equation (which determines his UT)
starting from our equation (2.19) of Sect. 2:
$$
W(\alpha)[H_F(\alpha)+H_I(\alpha)]W^{\dagger}(\alpha) = K(\alpha)
= K_0(\alpha)+K_I(\alpha)
\eqno(5.1)
$$
Remind that the free part $K_0(\alpha)$ of the total Hamiltonian $K(\alpha)$
(expressed in terms of the clothed operators $\alpha$) is equal to
$H_F(\alpha)$,
$H_F(\alpha)$ being given by Eq. (2.6) in which bare operators $a,b$ and
$d$ are
replaced by the clothed ones $a_c, b_c$ and $ d_c$.
Now, let us write Eq. (5.1)
in terms of interaction picture operators defined, e.g., as
$$
\alpha_p(t)=e^{iK_0(\alpha)t}\alpha_p e^{-iK_0(\alpha)t}
\eqno(5.2)
$$
For this purpose multiply both parts of Eq. (5.1) by $\exp(iK_0t)$ from
the left
and by $\exp(-iK_0t)$ from the right and use the notation
$$
A(t)=e^{iK_0t}A(\alpha)e^{-iK_0t}=A(e^{iK_0t}\alpha e^{-iK_0t})=A(\alpha(t))
$$
Then we obtain,
$$
W(t)[H_F(t)+H_I(t)]W^{\dagger}(t)=K_0(t)+K_I(t)
\eqno(5.3)
$$

Multiply both parts of this equation by $W^{\dagger}(t)$ from the left and
use
$W^{\dagger}W=1$ and $K_0(t)=H_F(t)$ (see above). We get
$$
[H_F(t), W^{\dagger}(t)]=-H_I(t)W^{\dagger}(t)+W^{\dagger}(t)K_I(t)
\eqno(5.4)
$$
The l.h.s. of this equation is equal to
$-i\frac{\partial}{\partial t}W^{\dagger}(t)$ (see, e.g., Eq. (11.52)
in \cite{SCH61}). So, we find,
$$
H_I(t)W^{\dagger}(t)-i\frac{\partial}{\partial t}W^{\dagger}(t)=W^{\dagger}(t)K_I(t)
\eqno(5.5)
$$
This equation coincides with Heitler's equation written in \cite{HEIT54}
(Ch. 4,
between Eqs. (15.6) and $(15.7_1)$). The relation of Heitler's notations and
ours is
$$
S=W^{\dagger},\quad K=K_I, \quad H=H_I
$$
So, Heitler's basic equation is equivalent to our Eq. (2.19) or (5.1). But
his
goal differs from the goal of "clothing". He requires that $K_I$ must not
contain
interaction terms which give rise to virtual processes. By definition, the
latter can proceed in spite of the inequality of energies of the initial and
final states (here "energy" means an eigenvalue of the free part of the total
Hamiltonian). Our bad terms also lead to virtual processes
(e.g., $\Omega \rightarrow \pi \bar N N$ or $N \rightarrow N \pi$) but there
are
many other virtual processes not generated by bad terms (e.g.,
$\pi \pi \rightarrow \bar N N$
or $\pi N \rightarrow \pi \pi N$ at low initial energies).

Of course, Heitler's requirement can be imposed also in the Schroedinger picture
and this has been done by Sato et al. \cite{SATO92,SALE96}.

\bigskip
\centerline{\em 5.2 Heitler-Sato Approach Versus the Clothing One}
\bigskip

We shall show here that under a condition all bad terms produce virtual
processes.
As there are many virtual processes which are not induced by bad terms,
one may state that the Heitler-Sato condition is stronger than the bad terms
elimination requirement.

Let us remind the exact definition of bad terms.  They are either two-,
three-,\ldots
operator terms which contain only creation operators (and do not contain
destruction
operators) or three-, four-,\ldots operators  containing only one destruction
operator.
We call bad also the terms which are Hermitian conjugated to the
above-mentioned.
Using the notion of the class defined in Sect. 2 one may define the bad terms
as terms of the class $[n,0]$ and $[n,1]$, $n\geq 2$ and their H.c.
Note that two--operator terms of the kind $a^{\dagger}(k)a(k)$ are not
attached to the bad ones.

Bad interaction terms are responsible for the processes
$$
\begin{array}{llll}
no\ particles & \leftrightarrow & 2,3,\ldots & particles\\
one-particle  & \leftrightarrow & 2,3,\ldots & particles
\end{array}
\eqno(5.6)
$$
This property also may be considered as the bad terms definition.
The processes (5.6) will be called bad below.

Let us prove the Statement: All bad processes are virtual ones under
a condition on the masses of interacting particles.
Indeed, energy is evidently not conserved in the bad process
"vacuum $\rightarrow$
several particles": the initial energy is zero while the final state energy
cannot
be less than the sum of the final masses. Further, consider the bad process
$a_j \rightarrow a_1+a_2+\ldots$, in which the particle $a_j$ with the mass
$m_j$ converts
into particles with masses $m_1, m_2,\ldots$. In the particle $a_j$ rest
frame the
initial energy is $m_j$. The energy is not conserved trivially if sum of the
masses
$\Sigma_i m_i$ of the final particles exceeds $m_j$. The energy is conserved
if
$m_j>\Sigma_i m_i$ and final particles possess nonzero momenta. The set of
inequalities $m_j<\Sigma_i m_i$, $\forall_j$ is the very Statement condition
on the particle masses.

In the Yukawa model the bad processes
$N\rightarrow N\pi$, $N\rightarrow N\bar N N$,
$N\rightarrow N \pi \pi$, etc. are certainly virtual,
the process $\pi \rightarrow
\bar N N$ being virtual if $\mu < 2m$. Under this condition all the bad
processes generated with the Yukawa model are virtual.

Let us consider the simplest virtual processes of the Yukawa model which are
of the order $g^1$,
namely $\Omega \rightarrow N \bar N \pi$, $N\rightarrow N \pi$, $\pi \rightarrow  \bar N N$
(if $\mu < 2m$). They coincide with the (simplest) bad processes of
the order $g^1$.
The corresponding three-operator interaction terms are of the order $g^1$ and can be
removed by the unitary operator $W=\exp R$ where $R$ is a three-operator expression
of the order $g^1$ (see Sect.2).

\bigskip
\centerline{\bf 6 The Okubo Blockdiagonalization Method}
\bigskip

Here, following \cite {OKU54} we regard the UT $H \to H_U \equiv U\dg HU$ that
makes the Hamiltonian $H$ (generally speaking, an Hermitian operator) block
diagonal (cf. our brief discussion in Introduction). After the transformation
the primary $H$ eigenvalue problem is reduced to the diagonalization of
separate neardiagonal blocks of $H_U$.

It is well known that the transformation $H \to H_U$ can be interpreted either
as  the connection between the two matrices of one and the same operator
$H$ with respect to the different orthonormal bases, or as the relationship between the
two operators $H$ and $H_U$ unitarily connected via the operator U.
We shall start from the first point of view.

\bigskip
\centerline{\em 6.1 UT as Change of Basis}
\bigskip

Let $H_{n^\prime n} \equiv \langle n^\prime{\mid}{H}{\mid}n \rangle \,
(\forall n^\prime, n)$ be the $H$ matrix with respect to a complete set of
orthonormal vectors ${\mid}n \rangle $ and
$H_{\nu^\prime\nu} \equiv (\nu^\prime{\mid}{H}{\mid}\nu) \,
(\forall \nu^\prime, \nu)$ represents $H$ in another orthonormal basis
${\mid}\nu)$.
The indices $n(\nu)$ can take on discrete or/and continuous values. One has
(see, e.g., \cite{FANO71}, Ch.I)
$$
H_{\nu^\prime \nu} \equiv ( \nu^\prime{\mid}{H}{\mid}\nu ) =
{\bf S}_{n^\prime} {\bf S}_n \,(\nu^\prime{\mid}n^\prime \rangle
\langle n^\prime{\mid}{H}{\mid}n \rangle \langle n{\mid}\nu) \, \eqno(6.1)
$$
where $ {\bf S}_n$ denotes a sum or/and an integral over $n$.

The r.h.s. of Eq. (6.1) can be written as the matrix product
$$
{\bf S}_{n^\prime} {\bf S}_{n} \, (U\dg)_{\nu^\prime n^\prime} H_{n^\prime n} U_{n \nu}
\equiv (U\dg H U)_{\nu^\prime \nu} \, \eqno(6.2)
$$
where we have introduced the notation $U_{n \nu} \equiv \langle n{\mid}\nu)$.
Then
$$
(U\dg)_{\nu^\prime n^\prime} \equiv U_{n^\prime \nu^\prime}^{*} =
\langle n^\prime{\mid}\nu^\prime)^{*} = (\nu^\prime{\mid}n^\prime \rangle
$$

The transition matrix $U$ is unitary in the following sense:
$$
(U\dg U)_{\nu^\prime \nu} \equiv {\bf S}_{n} \,(\nu^\prime{\mid}n \rangle
\langle n{\mid}\nu) = \delta_{\nu^\prime \nu}\ \eqno(6.3a)
$$
and
$$
(U U\dg)_{n^\prime n} \equiv {\bf S}_{\nu}\, \langle n^\prime{\mid}\nu)
(\nu{\mid}n \rangle = \langle n^\prime{\mid}n \rangle = \delta_{n^\prime n}\
\eqno(6.3b)
$$
Recall that all these basis vectors are orthonormal.

Whenever $U$ is given one can express ${\mid}\nu)$ in terms of ${\mid}n \rangle$,
$$
{\mid}\nu) = {\bf S}_{n}\, {\mid}n \rangle\, U_{n \nu} \, \eqno(6.4)
$$

One should point out that the spectrum of indices $n$ enumerating the vectors
${\mid}n \rangle$ needs not to be identical to the spectrum of indices $\nu$
enumerating ${\mid}\nu)$. For example, let us consider the Hamiltonian $H_{os}$
of one--dimensional quantum oscillator, whose matrix in the
coordinate representation can be made diagonal by using the Hermite functions
$\psi_{\nu}(x)\, (\nu = 0,1,2, \ldots)$ :
$$
( \nu^\prime{\mid}H_{os}{\mid}\nu ) =
\int\limits_{- \infty}^\infty dx^\prime \int\limits_{-\infty}^\infty dx \,
\psi^{*}_{\nu^\prime}(x^\prime) \,
\langle x^\prime{\mid}H_{os}{\mid}x \rangle \,
\psi_{\nu}(x) \sim \delta_{\nu^\prime \nu} \,  \eqno(6.5)
$$
(cf. Eqs. (6.1)--(6.2)). In this case the transition matrix $\langle x{\mid}\nu) =
\psi_{\nu}(x)$ is not "square" because its columns(rows) are enumerated by the
integer(continuous) numbers. Such a matrix can be called rectangular. Note that in
a finite--dimensional space the transition matrix always is square.
This exemplifies that one cannot, in general, diagonalize $H$ by means of a
square matrix.

In this connection, one should note that the diagonalization scheme considered
by Okubo and  exposed below is not aimed at a perfect solution of $H$--eigenvalue
problem. Rather, it sets a more humble task, viz., to find (approximately)
some of its eigenvalues and eigenvectors. Okubo \cite{OKU54} suggested a
realization of the scheme via a square matrix $U$.

\newpage

\bigskip
\centerline{\em 6.2 Blockdiagonalization in Matrix Form}
\bigskip

In accordance with \cite{OKU54} we require that the $H$ matrix in the new
basis
$$
H_{\nu^\prime\nu} = (U\dg\, H\, U)_{\nu^\prime \nu} \ \eqno(6.6)
$$
should have the blockdiagonal form,
$$
U\dg H U \equiv
K = \left( \begin{array}{cc}
           K_{11}& 0\\
           0&   K_{22}\end{array}\right) \ \eqno(6.7)
$$
with the two blocks $K_{11}$  and $K_{22}$ the meaning of which will be
clarified a little later. This requirement must determine the matrix $U$
that in its turn enables one to construct the new basis vectors ${\mid} \nu)$
(see Eq.\,(6.4)).

Now, following Okubo, we shall confine ourselves to finding
such a square matrix $U$ that ${\mid}\nu ) = U\,{\mid}n \rangle $. In other
words , it is assumed that one can set an one--to--one correspondence between
the indices $n$ and $\nu $.

First of all, let us turn to the definition of $K_{11}$. It is a block of the matrix $K$
with elements that are enumerated
by indices $\nu_1$ (or $n_1$) which belong to a subset of all $\nu$ (or $n$)
values. The $K_{22}$ elements are enumerated by the remaining $\nu$ values
denoted through $\nu_2$. Let us give examples of the $\nu_1$ choice.

One may take one value of $\nu$ as $\nu_1$, viz., the index of the
vacuum state. Hence, $K_{11}$ has one element. If we are able to find $U$
which leads to Eq. (6.7) then we can construct a normalized $H$ eigenstate,
namely the physical vacuum ${\mid}0) = {\bf S}_n \,{\mid}n\rangle \, U_{n0}$. After this step
we consider the block $K_{22}$ as a starting matrix for the subsequent
blockdiagonalization, viz., to introduce a new set of $\nu_1 '$ (e.g., let
$\nu_1 '$ be indices of the one--particle $H_F$ eigenstates), to find a new
$U '$. At the next stage one can enumerate elements of the recurrent block 11
by indices of two--particle states, e.g., the states "two nucleons, no mesons"
(this is Okubo's example, see Sect. 2 in \cite{OKU54}). One more choice of
$K_{11}$ will be discussed in Subsect. 6.5.

The option of $\nu_1$ or $n_1$ allows one to divide $H$ into
the four blocks: $H_{11}$ with elements
$\langle n_1 '{\mid}H{\mid}n_1 \rangle $, $H_{22}$ (elements
$\langle n_2 '{\mid}H{\mid}n_2 \rangle $), $H_{12}$ (elements
$\langle n_1{\mid}H{\mid}n_2 \rangle $ ), and $H_{21} = H_{12}\dg$. The matrix
$U$ can be represented analogously. Keeping this in mind the l.h.s. of Eq. (6.7)
can be rewritten as the product of matrices composed of the blocks described
above (see, e.g., Ch. 0.7 in \cite{HOJO86} and Ch. 1.6 in \cite{LANK69}),
$$
\left( \begin{array}{cc}
        U_{11}\dg & U_{21}\dg\\
        U_{12}\dg & U_{22}\dg\end{array}\right)
\left( \begin{array}{cc}
        H_{11} & H_{12}\\
        H_{21} & H_{22}\end{array}\right)
\left( \begin{array}{cc}
        U_{11} & U_{12}\\
        U_{21} & U_{22}\end{array}\right)
= \left( \begin{array}{cc}
           K_{11}& 0\\
           0&   K_{22}\end{array}\right) \ \eqno(6.7 ')
$$

One can consider Eq. (6.$7^\prime$ ) as the equation for $U$. Okubo suggested
to seek its solution in
the class of unitary matrices of the kind (see {\it Note} underneath)
$$
U =  \left( \begin{array}{cc}
            U_{11} & U_{12}\\
            U_{21} & U_{22}\end{array}\right) =
\left( \begin{array}{cc}
       (I_1 + A\dg A)^{- \frac{1}{2}}    & - A\dg (I_2 + A A\dg)^{- \frac{1}{2}}\\
       A (I_1 + A\dg A)^{- \frac{1}{2}}  & (I_2 + A A\dg)^{- \frac{1}{2}} \end{array}\right) \ , \eqno(6.8)
$$
where $I_1(I_2)$ is the unit matrix for the subset of indices $n_1(n_2)$,
and $A$ is an rectangular matrix with elements $A_{n_2 n_1}$ of the $U_{21}$
kind, which should be determined. Under this convention the unit matrix for
the full set of indices has the block structure
$$
1 = \left( \begin{array}{cc}
           I_1 & 0\\
           0 &   I_2\end{array}\right) \ , \eqno(6.9)
$$
and we have the properties
$$
A I_1 = I_2 A = A \ , \eqno(6.10a)
$$
$$
[A\dg A, I_1] = 0 \ \eqno(6.10b)
$$

The matrices $A\dg A$ (of the $U_{11}$ kind) and $A A\dg$ (of the $U_{22}$ kind)
are square, hermitian and positively definite, so that
the square roots $\sqrt{I_1 + A\dg A}$ and $\sqrt{I_2 + A A\dg}$ can be defined
(see, e.g., Ch. 5.8 in \cite{FANO71}). It is readily verified that
the r.h.s. of (6.8) meets left unitarity $U\dg \,U = 1$. Other unitarity
condition $U U\dg = 1$ can be proved by means of equality $A f(A\dg A) =
f(A A\dg)A$ that is valid if $f(x)$ is a polynomial or a series of $x$.

The requirement for blockdiagonalization, $K_{21} = 0$ (or $K_{12} = 0$),
gives the equation
$$
H_{21} + H_{22} A - A H_{11} - A H_{12} A = 0 \ \eqno(6.11)
$$
for $A$ determination. This equation is equivalent to the condition (13) from
\cite{OKU54}.

Eq. (6.11) is nonlinear and it can be solved exactly only in a few simple
cases (see Sect.\,6 in \cite{OKU54} and Appendix $ {\rm C} $ ). For
realistic field models one has to develop a perturbative method in order to
find $A$ (see Subsect.\,6.5).

\underline{Note}. Generally speaking, the left unitarity for a matrix $U$
with the nonsingular block $U_{11}$  enables to express blocks
$U_{ij} (i,j = 1,2)$ through the matrix  $A$ such that $U_{21} = A U_{11}$
and blocks $S_{11} = S_1$ and
$S_{22} = S_2$ of an arbitrary matrix $S$,
$$
S = \left( \begin{array}{cc}
           S_1 & 0\\
           0 &   S_2\end{array}\right)
$$
with $S_1\dg S_1 = I_1$ and $S_2\dg S_2 = I_2$ (cf. Eqs. (10) in
\cite{OKU54}).
The corresponding matrix $U_{Okubo}$ (i.e., its representation after Okubo)
can be written as the product $US = U_{Okubo}$ where $U$ is given by (6.8).
However, since the UT via $S$ conserves the blockdiagonal structure of $K$
it is sufficient to consider only the form (6.8). Of course, Okubo's
representation is not most general form for unitary matrices. The simplest
exception is the $2 \times 2$ matrix
$$
\left( \begin{array}{cc}
        0 & 1\\
        1 & 0\end{array}\right).
$$

\bigskip
\centerline{\em 6.3 Blockdiagonalization with Projection Operators}
\bigskip

Here the UT $H' = U\dg\, H\,U$ is interpreted as a relation between
the different operators $H$ and $H'$\footnote{Unlike the UT $K = W\,H\,W\dg$
in Subsect. 2.3 that represents the same operator $H$.}, where $U$ is
a unitary operator.
By definition, the operator is a linear one--to--one mapping of a Hilbert space
$ {\cal H} $ onto itself, i.e., an isomorphism of $ {\cal H} $.
In particular, it transforms the orthonormal basis vectors $ {\mid} n \rangle $
$ \in {\cal H} $ into the vectors $ {\mid} n ) = U {\mid} n \rangle $ of
other orthonormal basis $ \in {\cal H} $. Therefore, we can write
$$
\langle n' {\mid} H' {\mid} n \rangle =
\langle n' {\mid} U\dg H U {\mid} n\rangle = (n' {\mid} H {\mid} n )\ ,
\eqno(6.12)
$$
Eq. (6.12) sets up close links  of $ H' = U\dg H U$ with the UT considered
in Subsects. 6.1 and 6.2.

One can deal with $H^{\prime} = U^{\dag} H U$ in the same manner as in
Subsect. 6.2 introducing matrices of $H$, $U$, $H^{\dag}$ with
respect to a basis $|n\rangle$ and representing them in block form,
eq. (6.7 '). But one can realize the Hamiltonian blockdiagonalization
without reference to any basis on ${\cal H}$. We shall introduce
(cf. \cite{OKU54})
the projection operators
$\eta_1$ and $\eta_2 = 1 - \eta_1$ onto a subspace $ {\cal H}_1
\subset {\cal H} $ and its complement $ {\cal H}_2 $ such that
$ {\cal H} = {\cal H}_1 \bigoplus {\cal H}_2 $\footnote{Of course, the
particular option of $ {\cal H}_1$  depends on the nature of the problem
(see Appendix ${\rm C} $ and Subsect. 6.5).}. Mathematically strict definitions
of projection operators and their properties in a Hilbert space can be found
in \cite{NEUM32} (Ch.2). Of great importance for us is the property
$$
\eta_i \eta_k = \eta_k \eta_i = \delta_{ik} \eta_i \ \eqno(6.13)
$$
\centerline{(i,k = 1,2)}
Then, for any operator $O$ in $ {\cal H} $ one can write
$$
O = (\eta_1 + \eta_2)\, O\, (\eta_1 + \eta_2) = \sum_{i,j = 1,2} O^{ij}
\ , \eqno(6.14)
$$
where
$$
O^{ij} = \eta_i\, O\, \eta_j \  \eqno(6.15)
$$
with
$$
\eta_i\, O^{kj} = \delta_{ik} O^{ij}, O^{kj}\,\eta_l = O^{kl} \delta_{jl}\
\eqno(6.16)
$$

In terms of such decompositions the product of the two operators
$A = \sum_{ij} A^{ij}$ and $B = \sum_{ij} B^{ij}$ can be written as
$$
A\, B = \sum_{ijk} A^{ij} B^{jk} \ , \eqno(6.17)
$$
so that
$$
(A\,B)^{ik} = \sum_{j} A^{ij} B^{jk} \  \eqno(6.18)
$$
By using the rule we find
$$
H' = U\dg HU = \sum_{imjk} (U\dg)^{mi} H^{ij}U^{jk}\  \eqno(6.19a)
$$
or
$$
H' = \sum_{imjk} (U^{im})\dg H^{ij}U^{jk}\ , \eqno(6.19b)
$$
Remind that the projectors $\eta_i $ are hermitian , i.e., $\eta_i\dg = \eta_i $.

One can set an one--to--one correspondence between the operators $H^{ij}$ and
the matrix blocks $H_{ij}$ introduced in Subsect. 6.2. Actually, each operator
$H^{ij}$\, $(i,j = 1,2)$ acts in the full Hilbert space $ {\cal H} $ but its
matrix with respect to the basis $\{ {\mid}n \rangle \}$ has merely one nonzero
block $H_{ij}$ with the three remaining blocks being zero. Analogous relation
takes place between $U^{ij}$ and $U_{ij}$.

Further, the operators $U^{ij}$ in Eqs. (6.19) can be expressed through an
arbitrary operator ${\cal A}$ of  the kind $21$, i.e.,
${\cal A} = \eta_2 {\cal A} \eta_1 $. For instance, one can write
(cf. Eq. (6.8))
$$
U^{21} = {\cal A} U^{11} = {\cal A}(1 + {\cal A}\dg {\cal A} )^{ - \frac{1}{2}}\eta_1
= {\cal A}(1 + {\cal A}\dg {\cal A} )^{ - \frac{1}{2}} \ , \eqno(6.20)
$$

Okubo's requirement for decoupling the two subspaces $ {\cal H}_1$ and
$ {\cal H}_2$ means
that it must  be ${H'}^{21} = 0$ or according to  Eq. (6.19a)
$$
\sum_{ij} (U\dg)^{2i} H^{ij}U^{j1} = 0\  \eqno(6.21)
$$
It leads to the nonlinear equation for ${\cal A} $
$$
\eta_2 \{ H + [ H, {\cal A}] - {\cal A} H {\cal A}\} \eta_1 = 0 \ \eqno(6.22)
$$
As a matter of fact, it is the same equation as Eq. (6.11), viz., one may
consider the latter as the record of Eq. (6.22) for the only nonzero block
$A$ of the operator $ {\cal A} $.

The solution of Eq. (6.22) , if it exists (cf. the discussion in
\cite{GLOMU81} ), yields the Hermitian operator
$$
{H'}^{11} = \eta_1 (1 + {\cal A}\dg {\cal A} )^{ - \frac{1}{2}} (1 + {\cal A}\dg)
H (1 + {\cal A}) (1 + {\cal A}\dg {\cal A} )^{ - \frac{1}{2}} \eta_1 \ ,
\eqno(6.23)
$$

The resultant operator $H'$ given by Eqs. (6.19) (see also Eq. (6.23)) has
a block diagonal structure that simplifies determination of its eigenvalues and
eigenvectors. However, $H' \ne H$ and therefore $H'$ eigenvectors are not
$H$ eigenvectors. Recall that our UT is aimed at solving the $H$ eigenvector
problem (at least, its partial solution).
But if we have succeeded in finding  ${H'}^{11}$ eigenvectors
one can easily get the corresponding $H$ eigenvectors. Actually, if we have
${H'}^{11} \Psi_1 = E \Psi_1$ $(\Psi_1 \in {\cal H}_1) $, then $ \Psi_1$ is
also ${H'}$ eigenvector,
$$
H'\Psi_1 = ({H'}^{11} + {H'}^{22}) \Psi_1 = E \Psi_1\  \eqno(6.24)
$$
It follows from $U\dg\,H\,U \Psi_1 = E \Psi_1 $ that  $U \Psi_1$ is $H$
eigenvector,
$$
H\, U\Psi_1 = E\, U\Psi_1 \ \eqno(6.25)
$$

The obtained relation between $H'$ and $H$ eigenvectors can be represented in
another form. Let us expand the $H'$ eigenvector in the basis vectors
$ {\mid} n \rangle $,
$$
{\mid} \Psi_1\rangle = {\bf S}_n {\mid} n \rangle c_n
$$
The corresponding $H$ eigenvector $U\Psi_1$ has the same expansion
coefficients $c_n$ with respect to the other basis $ {\mid} n ) = U{\mid} n \rangle $,
$$
U \Psi_1 = {\bf S}_n U {\mid} n \rangle c_n = {\bf S}_n {\mid} n ) c_n\
\eqno(6.26)
$$

Exact solutions of the decoupling equation (6.22) can be derived for a few
simple cases
(see, e.g., Sect.\,6 in \cite{OKU54}). One of them is the so-called scalar
field  model (see Ch. 12 in \cite{SCH61}). The respective solution is very
instructive having many attributes of more realistic field models.
It is given in Appendix ${\rm C}$.

\bigskip
\begin{center}{\em 6.4 New Creation--Destruction Operators within the UT
Method.\\ Comparison with the Clothing Procedure}
\end{center}
\bigskip
As before (see Sect.\,2) we consider that the original field Hamiltonian $H$
is a function (functional) of the bare creation--destruction operators.
We denote their set by the same symbol $a$, $a_p$ being one of them. Moreover,
one may assume that the unitary operator $U$ also is a function of $a$. Then,
$H' = U\dg H U$ may be written as
$$
H'(a) = U\dg (a) H(a) U(a)\ \eqno(6.27)
$$

In order to compare the Okubo approach and the  clothing procedure developed
in Sect.\,2 let us introduce the set $\tilde a$ of new creation--destruction
operators (with the denotation $\tilde a_p$ for one of them) defined as
$$
\tilde a_p = V(a) a_p V\dg (a) \quad \forall p \ , \eqno(6.28)
$$
or
$$
a_p = V\dg (\tilde a) \tilde a_p V(\tilde a) \ , \eqno(6.29)
$$
where $V(a)= V(\tilde a) \ne 1$ is an arbitrary unitary operator (not
necessarily coincident with $U(a)$).

Applying the UT $V$ to the both sides of Eq. (6.27), we have
$$
V(a)\,H'(a)\,V\dg (a) = H'(\tilde a) = U\dg(\tilde a)\,H(\tilde a)\,U(\tilde a)\
\eqno(6.30)
$$

Further, by means of (6.29) the total Hamiltonian can be expressed (cf.
the derivation of Eq.(2.17) ) in terms of the new operators $\tilde a$,
$$
H(a) = H(V\dg (\tilde a)\,\tilde a\, V(\tilde a) ) =
V\dg (\tilde a)\, H(\tilde a)\, V(\tilde a)\  \eqno(6.31)
$$
One should stress that the operators $H'(\tilde a)$ and $H(\tilde a)$ are
different from $H'(a)$ and $H(a)$, respectively, if $V \ne 1$. However, if
$V = U$, than $H'(\tilde a)$ turns out to be equal to the starting
Hamiltonian $H(a)$. Actually, in this case the r.h.s of Eq. (6.31) gets to be
equal to $H'(\tilde a)$ in accordance with Eq. (6.30), so that
$$
H(a) = H'(\tilde a) = U\dg (\tilde a) H(\tilde a) U(\tilde a)\ \eqno(6.31')
$$

The option  $V = U = W\dg $ gives rise to the clothed operators
$\tilde a = \alpha = W\dg a W $ (see Subsect. 2.3), and then
$H'(\alpha) = K(\alpha) = W H(\alpha) W\dg$.

Thus, our consideration shows how the UT $H \to H'=
U\dg H U $  can be reduced to a trans\-for\-ma\-ti\-on of clothing type.
Of course, there are distinctions between the Okubo approach and
the procedure shown in Sect. 2. They are due to those purposes which are
inherent to each of them. We shall return
to this point in Subsect. 6.6. However, let us note here that although the
operator $U_{\it Okubo}(a)$ determined by solving the decoupling equation
(6.22) and the operator $ U_{\it clothing} (a) = W\dg (a)$ are different
functions
of $a$ one cannot {\it \'a priori} exclude a resemblance or even the perfect
coincidence of some approximations to them.

Note one more definition of $\tilde a$  by the relations
$$
(n '{\mid} \tilde a_p{\mid}n ) = \langle n '{\mid}\,a_p{\mid}n \rangle  \quad
\forall n', n , p \ , \eqno(6.32)
$$
where $ {\mid}n ) = U\,{\mid}n \rangle $ (cf. the beginning of Subsect. 6.3). In
particular, Eq.(6.32) means that the matrix elements of the new
meson destruction operator $\tilde a(\bk)$ with respect to the new vectors
${\mid}n)$ are equal to the corresponding matrix elements of the bare meson
destruction operator $a(\bk)$ with respect to the old vectors ${\mid}n \rangle$.
It is clear that this definition with $U = U\,(a)$ is equivalent to (6.28).

\newpage
\bigskip
\begin{center}{\em 6.5 Perturbative Construction of  Okubo's Unitary
Transformation.\\
Elimination of Mesonic Degrees of Freedom}
\end{center}
\bigskip

Here we present the approximate solution of Okubo's equation and determination
of the block $H'^{11}$  that have been given in \cite{KOSHE93}. Just as in
that
paper let us consider the system of fermions (nucleons) and mesons (pions)
with Yukawa--type interaction linear in the meson field (see Sect. 2). In this
model the basis states ${\mid}n \rangle$ are enumerated by the two indices
${\mid}n \rangle = {\mid}mf \rangle $, viz., $m(f)$ enumerates the meson
(fermion) states, ${\mid}0_m f \rangle $ being the state without mesons
(the $N \bar N$ contents may be arbitrary). In accordance with the Okubo
idea we decompose the full space ${\cal H}$ of meson--nucleon states into
two subspaces (sectors), namely, the fermion (nucleon) sector ${\cal H}_0$
which is composed of no--meson  states, being spanned onto the subset
$\mid 0_m f \rangle $, and its orthogonal complement ${\cal H}_{comp}$ that
consists of the states with nonzero meson number. The projection operator
$\eta_1 $ into ${\cal H}_0$ \footnote{In \cite{KOSHE93} the letter $P$ has
been used instead of $\eta_1$.} can be constructed as
$$
\eta_1 = {\bf S}_f \mid 0_m f \rangle \langle 0_m f \mid \ .
\eqno(6.33)
$$

By definition,
$$
a(\bk) \mid 0_m f \rangle = 0  \quad \forall f , \bk  \ , \eqno(6.34)
$$
and therefore
$$
a(\bk) \eta_1 = \eta_1 a\dg (\bk) = 0 \  \eqno(6.35)
$$

The operator of interest $H'^{11}$ has the following matrix structure
(cf. the note after Eqs. (6.19)):
$$
H'^{11} = \left( \begin{array}{cc}
                H'_{11}& 0\\
                      0& 0\end{array}\right) \ , \eqno(6.36)
$$
where the block $H'_{11}$ consists of the elements
$ \langle 0_m f '\mid H' \mid 0_m f \rangle $, $\forall f ',f$. At the same
time the matrix
$$
H'^{22} = \left( \begin{array}{cc}
                      0& 0\\
                      0& H'_{22}\end{array}\right) \ , \eqno(6.37)
$$
contains the elements $ \langle m 'f '\mid H' \mid m f \rangle $,
$\forall m '\ne 0_m \ne m$ , $\forall f ', f$. For brevity , henceforth
the index $0_m$ of the no--meson state will be replaced by $0$.

In order to simplify the subsequent equations let us separate the
matrix of any operator $O$ that acts in ${\cal H} = {\cal H}_0 \oplus
{\cal H}_{comp}$ into the subblocks $[O_{m 'm}]$ such that
$$
{[O_{m 'm}]}_{f'f} = \langle m 'f '\mid O \mid m f \rangle \  \eqno(6.38)
$$
So, the elements of these subblocks are marked by the fermionic indices $f$.
The subblocks can be called fermionic. The block $H_{11}$ coincides with
$[H_{00}]$, etc.
The rectangular matrix $A$, which determines $U$ in Eq. (6.8), consists of
the subblocks $[A_{m '0}]$ with $m ' \ne 0$, i.e., the submatrices $A_{\bk 0}$
with $m ' = \bk $ for  one--meson states marked by the momentum ${\bk}$ ,
the submatrices $[A_{\bk_1 \bk_2 ; 0}]$ with $m ' = (\bk_1 \bk_2)$ for
two--meson states , and so on.

Now, the key equation $H' = U\dg H U$ can be rewritten as the set of
equations for the subblocks $ [U_{m 'm}] $ ,
$$
[H'_{\mu ' \mu}] = S_{m '} S_{m} [U\dg _{\mu ' m '}] [H_{m 'm}] [U_{m \mu}]\ ,
\quad \forall \mu ', \mu \ , \eqno(6.39)
$$
where $\mu$ runs the same values as $m$ does.

In their turn the subblocks in question may be considered as the matrices of
operators which act onto fermionic degrees of freedom. Such operators have been
represented in \cite{KOSHE93} as functions of the bare creation--destruction
operators  but fermion ones only. Let us denote the set of them by the symbol
$\f $. Then, one can write, e.g.,
$$
[ U_{m 'm}(\f)]_{f'f} = \langle m 'f '\mid U \mid m f \rangle \ \eqno(6.40)
$$
Similarly we introduce the operators $H'_{m ' m}(\f)\, (H_{m ' m}(\f))$ which
correspond to $[H'_{m ' m}] ([H_{m ' m}])$.

Under such convention the relation $H' = U\dg H U$ can be reduced to a hybrid
form being expressed in terms of the functions of $\f$ and the matrix
elements with respect to the meson states,
$$
H'_{\mu ' \mu}(\f) = S_{m '} S_{m} U\dg _{\mu ' m '}(\f) H_{m 'm}(\f) U_{m \mu}(\f) \
\eqno(6.41)
$$

Such representation of the operators turns out to be convenient when finding
approximate solutions of Eq. (6.22) which can be rewritten as
$$
\eta_2 \{ H_I + [ H_F, {\cal A}] + [ H_I, {\cal A}] -
{\cal A} H_I {\cal A}\} \eta_1 = 0 \ \eqno(6.42)
$$
since $\eta_2 H_F \eta_1 = 0 $.

We consider its approximate solution assuming
$$
{\cal A} = {\cal A}^{(0)} + {\cal A}^{(1)} ,
{\cal A}^{(0)} \sim g^0 , {\cal A}^{(1)} \sim g^1 \    \eqno(6.43)
$$
The contribution  ${\cal A}^{(0)}$ is absent since with
interaction switched off ($g \to 0$) the Hamiltonian $H = H_F$
is already blockdiagonal. In this connection, remind that $H_I = V + M_{ren}$
where the mass counterterms $M_{ren} = M_{ferm} + M_{mes} $ are determined
by Eq. (2.7).

Such ${\cal A} = {\cal A}^{(1)}$ turns into zero only those terms of
Eq. (6.42) which are of the $g^1$--order, i.e.,
$$
\eta_2 \left( \left [ H_F,{\cal A} \right] + V \right) \eta_1 = 0 \ ,
\eqno(6.44)
$$
One can consider Eq. (6.44) as a relaxed form of Okubo's constraint
$H'^{21} = 0$ (or Eq. (6.42) ), which may be imposed instead of the latter.

Using the commutativity of $H_F$ and $\eta_1$ let us rewrite Eq. (6.44)
as
$$
\left [ H_F,{\cal A} \right] + \eta_2 V \eta_1 = 0 \ , \eqno(6.45)
$$
This equation is of the same type as Eq. (2.21) for the operator
$R$, viz., $[H_F, R] - V = 0 $, whose solution is given in App.{\rm A}.
Therefore, taking into account formula ({\rm A.4})  we find,
$$
{\cal A} = \imath \lim_{\epsilon \to 0+}
\int\limits_0^\infty V^{21} (t) \exp{ ( - \epsilon t) } dt\ .
\eqno(6.46)
$$

For interaction $V$ linear in the meson field we have $A_{m0} (\f) = 0$
if the index $m$ corresponds to two-- , three-- , \ldots meson states.
Let us consider the nonzero subblock $A_{\bk 0}$ , i.e., the subblock with
elements $\langle \bk f' \mid {\cal A} \mid 0 f \rangle $. With the help of
the representation ({\rm A.9}) for the interaction $V$ and relation
$$
\langle a\dg (\bk) \Omega \mid a\dg (\bk', t) \mid \Omega \rangle =
\delta (\bk - \bk') \exp{ \imath \omega_{\bk} t } \
$$
we obtain
$$
A_{\bk 0}(\f) = \imath \lim_{\epsilon \to 0+}
\int\limits_0^\infty V_{\bk 0} (\f (t))
\exp \left[ \imath ( \omega_{\bk} + \imath \epsilon ) t \right]  dt \ ,
\eqno(6.47)
$$
with
$$
V_{\bk 0} (\f (t)) = F\dg (t) V^{-\bk} F(t)\ , \eqno(6.48)
$$
where $\f (t) = \exp { \left[ \imath H_{F ferm}(\f) t \right] } \f
\exp { \left[ - \imath H_{F ferm}(\f) t \right] } $
is the subset of the fermion
operators in the interaction picture. We imply the division
$H_F = H_{F mes}(\m ) + H_{F ferm}(\f) $ into
the mesonic and nucleonic parts (see Eq.(2.8)), $\m$ being the subset
of the meson operators $a(\bk)$ and $a\dg(\bk)\, \forall \bk$. For notations
$F$ and $V^{-\bk}$ see App. {\rm A}.

At the same time the $(\bk 0)$ subblock of the operator
$R = {\cal R} - {\cal R}\dg$ determined by Eqs. ({\rm A.4}) and ({\rm A.9})
is
$$
R_{\bk 0}(\f) = - \imath \lim_{\epsilon \to 0+}
\int\limits_0^\infty V_{\bk 0} (\f (t))
\exp \left[ \imath ( \omega_{\bk} + \imath \epsilon ) t \right]  dt \ ,
\eqno(6.49)
$$

Comparing Eqs. (6.47) and (6.50), we get the solution for $A_{\bk 0}$,
$$
A_{\bk 0} = - R_{\bk 0} = {\cal R}\dg_{\bk 0} = F\dg R^{\bk} \dg F\
\eqno(6.50)
$$
in a compact form.

Here, we do not intend to write down explicit expressions for these subblocks
in
terms of the fermion creation(destruction) operators. The respective result
for $A^{(1)}_{\bk 0}(\f) $ with some extension to other Yukawa--type
meson--nucleon couplings can be found in \cite{KOSHE93} (see Eqs.(17)--(18)
therein).

After all, in order to derive an approximation to the operator $H'^{11}$, let
us employ
the expansion of $(1 + {\cal A}\dg {\cal A} )^{ - \frac{1}{2}} $
in the powers of ${\cal A}\dg {\cal A} \sim g^2$,
$$
(1 + {\cal A}\dg {\cal A} )^{ - \frac{1}{2}}
\approx 1 - {1 \over 2} {\cal A}\dg {\cal A} \  \eqno(6.51)
$$
Then, we obtain from Eq. (6.23) the expression
$$
H'^{11} = \eta_1 \{ H + {\cal A}\dg H + H {\cal A} + \
{1 \over 2} {\cal A}\dg [H, {\cal A}] - \
{1 \over 2} [H, {\cal A}\dg]{\cal A} \} \eta_1 + O(g^4) \ , \eqno(6.52)
$$
from which and condition (6.22) for ${\cal A}$ it follows that
$$
H'^{11} = H_{F ferm}(\f) \eta_1 + M^{(2)}_{ferm}(\f) \eta_1 +
{1 \over 2} \eta_1 \{ {\cal A}\dg H + H {\cal A} \} \eta_1 +  O(g^4) \ ,
\eqno(6.53)
$$
or neglecting the terms of order $g^4$,
$$
H'^{11} = H_{F ferm}(\f) \eta_1 + M^{(2)}_{ferm}(\f) \eta_1 +
{1 \over 2} \eta_1 \{ {\cal A}^{(1)}\dg V + V {\cal A}^{(1)} \} \eta_1 \
\eqno(6.54)
$$
(cf. Eq.(15) of \cite{KOSHE93}).
When deriving Eq. (6.53) we have used the equation
$$
\eta_1 H \eta_1 = H_{F ferm}(\f) \eta_1 + M^{(2)}_{ferm}(\f) \eta_1 +  O(g^4)\ ,
$$
whereas the change from (6.53) to (6.54) is based upon the properties
$$
\eta_2 H \eta_1 = \eta_2 M^{(2)}_{mes}(\m) \eta_1 + \eta_2 V \eta_1 + O(g^4)\,
$$
$$
\eta_1 H \eta_2 = \eta_1 M^{(2)}_{mes}(\m) \eta_2 + \eta_1 V \eta_2 + O(g^4)\,
$$
and the fact that
$$
\eta_1 M^{(2)}_{mes}(\m) {\cal A} \eta_1 = O(g^4)\ ,
$$
$$
\eta_1 V {\cal A} \eta_1 = \eta_1 V {\cal A}^{(1)} \eta_1 + O(g^4)\ ,
$$
for $V$ linear in the meson field.

The corresponding subblock $H'_{11}$ can be expressed through the subblocks
$A_{\bk 0}$ and $V_{\bk 0}$,
$$
H'_{11} = H_{F ferm}(\f) + M^{(2)}_{ferm}(\f) +
{1 \over 2} \int \{ A\dg\,_{0 \bk}(\f) V_{\bk 0}(\f) + H.c. \} d\bk =\
$$
$$
H_{F ferm}(\f) + M^{(2)}_{ferm}(\f) +
{1 \over 2} \int \{ F\dg R^{\bk} F \cdot F\dg V^{-\bk} F + H.c. \} d\bk
\eqno(6.55)
$$
where we have employed Eq. (6.50) and Eq. (6.48) at $t = 0$ and their H.c.

\newpage
\bigskip
\begin{center}{\em 6.6 Clothing Procedure vs Okubo Approach }
\end{center}
\bigskip

Let us discuss in detail common and distinctive features of the two kinds of
UT. Partly, we have concerned this subject in Subsect. 6.4 ( see also
 App. ${\rm C}$ ).

The Okubo and clothing approaches differ in their goals. The UT by Okubo
is aimed at
to nullify the nondiagonal blocks $H^{\prime}_{2 1}$ and $H^{\prime}_{1 2}$
of the transformed Hamiltonian. The clothing and akin approaches (see Sect. 5)
require that some undesirable operator terms ("bad" or "virtual") must be
absent in the transformed Hamiltonian.

In the context it is worthwile to mention Nishijima's
modification \cite{NISH56}
of the Okubo idea. Instead of Eq. (6.8) he used the representation
$U = \ldots \exp{(\imath S_2)} \exp{(\imath S_1)} $, similar to that employed
within the clothing approach. But  hermitian operators $S_n$
($n=1,2,3,\ldots$) are to implement Okubo's requirement, viz.,
the operator $S_n$ must remove $K_{2 1}$ matrix elements of the $g^n$--order
(instead of removing some bad terms $\sim g^n$).

No wonder different approaches give, in general, different resulting
transformed Hamiltonians. Nevertheless, we want to show that
the both approaches can give some coincident results (even for the realistic
field model), being realized approximately (in a
perturbative way). In particular,
effective quasipotentials of the $g^2$ order for nucleon-nucleon interaction
turn out to be the same.

Before comparison of the resulting Hamiltonians we must remind that Okubo's
$H^{\prime}$ (see Subsect 6.5) does not coincide with the starting
Hamiltonian $H$ whereas the operator $K$ of the clothing approach does.
$H^{\prime}$ can be considered as a function of bare operators $a$ while
$K$ depends on clothed ones $\alpha$. However, we have shown in Subsect. 6.4
that $H$ coincides with the operator $H^{\prime}({\tilde a})$ which depends
upon new destruction-creation operators ${\tilde a}=U a U^{\dagger}$ in the
same
manner as $H^{\prime}$ depends on $a$. So, $H^{\prime}({\tilde a})$ and $K$
represent the same operator and can be compared to each other. Let us recall,
however, that
${\tilde a}$ may not coincide with clothed operators $\alpha$, and
$H^{\prime}$ and $K$ may not be the same function of their arguments. In what
follows we imply that $H^{\prime}$ means $H^{\prime}({\tilde a})$ and
$H^{\prime}_{1 1}$ is a function of fermion destruction-creation operators
${\tilde f}=U f U^{\dagger}$.

Now, we can prove the statement: $H^{\prime}_{1 1 }$ given by Eq. (6.55)
coincides with the fermionic part of $K_F+M^{(2)}_{ren}+\frac{1}{2}[R,V]$
(see Eq. (2.23)), i.e., with
$$
K_{ferm}=K_{F ferm}+M^{(2)}_{ferm}+({\rm fermionic\ part\ of\ }
\frac{1}{2}[R,V])\  \eqno(6.56)
$$

More exactly, $H^{\prime}_{1 1}$ and $K_{ferm}$ are the same function of
their arguments (fermionic operators ${\tilde f}$ and $f_c$, respectively).
Indeed, the last term in Eq. (6.56) is that part of $\frac{1}{2}
[R,V]\sim g^2$ which depends only on fermionic operators. We calculate
$\frac{1}{2}[R,V]$ in App.{\rm A}. Its fermionic part (denoted below as $FP$)
is contained in $\frac{1}{2}[{\cal R},{\cal V}^{\dagger}]+H.c.$, for
$[{\cal R}, {\cal V}^{\dagger}]$ see the last term in the r.h.s. of Eq.
{\rm (A.14)}.
We have
$$
FP\equiv\frac{1}{2}\int [ F^{\dagger}R^{{\vec k}}F\cdot
F^{\dagger} V^{-{\vec k}} F + H.c.] d{\vec k}\  \eqno(6.57)
$$
Comparing this expression  with the r.h.s. of Eq. (6.55), we arrive to
the above statement.

The fermion mass counterterm $M^{(2)}_{ferm}$ in Eqs. (6.55) or (6.56) must
cancel
two-operator fermionic contributions which arise after normal ordering of $FP$
(see Subsect. 2.5). One should note that there are no other such terms in
$\frac{1}{2}[R,V] $.
So, we may omit $M^{(2)}_{ferm}$ from (6.55) or (6.56) provided
$FP$ is replaced by the normally ordered counterpart $:FP:$ of (6.57).

Let us mention that similar evaluation of $M^{(2)}_{mes}$ (cf. Subsect. (2.5))
would take additional efforts within Okubo's approach (see \underline {Note}
at the end of this subsection).

Of course, we may explicitly express $FP$ given by Eq. (6.57) through
$b,d,b^{\dagger}$ and $d\dg$ opening  the abbreviations accepted for
$F^{\dagger}V^{{\vec k}}F$ and $F^{\dagger}R^{{\vec k}}R$ in App. {\rm A}
(we omit here the symbol tilde or the
subscript $c$ when handling with the fermion operators). Such representation
shows that $:FP:$
contains "bad" terms of the kind $b^{\dagger}d^{\dagger}b^{\dagger}
d^{\dagger}$, $b^{\dagger}d^{\dagger}b^{\dagger}b$ and $b^{\dagger}d^{\dagger}
d^{\dagger}d$. As stressed in Subsect. 2.4., they should be eliminated from
$K$ via the clothing transformation $W_4$.
Such terms are also unpleasant within Okubo's approach, viz., they prevent
the no-fermion state
$\Omega_f$ and one-fermion ones $b^{\dagger}\Omega_f$ and
$d^{\dagger}\Omega_f$ to be $H^{\prime}_{1 1}$ eigenvectors along with
two-fermion states of the kind (4.14).

In other words, the problem of finding
$H^{\prime}_{1 1}$
eigenvectors is not essentially easier than that for the starting Hamiltonian.
In the spirit of Okubo's approach the elimination of "bad" terms can be
implemented by
performing $UT$ of $H^{\prime}_{1 1}$ such that the transformed subhamiltonian
$(H^{\prime}_{1 1})^{\prime}=U^{\prime\,\dagger}H^{\prime}_{1 1}U^{\prime}$
would not contain matrix elements corresponding to the processes $\Omega_f
\rightarrow {\rm two\ pairs}$, $N\rightarrow N+{\rm pair, etc.}$. The
relevant projector $\eta^{\prime}_1$ may project on states without pairs.

As argued in Sect. 4, these additional transformations $W_4$ or $U^{\prime}$
do not alter the remaining "good" four-fermionic pieces of
$:FP:$ which are of the $g^2$ order (see, e.g., Eq. (4.18)).

So we obtain from $H^{\prime}_{1 1}$ or $K_{ferm}$ the same expression
$H_{F ferm}+G:FP:$, where $G:FP:$ is a "good" part of $:FP:$. Using the
notations of Sect. 4 (see Eqs. (4.1)-(4.3)), it may be written as
$(K_{ZA})_{ferm}$ or
$$
K_{2 ferm}+K(NN\rightarrow NN)+K({\bar N}{\bar N}\rightarrow{\bar N} {\bar N})
+ K(N{\bar N}\rightarrow N{\bar N})\   \eqno(6.58)
$$
The no-fermion, one-fermion and two-fermion deuteron-like eigenvectors of this
operator give some approximations to the corresponding $H$ eigenvectors.

\underline{Note}. The zeroth approximation operator $K_{ZA}$ considered in
Sect. 4 enables us to find also approximations to $H$ eigenvectors
which describe one-meson
states and meson-nucleon bound and scattering states. In the
Okubo-Korchin-Shebeko approach exposed in Subsect. 6.5 one should use for this
purpose the submatrix
$H^{\prime}_{2 2}$. However, the latter embodies, in general, undesirable
matrix elements which correspond to the processes one-meson $\rightarrow$ other
states (e.g., two meson + pair), etc. In order to simplify the finding of $H$
eigenvectors under discussion one needs additional Okubo's transformation of
the
kind $(H^{\prime}_{2 2})^{\prime}= {\tilde U}^{\dagger} H^{\prime}_{2 2}{\tilde U}$
such that $(H^{\prime}_{2 2})^{\prime}$ would have no undesirable matrix
elements.

\bigskip
\begin{center}{\bf 7 Conclusions}
\end{center}
\bigskip

A considerable part of this work is devoted to development of the UT method
using the so--called clothing procedure in RQFT. This procedure has two
aspects. On the one hand,
we express with its aid a total Hamiltonian $H$ for interacting fields in
terms of the new operators which correspond to the creation (destruction)
of clothed particles. The latter possess, by definition, the properties
of observed (physical) particles and are as a matter of fact ''quasiparticles''
within  our approach, if one draws a parallel with the method of canonical
transformations in quantum theory (see, e.g., \cite{DAV73}, \S\S 52, 84 ).
On the other hand, such representation for $H$, being implemented partially
or perfectly, enables us to formulate an approach to solution of the
$H$ eigenvalue problem.

We have discussed in detail various UT of $H$ in Sect. 6.
In this connection, we distinguish two kinds of UT's destined for
approximate determination of $H$ eigenstates. Both may be deduced from
the relation between the Hamiltonian $H$ matrices with respect to
two different sets of basis vectors .

The first kind may be written as $H^{\prime}=U^{\dg} H U $, where $H$ is
the input Hamiltonian which is subject to UT, the transformed operator
$H^{\prime}$ being not equal to the input one. However, {H} eigenstates
can be obtained from those of $H^{\prime}$ using $U$ (see Subsect. 6.3).

The second kind is determined as  $H=U^{\dg} \tilde H U$ ($U = W^{\dg} $).
Now, the transformed Hamiltonian is the input one, but differently
represented, while the operator $\tilde H \neq H$ is related to $H$ in a
simple way. These kinds of UT's are used in literature but
authors sometimes overlook that $H^{\prime}$ eigenstates do not coincide
with those of $H$.

Applying either of such UT's one may impose different constraints on the form
of the transformed Hamiltonian that leads to the definite recipes for
constructing the corresponding unitary operators. We have shown the
resemblances and distinctions of some known applications of the UT method:
the clothing procedure, the approach by Heitler and Sato, and
Okubo's blockdiagonalization (Sects. 5 and 6). One should note that
in all applications the unitary operators $U$ is determined approximately
(with the exception of simple models).                                    .

Our consideration of the problem of bound and scattering states differs from
akin approaches (cf., \cite{NISH56}, \cite{SATO92}) with the following
distinctive feature. Our interactions between clothed particles
( as an illustration, the $\pi N$ and $NN$ quasipotentials) are parts of
a single operator $K_{ZA}$ which can
be regarded as a zeroth approximation to the total Hamiltonian. If
the $NN$ system is considered then $K_{ZA}$ generates the $NN$
Hamiltonian, whereas for the $\pi$N system $K_{ZA}$ gives
rise to the $\pi$N Hamiltonian. This makes clear the relation of the
eigenstates  of such  Hamiltonian to the eigenstates of the original field
Hamiltonian: the former are some approximations to the latter.

One should point out that nonlocal properties of these quasipotentials
have a relativistic origin. In this context, we would like to note a growing
interest in similar relativistic effects in the modern theories of nuclear
forces(see, e.g., \cite{HAID96}, \cite{SAMMA98} and refs. therein ).
We show how the mass renormalization program is realized within the
approach developed here. The transformed Hamiltonian when expressed in terms
of clothed operators turns out to be dependent on renormalized (physical)
masses and not bare ones. Some tricks shown in App. {\rm A} can be useful
in future calculations of the radiative corrections (renormalizations ) to
bare (trial) masses for field models with a cutoff in the momentum
representation.

We prove that clothing UT's
remove the undesired ("bad") terms simultaneously from the Hamiltonian and
the generators of Lorentz boosts. Our proof is valid for any RQFT model
(see Sect. 3).

Our three--dimensional formalism is covariant in that sense that we give
definite prescriptions for the transformation properties of the clothed
states with respect to the Lorentz boosts.

With the help of a simple example, we demonstrate in App. {\rm B} that
the clothing transformation $W$ of Sect. 2  may happen to be implemented
not by an unitary operator in its usual mathematical sense.
We argue that $W$ ought to be understood as an element of some algebra
(lacking any operator representation), being unitary in an algebraic sense.

At last, we show  in App. {\rm C} original tricks and results concerning
a nonperturbative solution of Okubo's decoupling equation.

We believe that the concept of clothed particles and the approach exposed
here can be applied to different areas of the nuclear physics: from the
theory of nuclear structure to description of nuclear reactions including
the processes with meson production.

\newpage

\bigskip
\begin{center}{\em Acknowledgements}
\end{center}
\bigskip

This work was supported by a grant from Bundesministerium f\"ur Bildung,
Wissenshaft, Forshung and Technologie. One of us (A.S.) is very grateful
to W.Gl\"ockle for many genuine and therefore fruitful discussions during
the recent visit to Institut f\"ur Theoretische Physik II (Bochum).

\bigskip
\hspace{-1em}{\large Appendix A}
\bigskip

\bigskip
\centerline{\bf Three-operator clothing transformation for the Yukawa model.}
\centerline{\bf Four-operator interactions between clothed particles and their}
\centerline{\bf  normal ordering.}
\bigskip

The defining equations for the Yukawa model are given in Sect. 2. We
use throughout this article notations of \cite{BD64} assuming, in
particular, $\gamma_{\mu}^{\dagger} = \gamma_0 \gamma_{\mu}\gamma_0\
(\mu =0,1,2,3),\ \hat{q}=q^{\mu} \gamma_{\mu},\ \gamma_5 =
i \gamma^0 \gamma^1 \gamma^2 \gamma^3 = \gamma_5^{\dagger}$ and
$\gamma_5^2 = 1$.

\bigskip

{\bf \underline{A.1.}}\hspace{2em}Three-operator clothing transformation
$W={\rm exp}R$, see Sect. 2.4, can be found by solving the equation
$[R,H_{\rm F}] + V =0$. The antihermitian operator $R$ is assumed to be
of the form $R = {\cal R} - {\cal R}^{\dagger}$ with ${\cal R}$ given by
Eq. (2.22). The commutator $[R,H_{\rm F}]$ can be directly evaluated by
using the commutation relations (2.5). Then we obtain the equations for
the coefficients $R_{ij}$ involved in ${\cal R}$, see Eq. (2.22). Their
solutions are
$$
\begin{array}{l}
R_{11}^{\vsm{k}}(\ve{p'}r';\ve{p}r) =
V_{11}^{\vsm{k}}(\ve{p'}r';\ve{p}r)/(E_{\vsm{p'}}-E_{\vsm{p}}-\omega_{\vsm{k}})\ ,
\\
R_{12}^{\vsm{k}}(\ve{p'}r';\ve{p}r) =
V_{12}^{\vsm{k}}(\ve{p'}r';\ve{p}r)/(E_{\vsm{p'}}+E_{\vsm{p}}-\omega_{\vsm{k}})\ ,
\\
R_{21}^{\vsm{k}}(\ve{p'}r';\ve{p}r) =
V_{21}^{\vsm{k}}(\ve{p'}r';\ve{p}r)/(-E_{\vsm{p'}}-E_{\vsm{p}}-\omega_{\vsm{k}})\ ,
\\
R_{22}^{\vsm{k}}(\ve{p'}r';\ve{p}r) =
V_{22}^{\vsm{k}}(\ve{p'}r';\ve{p}r)/(-E_{\vsm{p'}}+E_{\vsm{p}}-\omega_{\vsm{k}})\ .
\end{array}\eqno({\rm A.1})
$$
Here we have used the notations
$$
\left[ \matrix{
V_{11}^{\vsm{k}}(\ve{p'}r';\ve{p}r) & V_{12}^{\vsm{k}}(\ve{p'}r';\ve{p}r)\cr
V_{21}^{\vsm{k}}(\ve{p'}r';\ve{p}r) & V_{22}^{\vsm{k}}(\ve{p'}r';\ve{p}r)\cr
}\right] =
i { {g}\over {(2\pi)^{3/2} } }
{ {m}\over {\sqrt{2\omega_{\vsm{k}}E_{\vsm{p'}}E_{\vsm{p}} } } }
\delta ( \ve{p}+\ve{k}-\ve{p'} )
$$
$$
\times\
\left[ \matrix{
\bar{u}(\ve{ p'},r')\gamma_5 u(\ve{p},r) & \bar{u}(\ve{ p'},r')\gamma_5 v(\ve{-p},r)\cr
\bar{v}(\ve{-p'},r')\gamma_5 u(\ve{p},r) & \bar{v}(\ve{-p'},r')\gamma_5 v(\ve{-p},r)\cr
}\right]
\equiv V^{\vsm{k}}(\ve{p'}r';\ve{p}r)\ . \eqno({\rm A.2})
$$

Eqs. (A.1) have meaning if the denominators in their r.h.s. are not zero
(note that $\ve{p}+\ve{k}-\ve{p'} =0$ according to (A.2)). One can show
that this is the case under the condition
$$
\mu < 2m \ .\eqno({\rm A.3})
$$
The physical sense of this condition is discussed in Subsect. 2.4.

Alternatively, the solution $R$ of Eq. (2.21) can be represented as
$$
R = -i \lim_{\varepsilon\rightarrow 0+}\int_0^{\infty}
V(t) {\rm e}^{-\varepsilon t} {\rm d}t \ ,\eqno({\rm A.4})
$$
where $V(t)= {\rm exp}(iH_{\rm F}t) V {\rm exp}(-iH_{\rm F}t)$ is
the interaction operator in the Dirac picture. Obviously, $V(t)$ is
determined by Eq. (2.16) where the operators $a_{\rm c} (\ve{k}),\
b_{\rm c} (\ve{p},r)$ and $ d_{\rm c} (\ve{p},r)$ are replaced by
$a_{\rm c} (\ve{k}){\rm exp}(-i\omega_{\vsm{k}}t),\
b_{\rm c} (\ve{p},r){\rm exp}(-iE_{\vsm{p}}t)$ and
$ d_{\rm c} (\ve{p},r){\rm exp}(-iE_{\vsm{p}}t)$, respectively.
Therefore, the evaluation of $R$ is reduced to integrals of the kind
$$
\int_0^{\infty} {\rm e}^{i(x+i\varepsilon)t}\ {\rm d}t =
{{i}\over {x+i\varepsilon } }\ ,\eqno({\rm A.5})
$$
where $x$ is any of the denominators in (A.1) and $\varepsilon >0$
\footnote{According to the prescription given by Eq.(A.4)
the positive parameter
$\varepsilon$ should be put equal to zero at the end of the
calculations.}. The limit $\varepsilon\rightarrow 0+$ in Eq. (A.4)
exists, and it is finite if the inequality (A.3) takes place, i.e.,
if $x\neq 0$. This evaluation of $R$ shows readily that the solution
given by Eq. (A.4) coincides with that determined by Eqs. (A.1).
Also, one can directly verify that (A.4) meets Eq. (2.21). In fact,
we have
$$
[R,H_{\rm F}] = -i \lim_{\varepsilon\rightarrow 0+}\int_0^{\infty}
{\rm e}^{-\varepsilon t} [V(t),H_{\rm F}]\ {\rm d}t
$$
$$
= \lim_{\varepsilon\rightarrow 0+}\int_0^{\infty} {\rm e}^{-\varepsilon t}
{ {\partial}\over {\partial t} }V(t)\ {\rm d}t\ =\ -V\ .
\eqno({\rm A.6})
$$
The last equality follows (under the condition $\mu < 2m$) if one
calculates ${ {\partial}\over {\partial t} }V(t)$ and then integrates
with the help of (A.5). We use the form (A.4) in Sect. 3 when constructing
the Lorentz boosts in terms of the clothed operators.

\bigskip

{\bf \underline{A.2.}}\hspace{2em}Evaluation of $[R,V]$ is a tedious
exercise that can be simplified with the aid of a
more compact notation. Indeed, expressions (2.16) and (2.22) for $V$ and
$R$ have the identical operator structure, {\em viz.},
$$
b_{\rm c}^{\dagger}(\ve{p'},r') X_{11}^{\vsm{k}}(\ve{p'}r';\ve{p}r)
b_{\rm c}(\ve{p},r)\ +\
b_{\rm c}^{\dagger}(\ve{p'},r') X_{12}^{\vsm{k}}(\ve{p'}r';\ve{p}r)
d_{\rm c}^{Xdagger}(-\ve{p},r)
$$
$$
+\
d_{\rm c}(-\ve{p'},r') X_{21}^{\vsm{k}}(\ve{p'}r';\ve{p}r)
b_{\rm c}(\ve{p},r)\ +\
d_{\rm c}(-\ve{p'},r') X_{22}^{\vsm{k}}(\ve{p'}r';\ve{p}r)
d_{\rm c}^{\dagger}(-\ve{p},r)
\eqno({\rm A.7})
$$
with definite c-number coefficients $X_{ij}^{\vsm{k}}\ (i,j=1,2)$.

Now, let us rewrite (A.7) as a matrix product
$$
(b_{\rm c}^{\dagger}(\ve{p'},r'),\ d_{\rm c}(-\ve{p'},r') )\
\left[ \matrix{
X_{11}^{\vsm{k}}(\ve{p'}r';\ve{p}r) & X_{12}^{\vsm{k}}(\ve{p'}r';\ve{p}r)\cr
X_{21}^{\vsm{k}}(\ve{p'}r';\ve{p}r) & X_{22}^{\vsm{k}}(\ve{p'}r';\ve{p}r)\cr
}\right] \
\left( \matrix{
b_{\rm c}(\ve{p},r)\cr
d_{\rm c}^{\dagger}(-\ve{p},r)\cr
}\right)
$$
$$
\equiv\
F^{\dagger}(\ve{p'},r') X^{\vsm{k}}(\ve{p'}r';\ve{p}r) F(\ve{p},r)\ ,
\eqno({\rm A.8})
$$
where along with the $2\times 2$ matrix $X^{\vsm{k}}$ composed of the
coefficients $X_{ij}^{\vsm{k}}$ (cf. Eq. (A.2)) we have introduced the operator
column $F$ and row $F^{\dagger}$,
$$
F(\ve{p},r) =
\left( \matrix{
b_{\rm c}(\ve{p},r)\cr
d_{\rm c}^{\dagger}(-\ve{p},r)\cr
}\right)\ ,\ \ \ \
F^{\dagger}(\ve{p'},r') =
(b_{\rm c}^{\dagger}(\ve{p'},r'),\ d_{\rm c}(-\ve{p'},r') )\ .
$$
The subsequent operations become even more concise after adopting the
convention
$$
\int {\rm d}\ve{p'}\
\int {\rm d}\ve{p}\
\sum_{r'r} F^{\dagger}(\ve{p'},r') X^{\vsm{k}}(\ve{p'}r';\ve{p}r) F(\ve{p},r)
\equiv F^{\dagger} X^{\vsm{k}} F\ .
\eqno({\rm A.8'})
$$

Under these notations, Eqs. (2.16) and (2.22) look as
$$
V =
\int {\rm d}\ve{k}\ F^{\dagger} V^{\vsm{k}} F\
[a_{\rm c} (\ve{k}) + a_{\rm c}^{\dagger} (\ve{-k})]
$$
$$
=
\int {\rm d}\ve{k}\ F^{\dagger} V^{\vsm{k}} F a_{\rm c}(\ve{k})\ +\ {\rm H.c.}
\ \equiv\ {\cal V} + {\cal V}^{\dagger}\ ,\eqno({\rm A.9})
$$
$$
R =
\int {\rm d}\ve{k}\ F^{\dagger} R^{\vsm{k}} F a_{\rm c}(\ve{k})\ -\ {\rm H.c.}
\ =\ {\cal R} - {\cal R}^{\dagger}\ ,
$$
where H.c. means taking the Hermitian conjugate of the first terms. Note that
$(V^{\vsm{k}} )^{\dagger} = V^{\vsm{-k}} $.

\bigskip

{\bf \underline{A.3.}}\hspace{2em}After these preliminaries, we have
$$
[R,V] = [{\cal R}-{\cal R}^{\dagger},V] = [{\cal R},V] +
[V^{\dagger}, {\cal R}^{\dagger} ] = [{\cal R},V] + {\rm H.c.}
\eqno({\rm A.10})
$$
Keeping this in mind, it is sufficient to evaluate
$$
[{\cal R},V] = [{\cal R},{\cal V}] + [{\cal R},{\cal V}^{\dagger}]\ .
\eqno({\rm A.11})
$$
By using the definitions from (A.9) and carrying out straightforward
operator algebra, we obtain step by step
$$
[{\cal R},{\cal V}] =
\int {\rm d}\ve{k}_1\
\int {\rm d}\ve{k}_2\
[ F^{\dagger} R^{\vsm{k}_1} F,\ F^{\dagger} V^{\vsm{k}_2} F ]\
a_{\rm c} (\ve{k}_1) a_{\rm c} (\ve{k}_2)
$$
$$
=
\int {\rm d}\ve{k}_1\
\int {\rm d}\ve{k}_2\
F^{\dagger} [ R^{\vsm{k}_1} , V^{\vsm{k}_2} ] F
a_{\rm c} (\ve{k}_1) a_{\rm c} (\ve{k}_2)\ ,
\eqno({\rm A.12})
$$
where in accordance with Eq. (A.8$'$)
$$
F^{\dagger} [ R^{\vsm{k}_1} , V^{\vsm{k}_2} ] F  =
\int {\rm d}\ve{p'}\
\int {\rm d}\ve{p}\
\sum_{r'r} F^{\dagger}(\ve{p'},r') [ R^{\vsm{k}_1} , V^{\vsm{k}_2} ]
(\ve{p'}r';\ve{p}r)\ F (\ve{p},r) \ ,
$$
and it is implied that\\
\medskip
$
[ R^{\vsm{k}_1} , V^{\vsm{k}_2} ](\ve{p'}r';\ve{p}r)
$
$$
=
\int {\rm d}\ve{q}\
\sum_{s} [ R^{\vsm{k}_1}(\ve{p'}r';\ve{q}s) V^{\vsm{k}_2}(\ve{q}s;\ve{p}r) -
           V^{\vsm{k}_2}(\ve{p'}r';\ve{q}s) R^{\vsm{k}_1}(\ve{q}s;\ve{p}r)\ ]\ .
\eqno({\rm A.13})
$$
In the above calculations it has been convenient to employ the identity
$$
[AB,CD] =A\{B,C\}D -\{A,C\}BD -C\{D,A\}B +CA\{D,B\}
$$
for four operators $A$, $B$, $C$ and $D$.

Further, after applying another useful relation
$$
[AB,CD] =A[B,C]D +[A,C]DB +AC[B,D] +C[A,D]B\ ,
$$
a similar derivation for $[{\cal R},{\cal V}^{\dagger}]$ results in
$$
[{\cal R},{\cal V}^{\dagger}] =
\int {\rm d}\ve{k}_1\
\int {\rm d}\ve{k}_2\
\{ F^{\dagger} [ R^{\vsm{k}_1} , V^{-\vsm{k}_2} ] F
a_{\rm c}^{\dagger} (\ve{k}_2) a_{\rm c} (\ve{k}_1) +
\delta ( \ve{k}_1 - \ve{k}_2 )
F^{\dagger} R^{\vsm{k}_2} F \cdot F^{\dagger} V^{-\vsm{k}_1} F \} \ .
\eqno({\rm A.14})
$$

Now, we see that the $g^2$-order commutator $[R,V]$ brings in the total
Hamiltonian $K$, see Eq. (2.23), the interaction terms which describe the
following real physical processes:\\
a) the $\pi\pi\rightarrow f\bar{f}$ creation and the $f\bar{f}\rightarrow \pi\pi$
annihilation from $[{\cal R},{\cal V}]$ (see Eq. (A.12)) and its H.c.;\\
b) the $\pi f \rightarrow \pi f$, $\pi \bar{f} \rightarrow \pi \bar{f}$,
$f f \rightarrow f f$, $f \bar{f} \rightarrow f \bar{f}$ and
$\bar{f} \bar{f} \rightarrow \bar{f} \bar{f}$ scatterings from
$[{\cal R},{\cal V}^{\dagger}]$ and its H.c. (see (A.14)).

In addition to these contributions, $[R,V]$ contains interactions, which
have nonvanishing matrix elements between the vacuum $\Omega$ and two-particle
states (e.g., those responsible for the virtual process
$\Omega\rightarrow\pi\pi $), and between $\Omega$ and four-particle
states (e.g., for transitions $\Omega\rightarrow\pi\pi f\bar{f}$).
There are also interactions responsible for the transitions one particle
$\rightarrow$ three particles and one particle $\rightarrow$ one particle.
Except the latter, all these terms are ``bad'', i.e., they hinder $\Omega$
and one-particle states to be $K$ eigenstates.

For example, let us consider the term
$d_{\rm c} d_{\rm c}^{\dagger} a_{\rm c} a_{\rm c} $ which enters in
$[{\cal R},{\cal V}]$. It has nonzero matrix element
$\langle\Omega\mid d_{\rm c} d_{\rm c}^{\dagger} a_{\rm c} a_{\rm c}
\mid\pi\pi\rangle $ that becomes evident after normal ordering,
$$
d_{\rm c}(-\ve{p'},r') d_{\rm c}^{\dagger}(-\ve{p},r)
a_{\rm c}(\ve{k}_1 ) a_{\rm c}(\ve{k}_2 ) =
- d_{\rm c}^{\dagger}(-\ve{p},r) d_{\rm c}(-\ve{p'},r')
  a_{\rm c}(\ve{k}_1 ) a_{\rm c}(\ve{k}_2 ) +
\delta_{r'r}\delta (\ve{p'} - \ve{p} )\
a_{\rm c}(\ve{k}_1 ) a_{\rm c}(\ve{k}_2 )\ .
$$
This illustration shows that normal ordering is a constructive tool
in the framework of our clothing procedure. As in many applications
of the method of second quantization (e.g., in field theories of
the evolution operator or the $S$-matrix), this operation enables
us to classify the separate contributions to the original
Hamiltonian at every stage of the clothing procedure.

\bigskip

{\bf \underline{A.4.}}\hspace{2em}Now, we shall discuss in detail
the origin of two-operator meson terms which stem from the
commutator $[R,V]$. They are essential elements in
our treatment of the particle mass renormalization (see Sect. 2.5).
These terms appear after reshuffling the operators of the
expressions $F^{\dagger} [ R^{\vsm{k}_1} , V^{\vsm{k}_2} ] F
a_{\rm c}(\ve{k}_1 ) a_{\rm c}(\ve{k}_2 )$ (see Eq. (A.12)) and
$F^{\dagger} [ R^{\vsm{k}_1} , V^{-\vsm{k}_2} ] F
a_{\rm c}^{\dagger}(\ve{k}_2) a_{\rm c}(\ve{k}_1 ) $
(see Eq. (A.14)) into normal order.

In the first case, it touches upon the terms of the $dd^{\dagger}$-kind and
leads to
$$
F^{\dagger} [ R^{\vsm{k}_1} , V^{\vsm{k}_2} ] F =
: F^{\dagger} [ R^{\vsm{k}_1} , V^{\vsm{k}_2} ] F : +
{\rm Tr} [ R^{\vsm{k}_1} , V^{\vsm{k}_2} ]_{22} \ ,
\eqno({\rm A.15})
$$
where
$$
{\rm Tr} [ R^{\vsm{k}_1} , V^{\vsm{k}_2} ]_{22} \equiv
 \int {\rm d}\ve{p}\ \sum_{r} [ R^{\vsm{k}_1} , V^{\vsm{k}_2} ]_{22}
(\ve{p}r;\ve{p}r)\ ,\eqno({\rm A.16})
$$
\medskip
$
[ R^{\vsm{k}_1} , V^{\vsm{k}_2} ]_{22}(\ve{p'}r';\ve{p}r)
$
$$
\equiv
\int {\rm d}\ve{q}\ \sum_{s}
[ R_{2j}^{\vsm{k}_1}(\ve{p'}r';\ve{q}s) V_{j2}^{\vsm{k}_2}(\ve{q}s;\ve{p}r) -
  V_{2j}^{\vsm{k}_2}(\ve{p'}r';\ve{q}s) R_{j2}^{\vsm{k}_1}(\ve{q}s;\ve{p}r)\ ]\ .
\eqno({\rm A.17})
$$
The trace ${\rm Tr}$ is evaluated using the properties of the
$\gamma$-matrices and Dirac spinors. We find with the solutions (A.1)
$$
{\rm Tr} [ R^{\vsm{k}_1} , V^{\vsm{k}_2} ]_{22} =
2{ {t_{\vsm{k}_1}}\over {\omega_{\vsm{k}_1}} } \delta (\ve{k}_1 + \ve{k}_2)\ ,
\eqno({\rm A.18})
$$
$$
t_{\vsm{k}} = { {g^2}\over {4(2\pi )^3} } \int {\rm d}\ve{p}\
{ {m^2}\over {E_{\vsm{p}} E_{\vsm{p}-\vsm{k}} } }
\left\{
{ {{\rm Sp} \left[ \Lambda_{-} (-\ve{p}) \Lambda_{-} (\ve{p}-\ve{k})\right]}\over
{E_{\vsm{p}} + \omega_{\vsm{k}} + E_{\vsm{p}-\vsm{k}} } } +
{ {{\rm Sp} \left[ \Lambda_{-} (-\ve{p}) \Lambda_{+} (\ve{-p} + \ve{k})\right]}\over
{E_{\vsm{p}} + \omega_{\vsm{k}} - E_{\vsm{p}-\vsm{k}} } }
\right.
$$
$$
\left.
+\
{ {{\rm Sp} \left[ \Lambda_{-} (-\ve{p}) \Lambda_{-} (\ve{p}-\ve{k})\right]}\over
{E_{\vsm{p}} - \omega_{\vsm{k}} + E_{\vsm{p}-\vsm{k}} } } +
{ {{\rm Sp} \left[ \Lambda_{-} (-\ve{p}) \Lambda_{+} (-\ve{p}+\ve{k})\right]}\over
{E_{\vsm{p}} - \omega_{\vsm{k}} - E_{\vsm{p}-\vsm{k}} } }
\right\}
\eqno({\rm A.19})
$$
with the notations $\Lambda_+ (\Lambda_- )$ for the projection
operators onto the nucleon positive (negative)-energy states:
$$
\Lambda_{\pm}(\ve{q}) = { {\hat{q}\pm m}\over {2m} }\ .
$$
While deriving Eq. (A.19), we have taken into account that
$$
\sum_s \bar{v}(\ve{p}s) O v(\ve{p}s) =
-{\rm Sp} \{ \Lambda_- (\ve{p}) O \} \ ,
$$
where $O$ is a combination of $\gamma$-matrices.

Of course, one can collect the similar terms with the same numerators
in the r.h.s. of Eq. (A.19) ( e.g., the first term with the third one
inside the curly brackets). However, we prefer other continuation that
enables us to get immediately an explicitly covariant form of $t_{\ve{k}}$.

First of all, we find
$$
{\rm Sp} \left[ \Lambda_{-} (-\ve{p}) \Lambda_{-} (\ve{p}-\ve{k})\right] =
\frac{E_{\vsm{p}} E_{\vsm{p}-\vsm{k}} + \ve{p}\cdot(\ve{p}-\ve{k}) + m^2 }{m^2} \ ,
$$
$$
{\rm Sp} \left[ \Lambda_{-} (-\ve{p}) \Lambda_{+} (-\ve{p}+\ve{k})\right] =
\frac{E_{\vsm{p}} E_{\vsm{p}-\vsm{k}} - \ve{p}\cdot(\ve{p}-\ve{k}) - m^2 }{m^2} \ ,
$$

Substituting these expressions into (A.19) and uniting therein the first term
with the second one and the third term with the fourth one it can be shown that
$$
t_{\vsm{k}} = { {g^2}\over {2(2\pi )^3} } \int \frac{{\rm d}\ve{p}}
{E_{\vsm{p}}}\
\left\{ \frac{p_{-}k}{ {\mu}^2 + 2p_{-}k} + \frac{pk}{ {\mu}^2 - 2pk} \right\} ,
\eqno({\rm A.20a})
$$
or
$$
t_{\vsm{k}} = { {g^2}\over {2(2\pi )^3} } \int \frac{{\rm d}\ve{p}}{E_{\vsm{p}}}\
\left\{ 1 + \frac{{\mu}^4}{4(pk)^2 - {\mu}^4 } \right\} . \eqno({\rm A.20b})
$$
with $p_{-} =  (E_{\vsm{p}}, -\ve{p})$ ,
$p = (E_{\vsm{p}}, \ve{p})$ and $k = (\omega_{\vsm{k}}, \ve{k}) $.

Since this integral is a Lorentz--scalar, one can write
$$
t_{\vsm{k}} = t_{\vsm{k} = 0} =
{ {g^2}\over {2(2\pi )^3} } \int \frac{{\rm d}\ve{p}}{E_{\vsm{p}}}\
\left\{ 1 + \frac{{\mu}^2}{4 E_{\vsm{p}}^2 - {\mu}^2 } \right\}  \eqno({\rm A.21})
$$
or
$$
t_{\vsm{k}} = { {g^2}\over {4{\pi }^2} } \left\{ I_1 +
\frac{{\mu}^2}{4} \left[ I_2 - \frac{\sqrt{4 m^2 - {\mu}^2} }{\mu}
\arctan \frac{\mu}{\sqrt{4 m^2 - {\mu}^2} } \right] \right\} \ , \eqno({\rm A.22})
$$
where the integrals
$$
I_1 = \int\limits_0^\infty \frac{x^2}{\sqrt{x^2 + m^2} }\,{\rm d}x \
$$
and
$$
I_2 = \int\limits_0^\infty \frac{{\rm d}x }{\sqrt{x^2 + m^2} } \
$$
are, respectively, quadratically and logarithmically divergent.

Now, the resulting contribution to $[{\cal R},{\cal V}]$
which is of the $a_{\rm c}a_{\rm c}$-kind can be written as
$$
2
\int {\rm d}\ve{k}_1\
\int {\rm d}\ve{k}_2\
\delta (\ve{k}_1 + \ve{k}_2) { {t_{\vsm{k}_1}}\over {\omega_{\vsm{k}_1}} }
a_{\rm c}(\ve{k}_1 ) a_{\rm c}(\ve{k}_2 ) =
2
\int {\rm d}\ve{k}\
{ {t_{\vsm{k}}}\over {\omega_{\vsm{k}}} } a_{\rm c}(\ve{k}) a_{\rm c}(-\ve{k})\ .
\eqno({\rm A.23})
$$
The commutator $[R,V]$ includes also the Hermitian conjugate of
$[{\cal R},{\cal V}]$. Therefore, we obtain from
$[{\cal R},{\cal V}] + {\rm H.c.}$ the following ``bad'' two-operator
meson contribution:
$$
2
\int {\rm d}\ve{k}\
{ {t_{\vsm{k}}}\over {\omega_{\vsm{k}}} }
\{ a_{\rm c}(\ve{k}) a_{\rm c}(-\ve{k}) +
a_{\rm c}^{\dagger}(\ve{k}) a_{\rm c}^{\dagger}(-\ve{k}) \}
\ .
\eqno({\rm A.24})
$$
In the case of $F^{\dagger} [ R^{\vsm{k}_1} , V^{-\vsm{k}_2} ] F
a_{\rm c}^{\dagger}(\ve{k}_2 ) a_{\rm c}(\ve{k}_1 )$ (see the
beginning of this subsection) after normal ordering of the
$dd^{\dagger}$-kind terms one has to deal with
${\rm Tr} [ R^{\vsm{k}_1} , V^{-\vsm{k}_2} ]_{22}$ that differs
from Eq. (A.16) only by the replacement $\ve{k}_2\rightarrow -\ve{k}_2$.
Therefore, we obtain from $[{\cal R},{\cal V}^{\dagger}]$ the following
expression bilinear in the meson operators:
$$
2 \int {\rm d}\ve{k}\
{ {t_{\vsm{k}}}\over {\omega_{\vsm{k}}} }
a_{\rm c}^{\dagger}(\ve{k}) a_{\rm c}(\ve{k})\ . \eqno({\rm A.25})
$$
The same expression stems from the H.c. of $[{\cal R},{\cal V}^{\dagger}]$,

Finally, uniting all these results one can write the entire contribution
from ${1\over 2}[R,V]$ to $K(\alpha)$, which is bilinear in the meson
operators,
$$
\int {\rm d}\ve{k}\
{ {t_{\vsm{k}}}\over {\omega_{\vsm{k}}} }
\{ 2 a_{\rm c}^{\dagger}(\ve{k}) a_{\rm c}(\ve{k}) +
a_{\rm c}(\ve{k}) a_{\rm c}(-\ve{k}) +
a_{\rm c}^{\dagger}(\ve{k}) a_{\rm c}^{\dagger}(-\ve{k}) \}
\ .
\eqno({\rm A.26})
$$

\bigskip
\hspace{-1em}{\large Appendix B}
\bigskip

\bigskip
\centerline{\bf Mathematical Aspects of the Clothing UT.}
\bigskip

By way of a simple example we shall show that $W$ used in the framework of
``clothing'' approach (see Sect. 2) may happen to be not an unitary operator
in the usual sense: unitary operator transforms vectors of a Hilbert space
into vectors of the same space, the scalar products being conserved. We argue
that $W$ need not be such an operator, viz.,
the clothing program can be described using an algebraic language as if
$W$ is an element of some algebra, being unitary in an algebraic sense. Our
example shows that such $W$ can have a representation by an operator that
transforms vectors of a Hilbert space ${\cal H}_0$ into vectors of another
Hilbert space $\cal H$  which is orthogonal to ${\cal H}_0$. In general,
the operator representation of $W$ turns out to be unnecessary because $W$ is
not used in calculations of probability amplitudes, expectation values and
other quantities which have physical interpretation.
We note that "clothing" allows us to choose a proper Hilbert space for
field model with the total Hamiltonian $H=H_0+H_I$. This space usually is
different from the space spanned on $H_0$ eigenvectors.

B.1. The clothing program can explicitly be carried out
in the following model. The scalar (meson) field $\phi(x)$ interacts with a
fixed\,(external) source, the Hamiltonian being $H_0=H^{mes}_0+g \int
\phi(\bx)f(\bx) d\bx$, where $H^{mes}_0$ is the free meson Hamiltonian
(see Eq.(2.1)). The Hamiltonian can be represented as
$$
H=\int \omega^0_{\bk} a\dg(\bk) a(\bk) d\bk
+
\int \omega_{\bk}^0 [v(\bk)\,a(\bk) + v^*(\bk)\,a\dg(\bk)] d\bk\   \eqno(\rm B.1)
$$

with\\
$$
v(\bk) = \frac{g}{\sqrt{2(2\pi\omega_{\bk}^0)^3}}{\tilde f}(\bk), \\
$$
where $\omega_{\bk}^0 = \sqrt{\bk^2+\mu^2_0}$ is the bare meson mass.
The Fourier transform ${\tilde f}(\bk)$ of the source function $f(\bx)$ is
a constant if the source is pointlike: $f(\bx) \sim \delta(\bx)$.

In accordance with the recipe (A.4) we find the corresponding clothing
transformation $W = \exp R \equiv W(a_c) = W(a)$ with its generator
$$
R = \int v(\bk) [a\dg(\bk) - a(\bk)] d\bk\ \eqno(\rm B.2)
$$
When deriving this expression, it is convenient to exploit the relation
$$
e^{\imath H^{mes}_0 t} a(\bk) e^{ - \imath H^{mes}_0 t} =
e^{- \imath \omega^0_{\bk} t} a(\bk)
$$
For brevity, we suppose that the form factor $v(\bk)$ is real ($ v^*(\bk) =
v(\bk)$).

Further, one can verify that
$$
a(\bk)= W a_c(\bk) W^{\dagger} = a_c(\bk) - v(\bk)\ \eqno(\rm B.3)
$$
and
$$
H = K(a_c ) = \int \omega_{\bk}^0 a\dg_c (\bk) a_c(\bk) - L\ , \eqno(\rm B.4)
$$
where $L = \int v(\bk)^2 d\bk $ is the c--number that shifts the H spectrum.

Thus, this transformation $W$ not only removes from $H$ the "bad" terms
(linear in $a$ and $a\dg$) but it reduces the primary eigenvalue problem
to a very simple one. In our model $K(a_c )$ does not contain
interaction terms: clothed particles (mesons) turn out to be free.

Now, we are interested in the clothed no--particle
state, i.e. the vector $\Omega$ such that
$$
a_c(\bk) \Omega=0, \quad \forall \bk \  \eqno(\rm B.5)
$$
with
$$
\langle \Omega \mid \Omega \rangle = 1.
$$

One can show that the vector $W^{\dagger}(a)\Omega_0$ obeys the condition
(B.5). In fact,
$$
a_c(\bk) W^{\dagger}(a) \Omega_0 = W^{\dagger} a(\bk) \Omega_0 = 0 \ \eqno(\rm B.6)
$$
So, one can put $\Omega = W^{\dagger}(a) \Omega_0$, i.e.,
$$
\Omega=\exp\{ - \int v(\bk)\,[a\dg(\bk) - a(\bk)] d\bk \} \Omega_0 = \
$$
$$
\   \exp( - L/2)\exp\{ - \int v(\bk) a^{\dagger}(\bk) d\bk \}\Omega_0\ \eqno(\rm B.7)
$$
Here, we have used the Hausdorff--Weyl formula,
$$
e^{A+B}=e^A e^B e^{-1/2[A,B]}\  \eqno(\rm B.8)
$$
for the two operators $A$ and $B$ such that $[A,B]$ commutes with $A$ and $B$.

Eq. (B.7) represents the "clothed" vacuum $\Omega$ as the superposition of
"bare" states
$$
\Omega_0, \quad a\dg(\bk_1) \Omega_0, \quad a\dg(\bk_1)\,a\dg(\bk_2) \Omega_0, \ldots \
\eqno(\rm B.9)
$$
Expansions
similar to (B.7) can be written for "clothed" one-particle state $a\dg_c(\bk) \Omega$
as well as for all vectors of the kind
$$
W^{\dagger}a\dg(\bk_1) \ldots a\dg(\bk_n) \Omega_0 =
a\dg_c(\bk_1) \ldots a\dg_c(\bk_n) W\dg \Omega_0 \ ,  \eqno(\rm B.10)
$$
each of them being the H eigenvector.

The states (B.9) are the $H_0$ eigenvectors and they form the basis of
the Fock (Hilbert) space ${\cal H}_0$. The set
$$
\Omega, \quad a\dg_c(\bk_1) \Omega, \quad a\dg_c(\bk_1)\,a\dg_c(\bk_2) \Omega, \ldots \
\eqno(\rm B.11)
$$
of the $H$ eigenvectors
is the basis of the Hilbert space which we shall call $\cal H$.
We see that in our model
with the cutoff function $v(\bk)$ which decreases rapidly enough to yield a finite
normalization factor ${\rm exp}{( - L/2)}$, the space $\cal H$ can be spanned
onto the vectors (B.9).

It is not the case in the model with a "soft" form factor $v(\bk)$. Indeed,
if $L = \infty$ (this occurs, e.g., for the pointlike source) then we obtain
the zero values for all $W\dg\Omega_0$ projections on the vectors (B.9).
Moreover, all vectors (B.10) are orthogonal to ${\cal H}_0$ if $L = \infty$.
We may conclude that all vectors (B.11) are zero if ${\cal H}_0$
is assumed to be the complete space of states. The operator $W\dg$ transforms
all ${\cal H}_0$ vectors into the zero one and therefore can not be called
the unitary operator because the latter, by definition, must
transform vectors from ${\cal H}_0$ into vectors from ${\cal H}_0$ and
conserve the scalar product in the space.

Note that in the case with $ L \rightarrow \infty$ there are different ways
for calculating the scalar product  $\langle \Omega \mid \Omega \rangle =
\langle W\dg \Omega_0 \mid W\dg \Omega_0 \rangle $. On the one hand, putting
$L = \infty$ in Eq.(B.7), we obtain $W\dg \Omega_0 = 0$ once ${\cal H}_0$ is
complete. Therefore, $\langle W\dg \Omega_0 \mid W\dg \Omega_0 \rangle = 0$.
On the other hand, calculating at first $\langle \Omega \mid \Omega \rangle$
at finite $L$, we obtain unity for any $L$, i.e., such limit
of $\langle W\dg \Omega_0 \mid W\dg \Omega_0 \rangle$ as
$L \rightarrow \infty$ is unity\footnote{In the context,
we find in \cite{SCH61}(Ch. XII) that
$\langle \Omega \mid \Omega \rangle = \infty$ at $L = \infty $. It follows
from the supposition made therein that the projection
$\Phi^{(0)} = \langle \Omega_0 \mid \Omega \rangle$ is not zero at $L = \infty$.
However, this supposition is wrong ( see Eq. (B.7) and Van Hove paper
\cite{VANHOVE52})}.

B.2. So, in the case $L = \infty$ the standard approach that is relied upon
the initial Hilbert space ${\cal H}_0$ does not allow to find $H$ eigenstates
(B.11).
The situation was clarified by Van Hove \cite{VANHOVE52}. He considered $H$
as an operator in the space
constructed as the infinite product of the Hilbert spaces for the
oscillators $\omega_k a\dg(k)\,a(k)$ (Van Hove assumes that $k$ runs discrete
values). This extended space is not a Hilbertian one, but it can be
decomposed into a direct sum of mutually orthogonal Hilbert spaces,
${\cal H}_0$
being one of them (e.g., see \cite{VONNEU38}--\cite{ALDU57}).

By using this space Van Hove has proved
that the $H$ eigenvector  which belongs to the least $H$ eigenvalue (it
coincides with the no--particle state $\Omega$ because of Eq.(B.4)) has
unit norm and is orthogonal to ${\cal H}_0$. Moreover, all $H$ eigenvectors
(B.11) are orthogonal to ${\cal H}_0$. The vectors (B.11) form the basis of
another Hilbert space which is orthogonal to ${\cal H}_0$. We have called
it ${\cal H}$ above.

Of course, the model under discussion is too simplified and the
resulting theory is equivalent to the free one, viz., $H$ contains no interaction
terms after "clothing"  ( see Eq. (B.4)). But it enables us to suspect that
in nontrivial cases the total Hamiltonian eigenvectors may happen to be
orthogonal to the initial
Hilbert space ${\cal H}_0$, i.e., "bare" states space or space of the eigenvectors
of the free part $H_0$ of the total Hamiltonian. On this ground one
may cast doubt about validity of the usual quantum postulate  that
$H$ as well as other observables can be defined initially as operators in
${\cal H}_0$. One can also anticipate that the "clothing" transformation may
not be unitary operator in ${\cal H}_0$.

B.3. In order to avoid the troubles described above we suggest the following
algebraic approach to the quantum field theory.
We consider all the operators occurred in quantum field theory as elements
of an algebra, devoid of operator representation. This means that
{\it ab initio} we do not introduce the notion of vectors, describing states of
the physical system. In this algebra besides the addition and
multiplication of elements an operation of involution $\dagger$ is defined, which
corresponds to the Hermitian conjugation in the operator language (see,
e.g., \cite{BLOT90}\,(\S 1.5) and \cite{HAAG92}\,(Ch. III)). The algebra
contains $A^{\dagger}$ along with the element $A$. All elements of the
algebra (in particular, the Hamiltonian) can be expressed in terms of some
basic algebraic elements. In the case of Yukawa model the latter are
$$
a(\bk), a^{\dagger}(\bk), b(\bp,r), b^{\dagger}(\bp,r),
d(\bp,r), d^{\dagger}(\bp,r)   \quad  \forall \bk, \bp, r \  \eqno(\rm B.12)
$$
The multiplication operation is noncommutative: $AB$ may not be equal to $BA$. The commutators
$[A,B]$ for basic elements are assumed to be known (see, e.g., Eq.(2.3),
then the commutators of any two elements can be calculated.

The "clothing" program can be formulated and realized using this algebraic
language. For example, the requirements i) and ii) from Sect. 2 can be
replaced by equivalent ones: the Hamiltonian $H$ when expressed in terms of
new basic elements $\alpha_p$ (instead of the starting elements
denoted as $a_p$ in Sect. 2) must not contain "bad" terms (see Subsect. 2.4).
The elements $\alpha_p$ and $a_p$ are connected by the isomorphic
transformation $\alpha_p=W\dg a_p W$, $W$ being a unitary element of the
algebra, i.e., such that $W^{\dagger}W = W\,W\dg = 1$. Let us stress that
$W^{\dagger}\,a_p\,W$ is calculated using purely algebraic means, namely
commutation relations and Eq.(2.15).

Of course, our theory must provide numbers, which can be compared with experimental
data (cross sections, expectation values, etc.). In quantum mechanics this is
accomplished by means of the realization of the above algebraic elements as
operators in a vector space. We define the space as follows.

After "clothing" we introduce the additional notion of a state $\Omega$
(cyclic state) such that
$$
a_c(\bk)\Omega=b_c(\bp,r)\Omega=d_c(\bp,r)\Omega=0 \quad
\forall \bk, \bp, r
$$
This state coincides with the $H$ eigenstate corresponding to the least
$H$ eigenvalue. We assume that the observable no-particle state is described
by $\Omega$. By assumption observable one-particle states are described by
the states
$$
a\dg_c(\bk)\Omega, \quad b\dg_c(\bp,r)\Omega, \quad d\dg_c(\bp,r)\Omega
$$
The vectors
$$
a\dg_c(\bk_1) \ldots a\dg_c(\bk_n)\Omega, \quad
b\dg_c(\bp_1, r_1) \ldots b\dg_c(\bp_n, r_n)\Omega, \ldots \quad  n\geq2
$$
can be chosen as the remaining basic vectors of our Hilbert space ${\cal H}$.
Using the vectors one can calculate quantities of physical interest
in the usual manner.

One may say that one of the goals of the "clothing" is to select a Hilbert
space which would be suitable for the given field Hamiltonian $H$. The
first-quantized quantum mechanics uses only one universal Hilbert space for
different Hamiltonians, the second-quantized theory needs distinct spaces
for different interactions (e.g., for different values of the coupling
constant $g$).

The approach described above needs not initial space ${\cal H}_0$. We need
not to consider $W$ as an operator in a space. Algebraically $W$ is defined
as an unitary element of the described algebra. Besides $W$ the algebra has
other elements which need not operator realization, the free part $H_0$ of
the total Hamiltonian being the example.

\bigskip
\hspace{-1em}{\large Appendix C}
\bigskip

\bigskip
\centerline{\bf The UT Method in Scalar Field Model}
\bigskip

In order to illustrate the key points of clothing and Okubo's approaches let us
consider another exactly solvable model \cite{WENT46,SCH61} in which a
neutral scalar
field is coupled with spinless fermions whose energy is independent of momentum.
The model Hamiltonian is $H = H_0 + V$,
$$
H = m_0\,B(0) + \int \omega_{ \bsk } a\dg(\bk) a(\bk) d\bk, \quad
\omega_{\bsk} = \sqrt{\bk^2+\mu^2} \ ,
\eqno(\rm C.1)
$$
$$
V = g \int \omega_{\bsk} \left[ B(\bk)\,a({\bk}) + {\rm H.c.} \right] \
h(\bk^2) {\rm d}\bk, \quad h (\bk ^2) = { {f(\bk ^2)}\over {\sqrt{2(2\pi\omega_{\bsk})^3}} } \
\eqno(\rm C.2)
$$
with
$$
B(\bk ) \equiv \int b\dg(\bp + \bk )\,b(\bp ) {\rm d}\bp = B\dg (-\bk ) \ ,
$$
where $a({\bk})$ and $b({\bp})$ are the destruction operators for bosons and fermions,
respectively, which meet the usual commutation rules (cf. Eqs. (2.5)):
$$
[ a({\bk}) , a\dg ({\bk'})] = \delta (\bk - \bk' ) \ ,
\eqno(\rm C.3a)
$$
$$
\{ b({\bp}) , b\dg ({\bp'}) \} = \delta (\bp - \bp' ) \ .
\eqno(\rm C.3b)
$$
The translational invariance of the Hamiltonian provides the momentum conservation.
The cut-off factor $f(x)$ is assumed to fall off rapidly enough for large $x$
to make
finite all the integrals that occur in the theory.

\medskip

\underline{C.1} Again the clothing transformation of this Hamiltonian can be
found in a closed form (cf. App. B). Indeed, calculating the respective
integral (A.4) and noticing that
$$
[ B (\bk ) , B ( \bk' ) ] = 0 \ \ \ \forall\ \  \bk \ , \ \bk' \ ,
\eqno({\rm C.4})
$$
we get
$$
R = - g \int [ B ({\bk}) a ({\bk}) - {\rm H.c.} ] \ h({\bk}^2) \ {\rm d}{\bk}
\equiv g (X\dg -X )
\eqno({\rm C.5})
$$
with
$$
X = \int h({\bk}^2) B ({\bk}) a({\bk}) {\rm d}{\bk} \ .
$$

Further, it is useful to keep in mind the following formulae
$$
[ X , X\dg ] = \int h^2 ({\bk}^2) B ({\bk}) B\dg ({\bk}) {\rm d}{\bk}
\eqno({\rm C.6})
$$
and
$$
[X , [ X , X\dg ] ] = 0 \ ,
\eqno({\rm C.7})
$$
whence due to Eq. (B.8) the transformation of interest can be written as
$$
W = {\rm exp} [g(X\dg -X)] = {\rm exp} (gX\dg )\ {\rm exp} (-gX)\
{\rm exp} (-{g^2\over 2}[X , X\dg]) \ .
\eqno({\rm C.8})
$$
Now, by using the relation (cf. Eq. (B.3))
$$
a ({\bk}) = W a_{\rm c} ({\bk}) W\dg = a_{\rm c} ({\bk}) - gh ({\bk}^2 ) B\dg_{\rm c} ({\bk}) \ ,
\eqno({\rm C.9})
$$
where the boson-type operator
$$
B_{\rm c} ({\bk}) = \int b\dg_{\rm c} (\bp + \bk )\ b_{\rm c} (\bp ) {\rm d}\bp
\eqno({\rm C.10})
$$
commutes with $W \equiv W (a, b) = W (a_{\rm c}, b_{\rm c} )$ (see Eq. (C.4)),
one can show that
$$
H \equiv H (a, b) = K (a_{\rm c}, b_{\rm c} ) = W H (a_{\rm c}, b_{\rm c} ) W\dg
$$
$$
=
m_0 B_{\rm c} (0) +
\int \omega_{\bsk} a\dg_{\rm c}(\bk ) a_{\rm c}(\bk ) {\rm d}\bk -
g^2 \int \omega_{\bsk} h^2 (\bk ^2 ) B\dg_{\rm c} (\bk ) B_{\rm c} (\bk ) {\rm d}\bk \ .
\eqno({\rm C.11})
$$
Reshuffling the fermion operators in normal order in the r.h.s. of Eq.
({\rm C.11}) we obtain
$$
\int \omega_{\bsk} h^2 (\bk ^2 ) B\dg _{\rm c} (\bk ) B_{\rm c} (\bk ) {\rm d}\bk =
B_{\rm c} (0) \int \omega_{\bsk} h^2 (\bk ^2 ) {\rm d}\bk
$$
$$
-
\int \omega_{\bsk} h^2 (\bk ^2 ) {\rm d}\bk \int {\rm d}\bp \int {\rm d}\bp' \
b\dg_{\rm c} (\bp + \bk )\ b\dg _{\rm c} (\bp' ) \ b_{\rm c} (\bp ) \ b_{\rm c} (\bp' + \bk ) \ .
$$
The first term in the r.h.s. of this equation has the same structure as
$m_0 B_{\rm c} (0)$ in Eq. ({\rm C.11}) and gives the radiative correction
(renormalization) to the bare fermion mass $m_0$,
$$
m = m_0 - g^2 \int \omega_{\bsk} h^2 (\bk ^2 ) {\rm d}\bk \ .
\eqno({\rm C.12})
$$
Meanwhile, the $H$ eigenvector belongs to the {H} eigenvalue $m$.

So,
$$
H = K (a_{\rm c}, b_{\rm c} ) = K_{\rm F} + K_{\rm I} \ ,
\eqno({\rm C.13})
$$
$$
K_{\rm F} = m B_{\rm c} (0) + \int \omega_{\bsk} a\dg_{\rm c}(\bk ) a_{\rm c}(\bk )
{\rm d}\bk \equiv K_{\it ferm} + K_{\it boson} \ ,
\eqno({\rm C.14})
$$
$$
K_{\rm I} = \int {\rm d} \bx \int {\rm d} \bx' \psi\dg_{\rm c} (\bx )\ \psi\dg_{\rm c} (\bx' )\
V_{\rm ff} (\mid \bx - \bx'\mid )\ \psi_{\rm c} (\bx )\ \psi_{\rm c} (\bx' ) \ ,
\eqno({\rm C.15})
$$
$$
V_{ff} (\mid \br \mid ) = - g^2 \int \omega_{\bsk} h^2 (\bk ^2 )
e^{\imath \bk \br } {\rm d}\bk \ ,  \eqno(\rm C.16)
$$
where in agreement with the secondary quantization prescriptions we have
introduced the $ \psi_{\rm c} $ - field for clothed fermions in the Schr\"odinger
picture assuming
$$
\psi_{\rm c} (\bx ) = { {1}\over {\sqrt{(2\pi )^3}} }
\int b_c (\bp ) e^{ \imath \bp \bx} {\rm d} \bp \,
$$
$$
\{ \psi_{\rm c} (\bx ), \psi\dg_{\rm c} (\bx' ) \} = \delta (\bx - \bx' ) .
$$
Therefore, $ V_{ff} (\mid \br \mid ) $ can be considered as a two--fermion
interaction potential.

One should point out that the new interaction Hamiltonian $K_I$ expressed
through clothed operators no longer contains any self--interaction and leads
merely to an interaction between pairs of fermions.

The $K_F$ eigenvectors
$$
\Omega, \, a\dg_c (\bk_1) \Omega,\, b\dg_c (\bp_1) \Omega, \,
a\dg_c (\bk_1) a\dg_c (\bk_2) \Omega, \, b\dg_c (\bp_1) b\dg_c (\bp_2) \Omega, \,
a\dg_c (\bk_1) b\dg_c (\bp_2) \Omega, \ldots \  \eqno(\rm C.17)
$$
with the running element
$$
\Omega_c (\bk_1 \ldots \bk_r ; \bp_1 \ldots \bp_s) =
a\dg_c (\bk_1) \ldots a\dg_c (\bk_r) b\dg_c (\bp_1) \ldots b\dg_c (\bp_s) \Omega
\quad (r,s = 1, 2, \ldots)\ \eqno(\rm C.18)
$$
form a basis (see App. $B$). \\
Another basis can be composed of bare vacuum $\Omega_0$ and
vectors
$$
\Omega_0 (\bk_1 \ldots \bk_r ; \bp_1 \ldots \bp_s) =
a\dg (\bk_1) \ldots a\dg (\bk_r) b\dg (\bp_1) \ldots b\dg (\bp_s) \Omega_0
\quad (r,s = 1, 2, \ldots)\ , \eqno(\rm C.19)
$$
i.e., of the $H_0$ eigenstates.\\
Note also that in the given model, due to the absence of
$f \bar f $ --processes, the vacuum $\Omega$ for the coupled fields coincides
(to an accuracy of a phase factor) with the vacuum $\Omega_0$ for the free
fields. In the context, with the help of Eq. ({\rm C.9}) we find,
$$
a\dg_c (\bk) \Omega = a\dg_c (\bk) \Omega_0 = a\dg (\bk) \Omega_0 \ ,
\eqno(\rm C.20)
$$
i.e., the bare and clothed one--boson states are the same.

As to the clothed one--fermion states $ b\dg_c (\bp) \Omega $, they are
complex superpositions of vectors ({\rm C.19}) with one fermion and arbitrary
boson configurations\,(a boson cloud surrounding the fermion). This follows
directly from the explicit expression for
$$
b\dg_c (\bp) = W\dg b\dg (\bp) W =
\int {\cal F} (\bp - \bp' ) b\dg (\bp') {\rm d}\bp'\ , \eqno(\rm C.21)
$$
$$
{\cal F} (\bq ) = { {1} \over {(2\pi)^3 } } \int e^{ - \imath \bq \bx }
{\rm exp} \{ - g \int {\rm exp}[ - \imath \bk \bx] \left( a\dg (\bk) - a\dg (-\bk) \right)
h(\bk^2) {\rm d} \bk \} {\rm d} \bx \ .
$$
The factor ${\cal F}(\bq) $ characterizes a boson distribution in the cloud.
In the free case with ( $g = 0$) $ {\cal F} (\bq) = \delta (\bq ) $.

\medskip

\underline{\rm C.2 } It follows from Eqs. ({\rm C.13})--({\rm C.15}) that
clothed mesons do not interact with nucleons, they are free. In particular,
no--meson states cannot pass in states with one, two, etc. mesons. This
means that $K(a_c, b_c) $ possesses the property of the operator
$H' = U\dg H U $ obtained from $H $ via Okubo's transformation considered
in Subsect. 6.5., viz., $H'_{12} = H'_{21} = 0 $. In other words, $H'$ has
vanishing matrix elements between the no--meson states and one--, two--,
\ldots meson states.
This property of $K(a_c, b_c) $ is specific for the given simple model.
In general, the clothing UT permits $K$ to contain such nondiagonal terms
that are responsible for the process $NN \to NN \pi $.

Since the model clothing UT $W $ fulfills Okubo's requirement it is
interesting to compare the block structure of $W\dg $ with that described
in Subsects. 6.2 and 6.3 for Okubo's UT. To do it in a compact form one
may calculate separate blocks of $W\dg $ decomposition of the type
determined by Eq. (6.14 ). In the context, recall that Okubo's UT can be
represented by a function $U(a,b) $ of bare creation (destruction) operators.
$W\dg $ can also be given by a function $W\dg (a,b)$ of the bare
operators since $W(a_c,b_c) = W(a,b)$ (cf. Eq. (2.13)).
Therefore, for Okubo--like decomposition of $W\dg $ one may use
the projector $\eta_1 $ onto bare no--meson states.

In order to carry out the comparison with a solution of the corresponding
decoupling equation (6.22) (see {\rm C.3} below ) we shall find the operator
$ {\cal A}^{21} $ defined by
$$
{W\dg}^{21} = {\cal A}^{21} {W\dg }^{11}\   \eqno(\rm C.22)
$$
It determines the basic element $ {\cal A}^{21} $ of Okubo's UT
(see, for instance, Eqs. (6.8) and (6.20) ).

We have
$$
{W\dg}^{11} = {\eta}_1 W\dg {\eta}_1 =
{\eta}_1 {\rm exp} \left( - g X\dg \right) {\rm exp}
\left( - \frac{g^2}{2} \left [ X,X\dg \right] \right)
{\rm exp} \left( g X \right) {\eta}_1\
$$
$$
\ = {\rm exp} \left( - \frac{g^2}{2} \left [ X,X\dg \right] \right) {\eta}_1 \ ,
\eqno(\rm C.23)
$$
$$
{W\dg}^{21} = {\eta}_2 W\dg {\eta}_1 = \
{\eta}_2 {\rm exp} \left( - g X\dg \right) {\rm exp}
\left( - \frac{g^2}{2} \left [ X,X\dg \right] \right) {\eta}_1 = \
{\eta}_2 {\rm exp} \left( - g X\dg \right) {\eta}_1 {W\dg}^{11}\
\eqno(\rm C.24)
$$
While deriving these formulae we have used the property
$$
a_c (\bk) {\eta}_1 = {\eta}_1 a\dg_c (\bk) = 0  \quad  \forall \bk
\eqno(\rm C.25)
$$
and its consequence
$$
{\rm exp} (g\,X) {\eta}_1 = {\eta}_1 = {\eta}_1 {\rm exp} ( - g\,X\dg )\
\eqno(\rm C.26)
$$
We also have taken into account the relation
$$
\left [ X,X\dg \right] {\eta}_1 = {\eta}_1 \left [ X,X\dg \right]\ .
\eqno(\rm C.27)
$$

It follows from ({\rm C.23}) and ({\rm C.24}) that
$$
{\cal A}^{21} = {\eta}_2 {\rm exp} \left( - g X\dg \right) {\eta}_1 =
\left( {\rm exp} \left( - g X\dg \right) - 1 \right) {\eta}_1 \
\eqno(\rm C.28)
$$
Note that we have managed to find the operator $ {\cal A}^{21} $
in an explicit form without solving Eq. ({\rm C.22}).

Unitarity relations for the Okubo--type blocks $ U^{ij}\equiv
(W^{\dagger})^{ij} (i,j = 1,2) $
with $ U^{21} = {\cal A}^{21} U^{11} $,
$$
{U^{11} }\dg ( 1 + {{\cal A}^{21} }\dg {\cal A}^{21} ) U^{11} = \
{U^{11} }\dg \left( 1 + {{\cal A}^{21} }\dg \right)
\left( 1 + {\cal A}^{21} \right) U^{11} = \eta_1 \ , \eqno({\rm C.29} )
$$
$$
U^{12} = - {{\cal A}^{21} }\dg U^{22} \ , \eqno({\rm C.30} )
$$
and
$$
{U^{22} }\dg ( 1 + {\cal A}^{21} {{\cal A}^{21} }\dg ) U^{22} = \eta_2 \
\eqno({\rm C.31} )
$$
can directly be verified.
For instance, we get step by step,
$$
- {{\cal A}^{21} }\dg U^{22} = \eta_1 \left( 1 -
{\rm exp} \left( - g X \right) \right) {\rm exp} \left [ g(X - X\dg) \right ]
\eta_2 =
$$
$$
\eta_1 {\rm exp} \left[ g(X - X\dg) \right] \eta_2 - \
\eta_1 {\rm exp} \left( - g X \right) {\rm exp} \left [ g(X - X\dg) \right ]
\eta_2 = {W\dg }^{12} \equiv U^{12} \ , \eqno({\rm C.32} )
$$
since in accordance with Eq. ({\rm C.27})
$$
\eta_1 {\rm exp} \left( - g X \right) {\rm exp} \left [ g(X - X\dg) \right ]
\eta_2 = \eta_1 {\rm exp} \left( - g X\dg \right)\
{\rm exp} \left( \frac{g^2}{2} \left [ X,X\dg \right] \right) {\eta}_2 = \
$$
$$
\eta_1 {\rm exp} \left( \frac{g^2}{2} \left [ X,X\dg \right] \right) {\eta}_2 = 0 \ ,
$$
and
$$
{U^{22} }\dg ( 1 + {\cal A}^{21} {{\cal A}^{21} }\dg ) U^{22} = \
{U^{22} }\dg U^{22} + {U^{12} }\dg U^{12} = \
$$
$$
\eta_2 {\rm exp} \left[ g(X\dg - X ) \right] \eta_2 {\rm exp} \left [ g(X - X\dg) \right ] \
\eta_2 + \eta_2 {\rm exp} \left[ g(X\dg - X ) \right] \eta_1 \
{\rm exp} \left [ g(X - X\dg) \right ] \eta_2 = \
$$
$$
\eta_2 W W\dg \eta_2 = \eta_2 \
\eqno({\rm C.33} )
$$
Q.E.D.

\underline{\rm C. 3} Let us try to solve Okubo equation (6.22) for this
model Hamiltonian. It is of great interest to attain some experience in
handling with similar nonlinear equations.
We prefer to start with the equation equivalent to Eq. (6.22) :
$$
{\eta}_2 \{ \left[ H_0, J_1 \right]  + V\,J_1 - J_1 V\,J_1 \} {\eta}_1 = 0\
\eqno(\rm C.34)
$$
for the operator $ J_1 = (1 + {\cal A}^{21}){\eta}_1 $ (cf. Eq. (29)' in
\cite{OKU54}).

One should point out the properties
$$
{\eta}_1 J_1 = \eta_1,\ \eqno(\rm C.35a)
$$
$$
J_1 (1 - {\eta}_1) = 0\   \eqno(\rm C.35b)
$$
and the condition
$$
\left[ H_0 , \eta_1 \right] = 0\   \eqno(\rm C.36)
$$

Further, introducing the interaction picture operators $O$,
$$
O(t) = {\rm exp}(\imath H_0 t) O {\rm exp}( - \imath H_0 t)\ ,
$$
and noticing that $ {\eta}_1 (t) = {\eta}_1 $ one can get the Riccati--type
differential equation for $J_1(t)$ (cf. Eq. (31) in \cite{OKU54} ),
$$
\imath {\eta}_2 \frac{d}{dt}J_1(t) {\eta}_1 = {\eta}_2 \{ V(t)J_1(t) -
J_1(t) V(t) J_1(t) \} {\eta}_1 \  \eqno(\rm C.37)
$$
In the scalar model we have
$$
V(t) = g \int \omega_{\bsk} \left[ B\dg(\bk)\,a\dg(\bk,t) + {\rm H.c.} \right]
h(\bk^2) {\rm d}\bk\  \eqno(\rm C.38)
$$
with
$$
a(\bk,t) = a(\bk) {\rm exp} ( - \imath \omega_{\bsk} t)\ ,
$$
so that
$$
[ a(\bk,t) , a\dg (\bk',t')] = \delta (\bk - \bk' )
{\rm exp} [ - \imath \omega_{\bsk} (t - t') ]\   \eqno(\rm C.39)
$$

Trying to solve Eq. (C.37), one should note the relation
$$
V(t) = - \imath g \frac{d}{dt} \left[ X\dg (t) - X (t) \right] \ ,  \eqno(\rm C.40)
$$
where
$$
X(t) = \int h({\bk}^2) B ({\bk}) a({\bk},t) {\rm d}{\bk} \ .
$$

Now, applying the Lagrange method well known in the theory of ordinary
differential equations (see, e.g., \cite{SMIR74}, \S 1.6) let us search
a solution of Eq. (C.37) in the form
$$
J_1(t) = {\rm exp} \left[ - g\,X\dg (t) \right] G(t)\ , \eqno(\rm C.41)
$$
where in accordance with Eqs.(\rm C.35)--(\rm C.36) we have
$$
{\eta}_1 G(t)={\eta}_1\ , \eqno(\rm C.42a)
$$
$$
G(t)(1 - {\eta_1}) = 0\    \eqno(\rm C.42b)
$$
Differentiating (\rm C.41) and taking into account that
$$
\left[ \frac{d}{dt} X\dg (t) , X\dg (t) \right] = 0\ ,
$$
we find
$$
\frac{d}{dt}J_1(t) = - g \left( \frac{d}{dt}X\dg(t)\right)\, {\rm exp} \left[ - g\,X\dg (t) \right] G(t)
+ {\rm exp} \left[ - g\,X\dg (t) \right] \frac{d}{dt}G(t)\ .
$$
Substitution of this expression into Eq. (\rm C.37) enables us to remove
in the r.h.s. of this equation
not the linear term $V(t)J_1(t)$ as a whole but its part
$ - \imath g \frac{d}{dt} X\dg (t) J_1(t) $ (see Eq. (C.40)). So, we get
the following equation for the operator function $G(t)$ ,
$$
{\eta}_2 {\rm exp} \left[ - g\,X\dg (t) \right] \frac{d}{dt}G(t) {\eta}_1
= g {\eta}_2 {\rm exp} \left[ - g\,X\dg (t) \right]
\{ \frac{d}{dt} X(t) + g \left[ X\dg (t),\frac{d}{dt} X(t) \right] \} G(t) {\eta}_1 \
$$
$$
\ - g {\eta}_2 {\rm exp} \left[ - g\,X\dg (t) \right] G(t) {\eta}_1
\{ \frac{d}{dt} X(t) + g \left[ X\dg (t),\frac{d}{dt} X(t) \right] \} G(t) {\eta}_1\ ,
\eqno(\rm C.43)
$$
At the point we have used the properties
$ {\eta}_1 {\rm exp} \left[ - g\,X\dg (t) \right] = {\eta}_1 $
and $ \eta_1 \frac{d}{dt} X\dg (t) = 0 $  and the relation
$$
{\rm exp} \left[ g\,X\dg (t) \right] \frac{d}{dt} X(t) {\rm exp} \left[ - g\,X\dg (t) \right]
= \frac{d}{dt} X(t) + g \left[ X\dg (t),\frac{d}{dt} X(t) \right]\ ,
$$
that follows from Eqs. (2.15) and (\rm C.4).

Further, we have
$$
\left[ X\dg (t),\frac{d}{dt} X(t) \right] =
\imath \int \omega_{\bsk} h^2 ({\bk}^2) B\dg ({\bk}) B ({\bk}) {\rm d}{\bk}\
\equiv C
$$
In other words, this time independent commutator acts merely on the fermionic
degrees of freedom. Note that $C$ commutes with the projector $ \eta_1 $.

Eq. (\rm C.43) can be satisfied if we put
$$
\frac{d}{dt}G(t) = g \{ \frac{d}{dt} X(t) + gC \} G(t)
$$
$$
- g G(t) \{ \frac{d}{dt} X(t) + gC \} G(t) \ .
\eqno(\rm C.44)
$$
It may be shown that one of its possible solutions is
$$
G_1 (t) = {\eta}_1\  \eqno(\rm C.45)
$$
It is evident that $G_1(t)$ meets necessary requirements (C.42)

The corresponding operator $ J_1(t)  =
{\rm exp} \left[ - g\,X\dg (t) \right] {\eta}_1 $ yields
$$
{\cal A}^{21} (t) = \{ {\rm exp} \left[ - g\,X\dg (t) \right] - 1 \} {\eta}_1\ ,
\eqno(\rm C.46)
$$
that is equivalent at $t = 0 $ to the result (\rm C.28).

\newpage

\topmargin 40mm

\begin{figure}[h]
\hskip -30mm
\unitlength=1.00mm
\linethickness {0.4pt}
\begin{picture}(100.00,50.00)

\put(90.00,110.00){\makebox(0,0)[cc]{{\Large $K_4=$}}}


\put(50.00,80.00){\circle{20.00}}
\put(50.00,80.00){\makebox(0,0)[cc]{$V_{NN}$}}

\put(45.00,85.00){\line(-5,2){15.00}}
\put(45.00,75.00){\line(-5,-2){15.00}}
\put(55.00,85.00){\line(5,2){15.00}}
\put(55.00,75.00){\line(5,-2){15.00}}

\put(30.00,95.00){\makebox(0,0)[cc]{$N$}}
\put(30.00,65.00){\makebox(0,0)[cc]{$N$}}
\put(70.00,95.00){\makebox(0,0)[cc]{$N^{\prime}$}}
\put(70.00,65.00){\makebox(0,0)[cc]{$N^{\prime}$}}

\put(50.00,60.00){\makebox(0,0)[cc]{${\bf a)}$}}

\put(90.00,80.00){\makebox(0,0)[cc]{{\Large $+$}}}


\put(130.00,80.00){\circle{20.00}}
\put(130.00,80.00){\makebox(0,0)[cc]{$V_{{\bar N}{\bar N}}$}}

\put(125.00,85.00){\line(-5,2){15.00}}
\put(125.00,75.00){\line(-5,-2){15.00}}
\put(135.00,85.00){\line(5,2){15.00}}
\put(135.00,75.00){\line(5,-2){15.00}}

\put(110.00,95.00){\makebox(0,0)[cc]{${\bar N}$}}
\put(110.00,65.00){\makebox(0,0)[cc]{${\bar N}$}}
\put(150.00,95.00){\makebox(0,0)[cc]{${\bar N}^{\prime}$}}
\put(150.00,65.00){\makebox(0,0)[cc]{${\bar N}^{\prime}$}}

\put(130.00,60.00){\makebox(0,0)[cc]{${\bf b)}$}}

\end{picture}

\hskip -30mm
\begin{picture}(100.00,50.00)


\put(50.00,80.00){\circle{20.00}}
\put(50.00,80.00){\makebox(0,0)[cc]{$V_{MN}$}}

\put(45.00,85.00){\line(-5,2){4.00}}
\put(39.00,87.50){\line(-5,2){4.00}}
\put(33.00,90.00){\line(-5,2){4.00}}

\put(45.00,75.00){\line(-5,-2){15.00}}

\put(55.00,85.00){\line(5,2){4.00}}
\put(61.00,87.50){\line(5,2){4.00}}
\put(67.00,90.00){\line(5,2){4.00}}

\put(55.00,75.00){\line(5,-2){15.00}}

\put(30.00,95.00){\makebox(0,0)[cc]{$\pi$}}
\put(30.00,65.00){\makebox(0,0)[cc]{$N$}}
\put(70.00,95.00){\makebox(0,0)[cc]{$\pi^{\prime}$}}
\put(70.00,65.00){\makebox(0,0)[cc]{$N^{\prime}$}}

\put(50.00,60.00){\makebox(0,0)[cc]{${\bf c)}$}}

\put(90.00,80.00){\makebox(0,0)[cc]{{\Large $+$}}}


\put(130.00,80.00){\circle{20.00}}
\put(130.00,80.00){\makebox(0,0)[cc]{$V_{M{\bar N}}$}}

\put(125.00,85.00){\line(-5,2){4.00}}
\put(119.00,87.50){\line(-5,2){4.00}}
\put(113.00,90.00){\line(-5,2){4.00}}

\put(125.00,75.00){\line(-5,-2){15.00}}

\put(135.00,85.00){\line(5,2){4.00}}
\put(141.00,87.50){\line(5,2){4.00}}
\put(147.00,90.00){\line(5,2){4.00}}

\put(135.00,75.00){\line(5,-2){15.00}}

\put(110.00,95.00){\makebox(0,0)[cc]{$\pi$}}
\put(110.00,65.00){\makebox(0,0)[cc]{${\bar N}$}}
\put(150.00,95.00){\makebox(0,0)[cc]{$\pi^{\prime}$}}
\put(150.00,65.00){\makebox(0,0)[cc]{${\bar N}^{\prime}$}}

\put(130.00,60.00){\makebox(0,0)[cc]{${\bf d)}$}}

\end{picture}

\hskip -30mm
\begin{picture}(100.00,50.00)


\put(50.00,80.00){\circle{20.00}}
\put(50.00,80.00){\makebox(0,0)[cc]{$V_{pair}$}}

\put(45.00,85.00){\line(-5,2){15.00}}
\put(45.00,75.00){\line(-5,-2){15.00}}

\put(55.00,85.00){\line(5,2){4.00}}
\put(61.00,87.50){\line(5,2){4.00}}
\put(67.00,90.00){\line(5,2){4.00}}

\put(55.00,75.00){\line(5,-2){4.00}}
\put(61.00,72.50){\line(5,-2){4.00}}
\put(67.00,70.00){\line(5,-2){4.00}}

\put(30.00,95.00){\makebox(0,0)[cc]{$N$}}
\put(30.00,65.00){\makebox(0,0)[cc]{${\bar N}$}}
\put(70.00,95.00){\makebox(0,0)[cc]{$\pi$}}
\put(70.00,65.00){\makebox(0,0)[cc]{$\pi$}}

\put(50.00,60.00){\makebox(0,0)[cc]{${\bf e)}$}}

\put(90.00,80.00){\makebox(0,0)[cc]{{\Large $+$}}}


\put(130.00,80.00){\circle{20.00}}
\put(130.00,80.00){\makebox(0,0)[cc]{$V_{pair}^{\dg}$}}

\put(125.00,85.00){\line(-5,2){4.00}}
\put(119.00,87.50){\line(-5,2){4.00}}
\put(113.00,90.00){\line(-5,2){4.00}}

\put(125.00,75.00){\line(-5,-2){4.00}}
\put(119.00,72.50){\line(-5,-2){4.00}}
\put(113.00,70.00){\line(-5,-2){4.00}}

\put(135.00,85.00){\line(5,2){15.00}}
\put(135.00,75.00){\line(5,-2){15.00}}

\put(110.00,95.00){\makebox(0,0)[cc]{$\pi$}}
\put(110.00,65.00){\makebox(0,0)[cc]{$\pi$}}
\put(150.00,95.00){\makebox(0,0)[cc]{$N$}}
\put(150.00,65.00){\makebox(0,0)[cc]{${\bar N}$}}

\put(130.00,60.00){\makebox(0,0)[cc]{${\bf f)}$}}

\end{picture}

\hskip -30mm
\begin{picture}(100.00,50.00)


\put(90.00,80.00){\circle{20.00}}
\put(90.00,80.00){\makebox(0,0)[cc]{$V_{N{\bar N}}$}}

\put(85.00,85.00){\line(-5,2){15.00}}
\put(85.00,75.00){\line(-5,-2){15.00}}
\put(95.00,85.00){\line(5,2){15.00}}
\put(95.00,75.00){\line(5,-2){15.00}}

\put(70.00,95.00){\makebox(0,0)[cc]{$N$}}
\put(70.00,65.00){\makebox(0,0)[cc]{${\bar N}$}}
\put(110.00,95.00){\makebox(0,0)[cc]{$N^{\prime}$}}
\put(110.00,65.00){\makebox(0,0)[cc]{${\bar N}^{\prime}$}}

\put(90.00,60.00){\makebox(0,0)[cc]{${\bf g)}$}}

\end{picture}
\vskip -45mm
\caption{Schematic representation of separate
contributions to the effective operator $K_4$:
a) $b_c\dg b_c\dg b_c b_c $, b) $d_c\dg d_c\dg d_c d_c$,
c) $b_c\dg a_c\dg b_c a_c$,
d) $d_c\dg a_c\dg d_c a_c$, e) $b_c\dg d_c\dg a_c a_c$,
f) $a_c\dg a_c\dg d_c b_c$, g) $b_c\dg d_c\dg d_c b_c$}
\label{fig1}
\end{figure}

\newpage

\topmargin -20mm

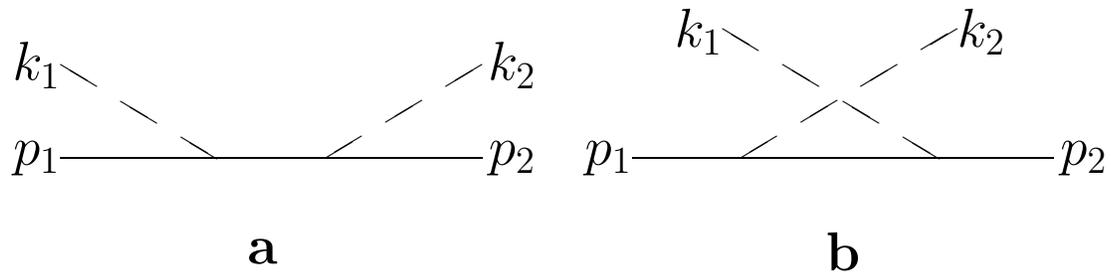
\begin{figure}[h]
%
\unitlength=0.8mm
\linethickness{0.4pt}
\begin{picture}(200.00,100.00)

\put(5.00,33.50){\line(1,0){70.00}}
\put(100.00,33.50){\line(1,0){70.00}}
%
%
\put(5.00,49.00){\line(5,-3){6.00}}
\put(15.00,43.00){\line(5,-3){6.00}}
\put(25.00,37.00){\line(5,-3){6.00}}
\put(49.00,33.50){\line(5,3){6.00}}
\put(59.00,39.50){\line(5,3){6.00}}
\put(69.00,45.50){\line(5,3){6.00}}
%
\put(115.00,55.00){\line(5,-3){6.00}}
\put(125.00,49.00){\line(5,-3){6.00}}
\put(135.00,43.00){\line(5,-3){6.00}}
\put(145.00,37.00){\line(5,-3){6.00}}
\put(118.00,33.50){\line(5,3){6.00}}
\put(128.00,39.50){\line(5,3){6.00}}
\put(138.00,45.50){\line(5,3){6.00}}
\put(148.00,51.50){\line(5,3){6.00}}

\put(1.00,33.50){\makebox(0,0)[cc]{{\LARGE $p_1$}}}
\put(96.00,33.50){\makebox(0,0)[cc]{{\LARGE $p_1$}}}
\put(80.00,33.50){\makebox(0,0)[cc]{{\LARGE $p_2$}}}
\put(175.00,33.50){\makebox(0,0)[cc]{{\LARGE $p_2$}}}
\put(1.00,49.00){\makebox(0,0)[cc]{{\LARGE $k_1$}}}
\put(111.00,54.50){\makebox(0,0)[cc]{{\LARGE $k_1$}}}
\put(80.00,49.00){\makebox(0,0)[cc]{{\LARGE $k_2$}}}
\put(158.00,54.50){\makebox(0,0)[cc]{{\LARGE $k_2$}}}

\put(38.50,18.00){\makebox(0,0)[cc]{{\LARGE \bf a}}}
\put(135.00,18.00){\makebox(0,0)[cc]{{\LARGE \bf b}}}

\end{picture}

\vskip 15mm
\caption{The $g^2$-order Feynman diagrams for $\pi N$ scattering:
a -- the $s$-pole graph; b -- the $u$-pole graph}

\label{fig2}

\end{figure}

\newpage

\topmargin -20mm

\begin{figure}[h]
\hskip -10mm
\unitlength=0.8mm
\linethickness{0.8pt}
\begin{picture}(200.00,100.00)

\put(45.00,80.00){\line(1,0){100.00}}
\put(45.00,35.00){\line(1,0){100.00}}
\put(95.00,35.00){\line(0,1){5.00}}
\put(95.00,45.00){\line(0,1){5.00}}
\put(95.00,55.00){\line(0,1){5.00}}
\put(95.00,65.00){\line(0,1){5.00}}
\put(95.00,75.00){\line(0,1){5.00}}

\put(40.00,80.00){\makebox(0,0)[cc]{\LARGE $p_1$}}
\put(40.00,35.00){\makebox(0,0)[cc]{\LARGE $p_2$}}
\put(100.00,57.00){\makebox(0,0)[cc]{\LARGE $k$}}
\put(150.00,80.00){\makebox(0,0)[cc]{\LARGE $p'_1$}}
\put(150.00,35.00){\makebox(0,0)[cc]{\LARGE $p'_2$}}

\put(95.00,20.00){\makebox(0,0)[cc]{${\bf a)}$}}
\end{picture}

\begin{picture}(100.00,100.00)

\put(45.00,80.00){\line(1,0){100.00}}
\put(45.00,35.00){\line(1,0){100.00}}
\put(95.00,35.00){\line(0,1){5.00}}
\put(95.00,45.00){\line(0,1){5.00}}
\put(95.00,55.00){\line(0,1){5.00}}
\put(95.00,65.00){\line(0,1){5.00}}
\put(95.00,75.00){\line(0,1){5.00}}

\put(40.00,80.00){\makebox(0,0)[cc]{\LARGE $p_2$}}
\put(40.00,35.00){\makebox(0,0)[cc]{\LARGE $p_1$}}
\put(100.00,57.00){\makebox(0,0)[cc]{\LARGE $k$}}
\put(150.00,80.00){\makebox(0,0)[cc]{\LARGE $p'_1$}}
\put(150.00,35.00){\makebox(0,0)[cc]{\LARGE $p'_2$}}

\put(95.00,20.00){\makebox(0,0)[cc]{${\bf b)}$}}
\end{picture}

\vskip 5mm
\caption{The one-pion-exchange Feynman diagrams for $NN$ scattering}

\label{fig3}
\end{figure}

\newpage

\centerline{\bf Figure Captions}

\bigskip
Fig. 1. Schematic representation of separate
contributions to the effective operator $K_4$:

a) $b_c\dg b_c\dg b_c b_c $, b) $d_c\dg d_c\dg d_c d_c$,
c) $b_c\dg a_c\dg b_c a_c$,
d) $d_c\dg a_c\dg d_c a_c$, e) $b_c\dg d_c\dg a_c a_c$,
f) $a_c\dg a_c\dg d_c b_c$,\\
g) $b_c\dg d_c\dg d_c b_c$

\bigskip
Fig. 2. The $g^2$-order Feynman diagrams for $\pi N$ scattering:

a -- the $s$-pole graph; b -- the $u$-pole graph

\bigskip
Fig. 3. The one-pion-exchange Feynman diagrams for $NN$ scattering


\begin{thebibliography}{100}

\bibitem{VLECK29} J.H. Van Vleck. Phys. Rev. {\bf 33} (1929) 467.

\bibitem{KEMBLE37} E.C.Kemble. Fundamentals of Quantum Mechanics
(Dover, N.Y., 1937), Ch. XI, Sect. 48c.

\bibitem{KLEIN74} D.J.Klein. J. Chem. Phys. {\bf 61} (1974) 786.

\bibitem{WENT46} G.Wentzel. Einf\"uhrung in die Quantheorie der Wellenfelder
(Wien, 1946).

\bibitem{HEIT54} W.Heitler. The Quantum Theory of Radiation (Clarendon,
Oxford, 1954). **

\bibitem{FST54} N.Fukuda, K.Sawada and M.Taketani. Prog. Theor. Phys. {\bf 12}
(1954) 156.

\bibitem{OKU54} S.Okubo. Prog. Theor. Phys. {\bf 12} (1954) 603. ***

\bibitem{TANI54} S.Tani. Prog. Theor. Phys. {\bf 12} (1954) 104.

\bibitem{NISH56} K.Nishijima. Prog. Theor. Phys. Suppl. {\bf 3} (1956) 138.

\bibitem{GAHY76} M.Gari and H.Hyuga. Z. Phys. {\bf A277} (1976) 291.

\bibitem{MESONS79} Mesons in Nuclei, eds. M.Rho and D.H.Wilkinson
(North - Holland, Amsterdam, 1979).

\bibitem{SHE87} A.V.Shebeko. In: Proc. Intern. Conf. on the Theory of
Few - Body and Quark - Hadronic Systems (Dubna,1987), p. 183.

\bibitem{KOMESHE95} V.V.Kotlyar, Yu.P. Mel'nik and A.V.Shebeko. Phys. Part.
Nucl. {\bf 26} (1995) 79.

\bibitem{GRESCH58} O.Greenberg and S.Schweber. Nuovo Cim. {\bf 8} (1958) 378. *

\bibitem{SCH61}  S.S.Schweber. An Introduction to Relativistic Quantum Field
Theory (Row , Peterson \& Co., New York, 1961).

\bibitem{FAD63} L.D.Faddeev. Dokl. Akad. Nauk USSR {\bf 152} (1963) 573.

\bibitem{FIV70} D.I.Fivel. J. Math. Phys. {\bf 11} (1970) 699.

\bibitem{SHIVI74} M.I.Shirokov and M.M.Visinescu. Rev. Roum. Phys.
{\bf 19} (1974) 461. *

\bibitem{THES97} A.V.Shebeko and M.I.Shirokov. In: Proc. European Conf. on
"Advances in Nuclear Physics and Related Areas" (Thessaloniki, 1997) (in press).

\bibitem{GRON97} A.V.Shebeko and M.I.Shirokov. Nucl.Phys. {\bf A631} (1998) 564c.

\bibitem{GLOMU81} W.Gloeckle and L.Mueller. Phys. Rev. {\bf C23} (1981) 1183.

\bibitem{SATO91}  T.Sato et al. J. Phys. {\bf G17} (1991) 303.

\bibitem{HAGLO92} B.Hamme and W. Gloeckle. Few-Body Syst {\bf 13} (1992) 1.

\bibitem{KO92} A.Yu.Korchin. In: Proc. Nat. Conf. on Physics of Few - Body and
 Quark - Hadronic Systems (Kharkov, 1992), eds. V.Boldyshev, V.Kotlyar and
 A.Shebeko (Kharkov, 1994), p.\ 105.

\bibitem{KOSHE93} A.Yu.Korchin and A.V.Shebeko. Phys. At. Nucl. {\bf 56} (1993)
1663. *

\bibitem{PLGARI92} D.Pluemper and M.F.Gari. Z. Phys. {\bf A343} (1992) 343.

\bibitem{EDGARI96} J.A.Eden and M.F.Gari. Phys. Rev. {\bf C53} (1996) 1510.

\bibitem{LESHE89} L.G.Levchuk and A.V.Shebeko. Sov. J. Nucl. Phys. {\bf 50}
(1989) 607.

\bibitem{LESHE95} L.G.Levchuk and A.V.Shebeko. Phys. At. Nucl. {\bf 58} (1995) 923.

\bibitem{LESHE96} L.G.Levchuk and A.V.Shebeko. In: Proc. of the 12th Intern.
Symposium on High - Energy Spin Physics, eds. C.W. de Jager, T.J.Ketel,
P.J.Mulders, J.E.J. Oberski and M.Oskam - Tamboezer (Amsterdam,1996),p.\ 558.

\bibitem{FUZH96} M.G.Fuda and Y.Zhang. Phys. Rev. {\bf C54} (1996) 495.

\bibitem{SALE96} T.Sato and T.-S.H.Lee. Phys. Rev. {\bf C54} (1996) 2660.


\bibitem{KADMIR72} V.G.Kadyshevsky, R.M.Mir--Kasimov and N.B.Skachkov.
                   Sov. J. Part. Nucl. {\bf 2} (1972) 635.

\bibitem{NAKAN88} N.Nakanishi. Progr. Theor. Phys. Suppl. {\bf 95} (1988) 1.

\bibitem{SATO92} T.Sato et al. Few-Body Syst. Suppl. {\bf 5} (1992) 254.


\bibitem{BEL40} F.J.Belinfante. Physika. {\bf 7} (1940) 305.

\bibitem{ZUB80} C.Itzykson and J.-B.Zuber. Quantum Field Theory
(McGraw-Hill, New York, 1980).

\bibitem{BD64} J.D.Bjorken and S.D.Drell. Relativistic Quantum Mechanics
(McGraw-Hill, New York, 1964).

\bibitem{DIR49}  P.A.M.Dirac. Rev.Mod.Phys. {\bf 21} (1949) 392.

\bibitem{SHE90} A.V.Shebeko. Sov.J.Nucl.Phys. {\bf 52} (1990) 970.

\bibitem{KOMESHE90} A.Yu.Korchin, Yu.P.Mel'nik and A.V.Shebeko. Few-Body
Systems {\bf 9} (1990) 211.

\bibitem{MATT57} P.T.Matthews. The Relativistic Quantum Theory of Elementary
Particle Interactions (Rochester, New York, 1957).

\bibitem{FUZH95} M.G.Fuda and Y.Zhang. Phys.Rev. {\bf C51} (1995) 23.

\bibitem{FOLKRA75}  L.L.Foldy and R.A.Krajcik. Phys.Rev. {\bf D12} (1975) 1700.

\bibitem{FANO71} G.Fano. Mathematical Methods of Quantum Mechanics
(McGraw-Hill, New York, 1971).

\bibitem{NEUM32} J. Von Neumann. Mathematische Grundlagen der Quantenmechanik
(Verlag Von Julius Springer, Berlin, 1932).

\bibitem{HOJO86} R.Horn and C.Johnson. Matrix Analysis
(Cambridge University Press, Cambridge, 1986).

\bibitem{LANK69} P.Lankaster. Theory of Matrices (Academic Press, New York,
1969).

\bibitem{VANHOVE52} L.Van Hove. Physica {\bf 18} (1952) 145.

\bibitem{VONNEU38} J.Von Neumann. Compositio Mathematica {\bf 6} (1938) 1.

\bibitem{HAAG55} R.Haag. Dan.Mat.Fys.Medd. {\bf 29} (1955) No. 12.

\bibitem{ALDU57} Albertoni and F.Duimio. Nuovo Cim. {\bf 6} (1957) 1193.

\bibitem{BLOT90} N.N.Bogolubov, A.A.Logunov, A.I.Oksak and I.T.Todorov.
General Principles of Quantum Field Theory\\
( Kluwer Academic Publ., Dordrecht, 1990 ).

\bibitem{HAAG92} R.Haag. Local Quantum Physics (Springer-Verlag, Berlin, 1992).

\bibitem{DAV73}  A.S.Davydov. Quantum Mechanics ("Nauka", Moscow, 1973).

\bibitem{HAID96} J.Haidenbauer and K.Holinde. Phys.Rev. {\bf C53} (1996) R25.

\bibitem{SAMMA98} F.Sammarruca and R.Machleidt. Few--Body Physics. {\bf 24}
(1998) 87.

\bibitem{SMIR74} V.I.Smirnov. Course of Higher Mathematics V.II
(Nauka, Moscow, 1974).


\end{thebibliography}
\end{document}